\documentclass[pre,longbibliography]{revtex4-1}

\usepackage[latin1]{inputenc}

\usepackage{amsmath,amssymb,latexsym,epsfig,graphics,epsf}
\usepackage{graphics,xcolor}
\usepackage{siunitx,hyperref}
\usepackage{float}

\newcommand{\be}{\begin{equation}}
\newcommand{\ee}{\end{equation}}
\newcommand{\fig}[1]{Fig.~\ref{#1}}
\newcommand{\Fig}[1]{Figure~\ref{#1}}
\newcommand{\eq}[1]{Eq.~\ref{#1}}
\newcommand{\Eq}[1]{Equation~\ref{#1}}

\newcommand{\note}[1]{{#1}}

\usepackage{soul,ulem}

\newcommand{\beginsupplement}{%
	\setcounter{table}{0}
	\renewcommand{\thetable}{S\arabic{table}}%
	\setcounter{figure}{0}
	\renewcommand{\thefigure}{S\arabic{figure}}%
	\setcounter{equation}{0}
	\renewcommand{\theequation}{s\arabic{equation}}%
}

\begin{document}
\title{Predicting nonlinear physical aging of glasses from equilibrium relaxation via the material time}
\author{Birte Riechers*, Lisa A. Roed, Saeed Mehri, Trond S. Ingebrigtsen, Tina Hecksher, Jeppe C. Dyre**, Kristine Niss**}
\affiliation{``Glass and Time'', IMFUFA, Department of Science and Environment, Roskilde University, P. O. Box 260, DK-4000 Roskilde, Denmark\\
{*\textrm{Present address:} Federal Institute of Materials Research and Testing (BAM), Unter den Eichen 87, 12205 Berlin, Germany\\
 ** Corresponding authors. Email: dyre@ruc.dk and kniss@ruc.dk.} }

\date{\today}
 \maketitle
 \section*{Abstract} {\bf The noncrystalline glassy state of matter plays a role in virtually all fields of materials science and offers complementary properties to those of the crystalline counterpart. The caveat of the glassy state is that it is out of equilibrium and therefore exhibits physical aging, i.e., material properties change over time. For half a century the physical aging of glasses has been known to be described well by the material-time concept, although the existence of a material time has never been directly validated. We do this here by successfully predicting the aging of the molecular glass 4-vinyl-1,3-dioxolan-2-one from its linear relaxation behavior. This establishes the defining property of the material time. Via the fluctuation-dissipation theorem, our results imply that physical aging can be predicted from thermal-equilibrium fluctuation data, which is confirmed by computer simulations of a binary liquid mixture.}

\section*{INTRODUCTION}

Physical aging deals with small property changes resulting from
molecular rearrangements
\cite{maz77,str78,kov79,Scherer1986,hod95}. While the aging of a
material is, in practice, often due to chemical degradation, physical
aging does not involve any chemical change. Understanding this type of
aging is crucial for applications of noncrystalline materials such as
oxide glasses \cite{Narayanaswamy1971,Scherer1986,mau16a,mic16}, polymers
\cite{str78,hod95,che07,gra12,can13,roth,mck17}, metallic glasses
\cite{ket15,rut17,kuc18,lut18,ket20}, amorphous pharmaceuticals \cite{vya07},
colloidal suspensions \cite{bon20}, etc. For instance, the performance
of a smartphone display glass substrate is controlled by details of the physical
aging during production \cite{mauro}, and some plastics eventually
become brittle as a result of physical aging \cite{and19}. Noncrystalline or partly noncrystalline states play a role in modern materials science, e.g., in connection with metal-organic frameworks
\cite{Fonseca2021} and high-entropy alloys \cite{Zhao2021}, and physical aging is also important in connection with active matter \cite{fie00,man20,jan21}. Last, it should be mentioned that aging under nanoconfined conditions differs from that of bulk materials \cite{pri09}. The lack of a fundamental understanding of the glassy state and its aging with time influences all branches of materials science, which explains the continued interest in the field from a theoretical point of view \cite{cug94,lut18,hol19,arc20,lul20,jan21}.
 
Describing and predicting physical aging has been a focus of glass
science for many years, yet the subject still presents important challenges
\cite{mck17,arc20}. In this work, we address the concept of a
material (``reduced'') time controlling aging, which was proposed by
Narayanaswamy in 1971 in a paper dealing with the physical aging of
oxide glasses \cite{Narayanaswamy1971}. A closely related formalism
describing polymer aging was developed a few years later by Kovacs and
coworkers \cite{kov79} and, in the 1990s, in the entirely different
context of spin glasses by Cugliandolo and Kurchan \cite{cug94}. The
material-time concept rationalizes several notable aging phenomena
\cite{Tool1946,Narayanaswamy1971,Scherer1986,Dyre2015,McKenna2020}. It
is used routinely in both basic research and applications. The
material-time formalism is generally recognized to describe well the
physical aging of systems subjected to relatively small temperature
variations, but the existence of a material time has never been validated
in direct experiments. We do this here in long-time
experiments on a glass-forming molecular liquid \note{by demonstrating the
fundamental prediction that linear response aging data determine the
nonlinear aging behavior in the intermediate regime involving
temperature variations of a few percent}.

\begin{figure}[bphp!]  \centering
\includegraphics[width=15cm]{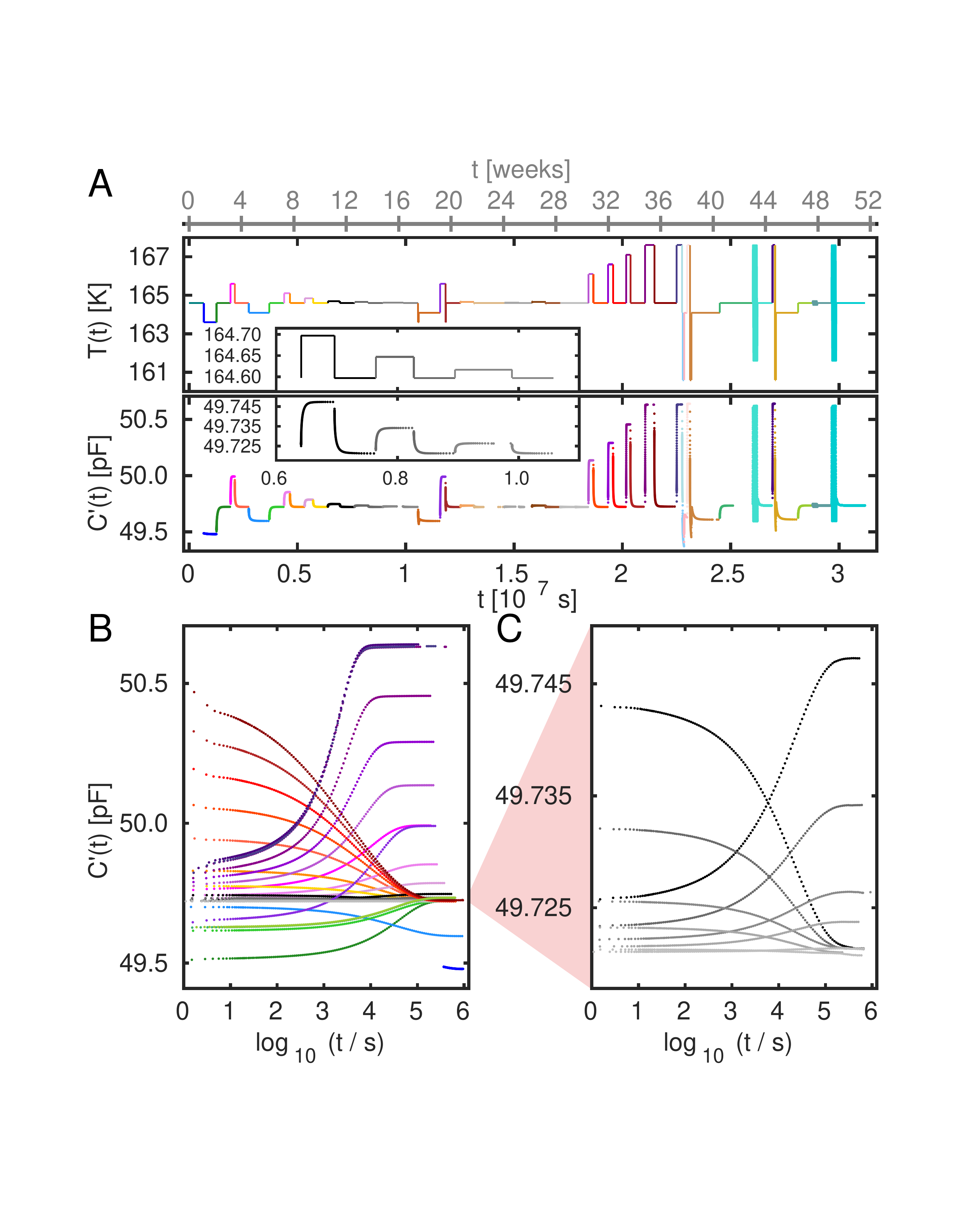}
  \caption{Overview of the temperature protocol and the raw data of
the full experiment on VEC.  A The experimental protocol realized by
temperature modulations around the reference temperature 164.6 K
(upper panel) and the real part of the measured capacitance $C'({\rm
10\,kHz})$ (lower panel), both plotted as functions of time on a
linear scale. Jumps larger than 100~mK are colored while jumps of
100~mK or less are depicted on a grey scale; a selection of the latter
is shown in the inset. The sinusoidal temperature modulations that are also studied
 (see below) are not resolved in this figure where they appear as
turquoise thick vertical lines.  B The capacitance $C'({\rm 10\,kHz})$
data plotted as functions of the logarithm of the time $t$ that has
passed after each jump.  C Magnification on the jumps of
magnitude 100~mK or less.}\label{fig:raw}
\end{figure}

Physical aging is a complex phenomenon as it is both
nonexponential and nonlinear. The simplest and best controlled aging
experiment is based on the temperature jump protocol: The sample is initially in
a state of thermal \note{equilibrium}, then its temperature is changed
instantaneously, i.e., rapidly compared to the response time scale of
the material, and the full approach to equilibrium at the new
temperature is monitored as a function of time
\cite{Hecksher2010}. This procedure requires a setup that allows for fast
temperature changes and has a precise temperature control with a minimal
long-time drift. Moreover, accurate measurements are needed because
the long time tail of physical aging, as well as the entire aging response to
a small temperature step, involves only minute changes of material
properties.

Our experimental setup is based on a Peltier element in
direct contact with a plane-plate capacitor. The setup
keeps temperature constant over months with less than 1 mK variation,
and the samples are so thin (50~$\mu$m) that the temperature may be
changed within a few seconds to a new, constant value. Dielectric
properties are monitored using an ultra-precision Andeen-Hagerling capacitance bridge. More details on the setup are provided in Materials and Methods and in \onlinecite{Igarashi2008b,Hecksher2010,Niss2012,Niss2017,Niss2020}.

\section*{RESULTS}

We performed several temperature jump experiments around a reference
temperature on the glass-forming liquid 4-vinyl-1,3-dioxolan-2-one
(VEC) and monitored, after each jump, both the real and the imaginary
part of the capacitance at 10~kHz as the system gradually equilibrates
\cite{lun05,ric15}. The real part of the VEC data is presented in
Fig.~\ref{fig:raw}; the imaginary part of the data can be found in the
Supplementary Materials in which we also give analogous data for
N-methyl-$\epsilon$-caprolactam (NMEC). Capacitance can be measured
very precisely and is an excellent probe in aging experiments
\cite{Hecksher2010,Lunkenheimer2010,Paluch2013}. For samples of
molecules with a low dipole moment, the real part of the capacitance
provides a direct measure of the density \cite{Niss2012}. The VEC and NMEC
molecules have large dipole moments, which implies that
rotational polarizations contribute substantially to the capacitance
even at high frequencies \cite{Jakobsen2005,Niss2012}. Although this
means that the simple connection to density is lost, the capacitance
still provides a precise probe of the state of the sample during
aging.

\begin{figure}[bphp!]
\includegraphics[width=16cm]{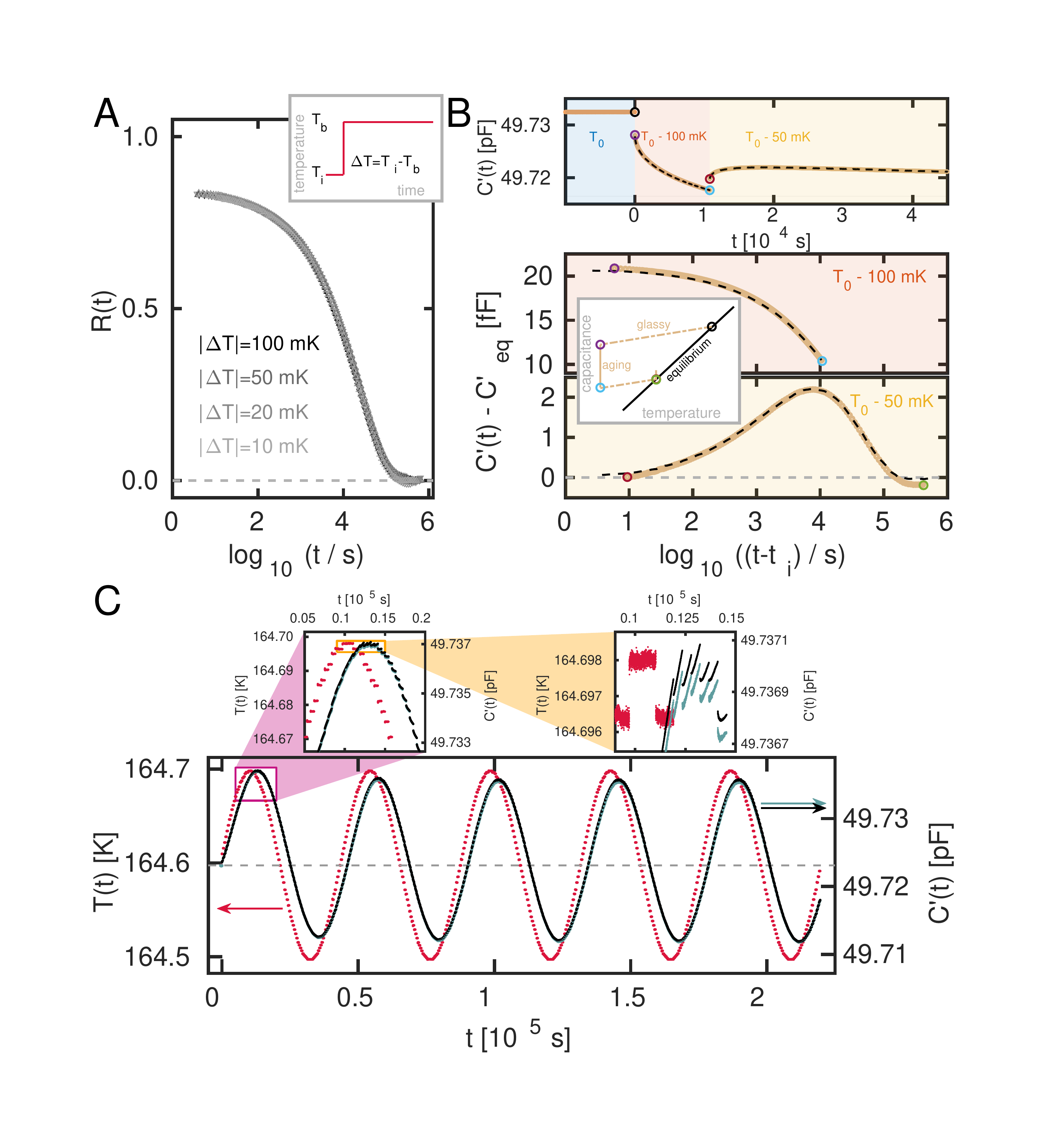}
\caption{Data for the real part of the capacitance, $C'({\rm
10\,kHz})$, from small-amplitude temperature modulation experiments on
VEC along with predictions demonstrating that the response is linear.
A Normalized relaxation function (\eq{R_def}) of single temperature
jumps of amplitude 10~mK-100~mK around the reference temperature $T_0
= 164.6~K$. All data collapse as predicted for linear relaxation.  B
Data from a small-amplitude temperature double jump starting at $T_0 =
164.6~K$ and jumping first by -100~mK and then by +50~mK (colored
curves) along with the prediction according to \eq{eq:doublelin}
(black dashed lines). The upper panel shows the full experiment on a
linear time scale, the lower panel shows the data on a logarithmic
time scale that sets the time of the beginning of each temperature
jump to zero. The inset illustrates the temperature protocol.  C Sinusoidal
small-amplitude temperature protocol (red points) and data (turquoise
points behind the black points). The prediction based on the linear-response formalism
(\eq{eq:lingen}) is shown as black points. The deviations between
prediction and data that can barely be discerned in the main figure 
can be seen in the magnification. }\label{fig:lin}
\end{figure}

\begin{figure}[bphp!]  
\includegraphics[width=16cm]{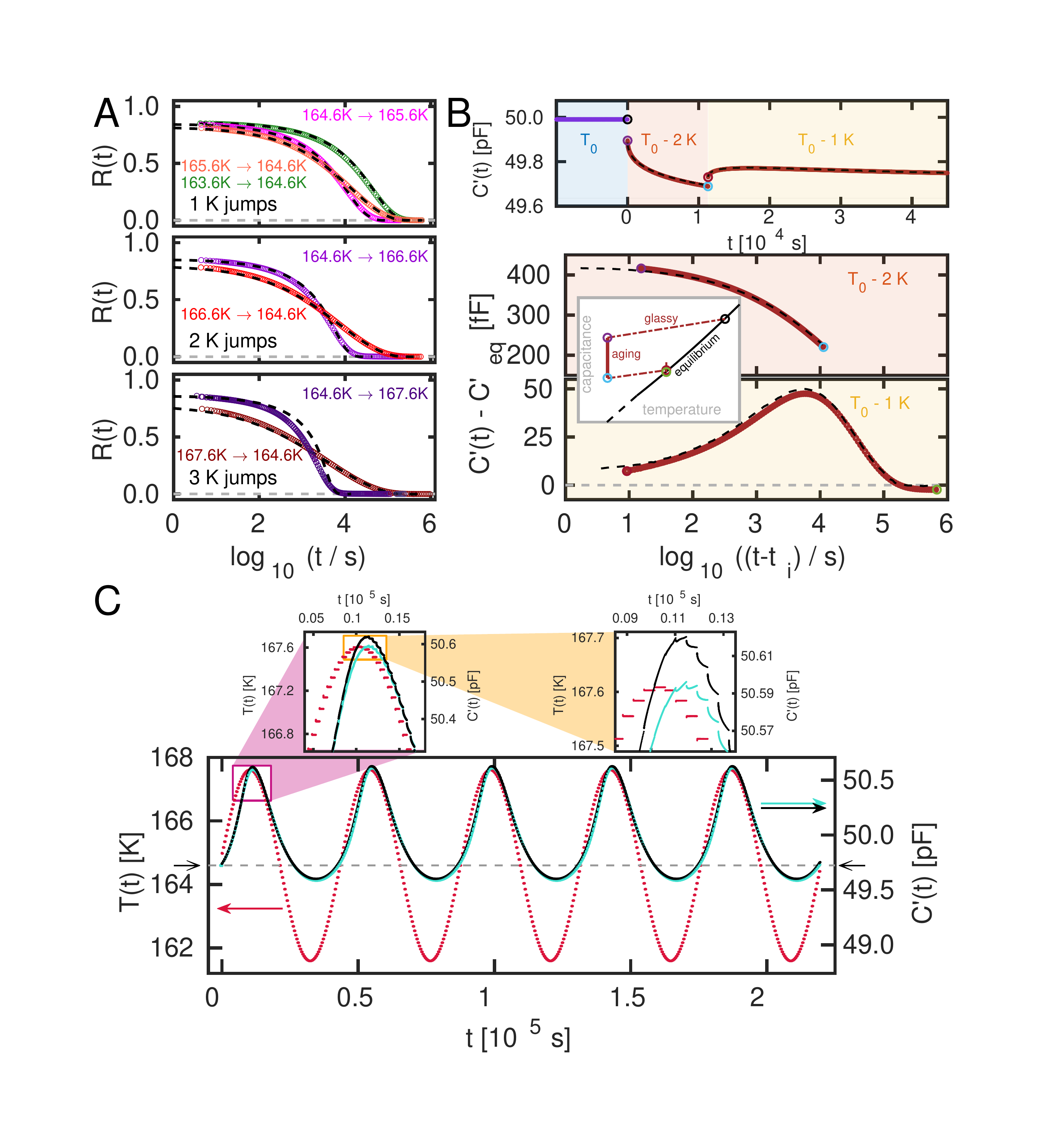}
  \caption{Data from the real part of the capacitance, $C'({\rm
10\,kHz})$, from large-amplitude temperature-modulation experiments on
VEC along with predictions based on the measured linear response
$R_{\rm lin}(t)$, i.e., using \eq{eq:nonlin1} in combination
with \eq{eq:gamma1} and \eq{eq:gamma2}. Colored curves are
data and black dashed lines are predictions.  A Normalized relaxation
functions (\eq{R_def}) of single temperature jumps with amplitudes
ranging from 1~K to 3~K.  B Data from a double jump starting at $T_0 =
167.6~K$ and jumping by -2~K and +1~K (colored curves). The upper
panel shows the full experiment on a linear time scale, the lower
panel shows the data on a logarithmic time scale, setting the time of
the beginning of each temperature jump to zero. The inset illustrates
the temperature protocol.  C Sinusoidal temperature protocol (red points) and data
(turquoise points) shown together. The horizontal dashed line
marks the equilibrium capacitance at the starting temperature
164.6~K. The prediction of the material-time formalism is shown as
black points. The response is highly nonlinear, resulting in a
non-sinusoidal curve that is far from symmetric around the horizontal
dashed line. The deviations between prediction and data that can
barely be discerned in the main panel are visible in the magnification.} 
\label{fig:nonlin}
\end{figure}

The reference temperature for the VEC experiment is 164.6~K at which
the \note{main (alpha)} relaxation time is roughly 12~hours (see the
Supplementary Materials). This is large enough for the setup to
thermalize after a temperature jump before any significant relaxation
has taken place in the sample. Figure \ref{fig:raw}A shows our temperature protocol
with the 10~kHz real part of the capacitance measured as
a function of time. The first 36~weeks of the
experiment were devoted to single temperature jumps with size
varying from 10~mK to 3~K, carefully equilibrating the sample after
each jump before the next one was initiated. The last 15~weeks were
spent on temperature variations involving double jumps and sinusoidal
modulations. The latter are not resolved in this figure, where they
appear as thick turquoise vertical lines; we return to these protocols
later (Figs.~\ref{fig:lin} and
\ref{fig:nonlin}). Figure~\ref{fig:raw}B shows the data for the single
jumps plotted as a function of the logarithm of the time that has passed after each
jump was initiated. Note that these curves have very different
shapes, demonstrating that even fairly small temperature jumps
lead to a notable nonlinear response. This is a hallmark of
physical aging, reflecting the ``asymmetry of approach'' that jumping
to the same final temperature from a higher temperature results in a
faster and more stretched response than the same size jump coming from
below \cite{Kovacs1963,Narayanaswamy1971,Scherer1986,mck17}. Figure
\ref{fig:raw}C focuses on the smaller jumps that are not resolved in
Fig.~\ref{fig:raw}B.

The response to a temperature variation is usually
highly nonlinear. Nevertheless, any response is expected to have a
small-amplitude limit at which the measured quantity, $X(t)$, depends
linearly on the external perturbation. Even in this limit, the
measured signal, in general, depends on the temperature history. 
This means that $X(t)$ in the linear limit is given by a convolution of the change in temperature with the normalized linear time-domain response function $R_{\rm lin}(t)$ in the following
manner
\be X(t)-X_{\rm eq}(T)=-\alpha_X\int_{-\infty}^t R_{\rm
lin}(t-t')\frac{d T}{d t'} dt'\,.
\label{eq:lin1} \ee
Here, the temperature $T$ is in general a function of the time
$t$, $X_{\rm eq}(T)$ is the equilibrium value of the measured property
at temperature $T$, and $\alpha_X={d X_{\rm eq}}/{d T}$ quantifies its
temperature dependence. This linear description is also known as
the Boltzmann superposition principle. In the case of a temperature jump at time zero from the
initial temperature $T_i$ to the ``bath'' temperature $T_b$, one has ${d
T}/{d t'}=-\Delta T\delta(t')$ in which (following the convention in the field) $\Delta T= T_i-T_b$ 
and $\delta(t')$ is the Dirac delta function. \Eq{eq:lin1} implies that the time-dependent response is 
$\alpha_X\Delta T R_{\rm lin}(t)$. Defining $\Delta X=X_{\rm
eq}(T_i)-X_{\rm eq}(T_b)$ and noting that $\Delta X=\alpha_X \Delta
T$, the response is given by
\be X(t)-X_{\rm eq}(T_b)=\Delta X R_{\rm lin}(t) \,.
\ee
We have previously worked with this linear limit for temperature
jumps down to 100~mK \cite{Niss2012,Niss2017,Hecksher2019,Niss2020};
the data of the present paper take this a step further by involving
temperature jumps as small as 10~mK, as well as by optimizing the
protocol to make it possible to properly resolve both the long- and
the short-time plateaus of the linear aging curve.

Linearity is investigated, in general, by considering the normalized relaxation
function of the quantity $X$, denoted by $R(t)$, which for a
jump to temperature $T_b$ at $t=0$ is defined by
\be\label{R_def} R(t) \,=\,\frac{X(t)-X_{\rm eq}(T_b)}{\Delta X}\,.
\ee
$R(0)=1$ and $R(t)$ goes to zero as the system
equilibrates at $T_b$ at long times. Whenever the data are in the linear regime, the
relaxation function is the response function of \eq{eq:lin1},
$R(t)=R_{\rm lin}(t)$, i.e., relaxations following all 
temperature jumps have the same time-dependent normalized
relaxation function in the linear limit.

Figure \ref{fig:lin} shows $R(t)$ for temperature jumps of magnitude
10 to 100~mK to and from the reference temperature, where
$X=C'(10~\text{kHz})$ is the real part of the capacitance at 10 kHz of
VEC. Similar data are shown for the imaginary part and for NMEC in the
Supplementary Materials. The short-time plateau of $R(t)$ is below the
theoretical value $R(0)=1$. This is because there is a fast response
that cannot be resolved by our setup, a common finding in studies of
physical aging \note{that was discussed in detail in previous
  works \cite{Niss2017,Hecksher2019}}. \note{We believe that the fast
  response mainly happens on the phonon time scale due to vibrational
  and librational equilibration. In addition,} one or more beta
relaxations may take place at times shorter than the experimental
cutoff of about 4s (the time it takes to change temperature and
equilibrate the setup at the new temperature). \note{However, in the
  dielectric spectra (fig. S1), the beta relaxation is only seen as a
  small shoulder, which indicates that it probably only accounts for a few
  percent of the initial decay in $R(t)$.}

All the normalized relaxation functions observed for these small
temperature jumps around the reference temperature collapse within the
experimental uncertainty, as predicted for a linear response. True
linearity is a theoretical limit, which means that higher precision
and better resolution would reveal tiny differences between the
relaxation curves. For the data of Fig. ~\ref{fig:lin}, the uncertainty
is of the same order of magnitude as the symbol size and no
differences are resolved, meaning that the measured curves for all
practical purposes represent the linear-response function
$R_{\rm lin}(t)$. In the following, we demonstrate how the
linear response function can be used to predict the response for
different temperature protocols resulting in linear
(Fig.~\ref{fig:lin}) and nonlinear (Fig.~\ref{fig:nonlin})
aging responses.

A simple generalization of the temperature-jump experiment is to
introduce a second jump before the system has equilibrated fully in
response to the first one, a so-called double-jump experiment. If the
temperature changes are both small enough to be within the linear
range, then the measured value of $X(t)$ after the second jump is a sum of
the responses to the individual jumps. For two temperature jumps
corresponding to changes in the measured property $X$ by $\Delta X_1$
and $\Delta X_2$ performed at times $t_1$ and $t_2$ ($t_1<t_2$),
respectively, one has
\be X(t)=\Delta X_1 R_{\rm lin}(t-t_1)+\Delta X_2 R_{\rm
lin}(t-t_2)+X_{\rm eq}(T_2)\,\,\,\textrm{for} \,\,\,t>t_2
\label{eq:doublelin} \ee
where $T_2$ is final temperature after the second jump.

We test \eq{eq:doublelin} for the Ritland-Kovacs crossover protocol
\cite{Scherer1986,Kovacs1963,rit56,son20} consisting of two consecutive
temperature jumps with an opposite sign determined such that the
observable $X$ has its equilibrium value right after the second
jump. Figure~\ref{fig:lin}B illustrates this protocol and shows the
observations after a $-100$~mK jump followed by a $+50$~mK jump for
VEC. The data reproduce the crossover effect that $X(t)$ exhibits a
peak after the second jump \cite{Scherer1986,Kovacs1963}. This bump is
a manifestation of the memory present for any
nonexponential linear response \cite{Scherer1986}. Along with the
data, the predictions based on \eq{eq:doublelin} and the measured
$R_{\rm lin}(t)$ from the 50~mK jump in Fig.~\ref{fig:lin}A are also
shown. The prediction collapses almost exactly with the double-jump
data. These small-amplitude double jump results provide an extra
confirmation that we have reached the linear limit of physical
aging. Similar data are presented for NMEC in the
Supplementary Materials, which also provides data for more VEC small (linear) jumps.

Moving on from the double temperature jump, we note that \eq{eq:lin1}
predicts the response to any temperature perturbation small enough to
be linear. Because we do not have an analytic expression for $R_{\rm
lin}(t)$, the integral is calculated by generalizing the sum in
\eq{eq:doublelin}:
\be X(t)= \sum_{i=1}^N \Delta X_i R_{\rm lin}(t-t_i)+ X_{\rm eq}(T_N)
\,\,\,\textrm{for}\,\,\,t>t_N \label{eq:lingen}\ee
where $T_N$ is the final temperature after $N$ jumps. In
Fig. \ref{fig:lin}C, we show how \eq{eq:lingen} predicts the output of
a small sinusoidal temperature perturbation. The frequency of the
perturbation is $2.3 \cdot 10^{-5}$~Hz, which is the inverse of the
estimated equilibrium relaxation time of the sample at the reference
temperature 164.6 K. The amplitude is 100~mK, i.e., within the linear
regime of single jumps. The prediction follows the data with a high
accuracy, including both the transient behavior (seen, e.g., in a
first peak that is higher than the second) and the phase shift. Tiny
deviations between prediction and data can be seen in the inset, which
also shows how the temperature protocol is composed of 2~mK
temperature steps.

After establishing the linear aging limit and showing how linear
temperature-jump data can be used to predict the response of other
linear temperature protocols, we now turn to the main result of this
paper, a proof of the existence of a material time for VEC. The
radically new idea in the 1970s \cite{Narayanaswamy1971,kov79} was
that aging becomes linear when it is described in terms of the
material time $\xi(t)$ instead of the laboratory time $t$. 
One assumes the so-called time aging-time superposition, meaning that
the spectral shape of $R_{\rm lin}$ is independent of the state of the
sample. As a consequence of these assumptions, \eq{eq:lin1},
\eq{eq:doublelin}, and \eq{eq:lingen} describe also nonlinear
experiments by replacing the laboratory time with the material
time, i.e.,
\be X(\xi)= \sum_{i=1}^N \Delta X_i R_{\rm lin}(\xi-\xi_i)+ X_{\rm
eq}(T_N) \,\,\,\textrm{for}\,\,\,\xi >\xi_N\,.
 \label{eq:nonlin1} 
 \ee
 The material time is ``measured'' by a clock with a rate that
reflects the state of the sample, and the nonlinearity of physical
aging is a consequence of this fact
\cite{Narayanaswamy1971,kov79,cug94,cha02}. The material time may be thought of as
analogous to the proper time in the theory of relativity, which is the
time recorded on a clock following the observer. Although a microscopic definition
of the material time remains elusive, this concept is generally recognized to form the basis of a good
description of physical aging involving relatively small temperature
variations \cite{Scherer1986}. The very fundamental assumption of the
formalism, however, that nonlinear aging phenomena can be predicted
from the linear aging limit, has never been validated. In the
following, we do so by showing how the measured linear response
determines the response to nonlinear temperature protocols for 
VEC and, in the Supplementary Materials, for NMEC.

Using \eq{eq:nonlin1} requires a connection between the laboratory
time $t$ and the material time $\xi$. This is obtained by introducing
the time-dependent aging rate $\gamma(t)$ defined
\cite{Narayanaswamy1971,kov79,Scherer1986,hod95,mic16,can13,svo13} by
\be \gamma(t)=d\xi(t)/dt.
\label{eq:gamma1} \ee
In equilibrium, the aging rate equals the relaxation rate $\gamma_{\rm
eq}$ defined as the inverse of the equilibrium relaxation time. Thus, a
linear experiment is the limiting case for which the
aging rate is constant and the material time is proportional to the
laboratory time, $\xi_{\rm lin}(t) =\gamma_{\rm eq}t$.

Different strategies have been used to estimate $\gamma(t)$ during
aging, often via the so-called fictive temperature
\cite{Tool1946,rit56,Scherer1986,McKenna2020}. We here adopt the
single-parameter-aging ansatz \cite{Narayanaswamy1971,Scherer1986,gra06,Hecksher2015a,Roed2019} according to which the aging rate
is controlled by the measured quantity $X(t)$ itself. In the simplest realization, single-parameter aging is characterized by \cite{Hecksher2015a}
\be\label{eq:gamma2} \log (\gamma(t))-\log (\gamma_{\rm eq}(T))
=\Lambda(X(t)-X_{\rm eq}(T))\,.  \ee
Here, $\gamma_{\rm eq}(T)$ and $X_{\rm eq}(T)$ are the equilibrium
values of $\gamma$ and $X$ at the temperature $T$, and $\Lambda$ is a
constant that depends only on the substance and the
monitored property $X$. It should be noted that \eq{eq:gamma2} is
arrived at by first-order Taylor expansions and, for this reason, can only be expected
to apply for relatively small temperature variations.

The material-time description in \eq{eq:nonlin1}, combined with
\eq{eq:gamma1} and \eq{eq:gamma2}, gives a unique prediction for
$X(t)$ for any temperature protocol. \Eq{eq:nonlin1} predicts the
value of $X(\xi)$ while Eqs.~\ref{eq:gamma1} and \ref{eq:gamma2}
connect the material and laboratory times by stretching or compressing
the time-scale axis. The input needed for the prediction is
$R_{\rm lin}(t)$ as determined in Fig.~\ref{fig:lin}A, the equilibrium
values of the rate $\gamma_{\rm eq}(T)$ and of the measured property
$X_{\rm eq}(T)$, and the parameter $\Lambda$. We have equilibrium
measurements of $X_{\rm eq}(T)$ down to 163.6~K and have extrapolated
values to lower temperatures (see the Supplementary Materials). The
values used for $\gamma_{\rm eq}(T)$ are extrapolations from a fit of
relaxation times derived from dielectric spectra, which, down to
163.6~K, are proportional to the aging rates (see the Supplementary
Materials).

The parameter $\Lambda$ is determined by the method described in
Ref. \onlinecite{Hecksher2015a} from the two temperature-jump experiments
of magnitude $\pm 1$~K to the reference temperature 164.6~K (see the Supplementary
Materials). This $\Lambda$ value was used for 
predicting all other nonlinear responses. Figure~\ref{fig:nonlin}A shows $R(t)$ data from the
nonlinear single temperature jumps. It is seen that the short-time
plateaus of $R(t)$ for the different jumps do not coincide. This is due
to a difference in the short-time relaxation deriving from
\note{the response on the phonon time scale and,} possibly, also from
one or more beta relaxations. To predict the aging, we have
adjusted for this difference in a manner where the short-time decay of
$R(t)$ depends on both the initial and final temperatures (see the
Supplementary Materials).

\Fig{fig:nonlin} reports the main results of the paper: data from
nonlinear temperature protocols along with predictions based on the
linear temperature-jump data. The nonlinear protocols mirror the
linear protocols of \fig{fig:lin}. \Fig{fig:nonlin}A shows single
temperature jumps, \fig{fig:nonlin}B shows a 2~K and 1~K double
jump, and \fig{fig:nonlin}C shows a sinusoidal temperature modulation
with amplitude 3~K and the same frequency as the linear sinusoidal
protocol of \fig{fig:lin}.

The single jumps in Fig. \ref{fig:nonlin}A exhibit the asymmetry of
approach characteristic of nonlinear aging \cite{Kovacs1963,mck17}:
``self-acceleration'' of up jumps where the relaxation rate speeds up
as equilibrium is approached and ``self-retardation'' of down jumps
\cite{can13}. The material-time formalism captures well this
asymmetry (black dashed lines), and the measured data are predicted with a high accuracy
for all down jumps and for up jumps up to 2~K. However, there is a
clearly visible deviation for the largest (3 K) up jump and in the
Supplementary Materials it is documented that deviations in fact emerge
already for a 2.5~K up jump. Thus, the formalism breaks down for large amplitude up jumps. \note{This may be related to only going to first order in the Taylor expansion in
Eq. \ref{eq:gamma2}, but it could also}
be caused by the sample reaching equilibrium by other
mechanisms than the one involved in smaller jumps. This may be similar to what is seen in the case of very large up jumps (30 to 70~K) performed on ultrastable vapor-deposited
glasses where it has been shown that equilibrium is reached by
heterogeneous growth of mobile domains
\cite{Sepulveda2014}. \note{Alternatively, the deviations between data
  and predictions could be caused by beta processes playing a role in aging, 
  as has been seen for polymers deep in the glass state \cite{Cangiaolosi2013,Monnier2021}}.

The predictions agree well with the data of the nonlinear double
jumps shown in Fig. \ref{fig:nonlin}B. This demonstrates that the
material-time formalism works well also in this situation\note{; we
note that Ref. \onlinecite{lun05} presents an alternative approach for
predicting the nonlinear aging response from linear data}. The data
shown in \fig{fig:nonlin}B are all from measurements in the
temperature range where we have access to measured values of $X_{eq}$ and to
the fast contribution of $R(t)$, while $\gamma_{eq}$ used for the prediction
is derived from an extrapolation of higher-temperature dielectric relaxation times. 
The parameter $\Lambda$ is the same as for the single jumps, and the test of the
nonlinear double-jump prediction is therefore performed with no free
parameters. In contrast, in the classical Ritland-Kovacs crossover
experiment \cite{rit56,Kovacs1963,son20}, the first down jump goes deep into
the glass state where the properties of the equilibrium liquid are not
known. In the Supplementary Materials we show data for a large down
jump (7~K); the predictions using extrapolated parameters demonstrate
qualitatively good results, although the formalism is not able to 
predict the time scale of aging in the temperature regime where
equilibrium cannot be reached.

Last, \fig{fig:nonlin}C shows the response of the nonlinear
sinusoidal temperature modulation along with the predictions. The
lowest temperatures in the modulation are in a range where the
parameters are extrapolated. Again there are no free parameters in the
prediction.  The nonlinearity is seen as a sizable asymmetry in
the peak shape: When the temperature is high, there is a substntial
response whereas the liquid responds much less to a decreased
temperature. The gray horizontal dashed line corresponds to the
equilibrium capacitance at the starting temperature 164.6~K. The
asymmetry of the response is very well captured by the
prediction. \note{Because a large part of a sinusoidal is close to
  linear in time, and thus similar to a temperature ramp over several
  Kelvin, the aging in connection with a standard differential scanning calorimetry
  cooling or heating protocol is likewise expected to be predicted
  accurately.}  Deviations between prediction and data can be seen in
the magnifications of \fig{fig:nonlin}C and are most likely related to the
first-order nature of \eq{eq:gamma2}.

\begin{figure}[bphp!]  \centering
\includegraphics[width=17cm]{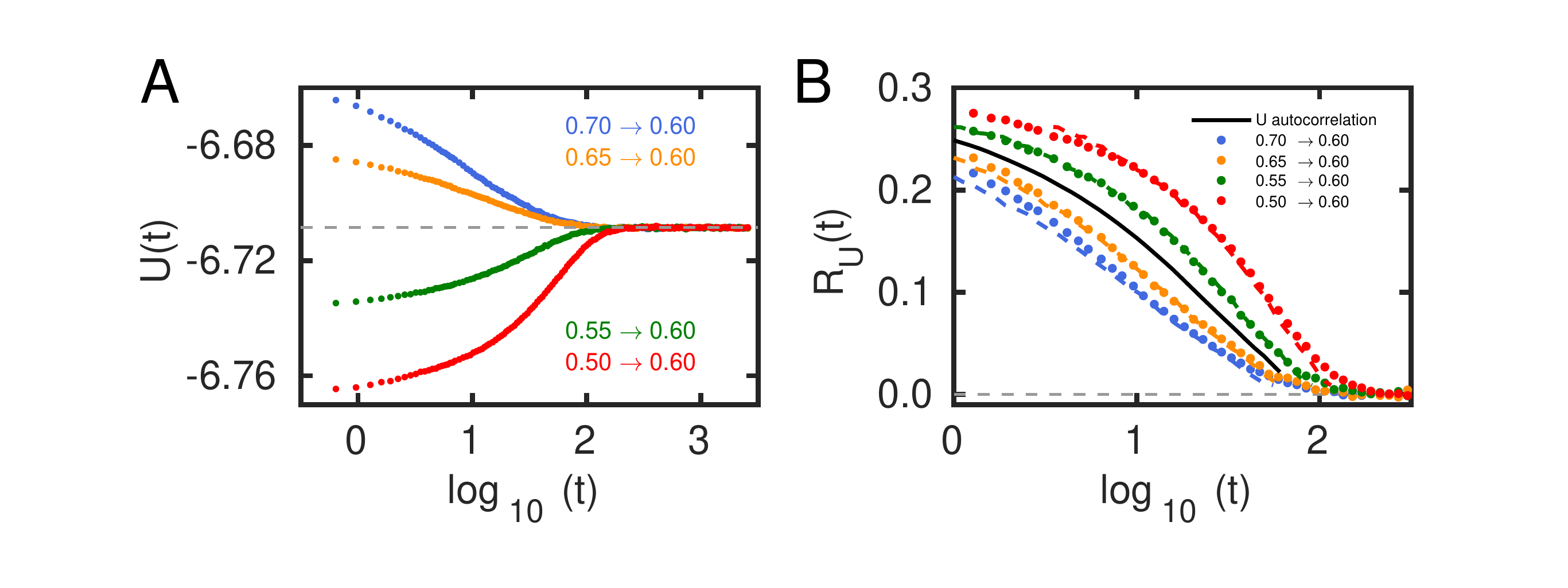}
	\caption{Results from computer simulations of a binary model
liquid monitoring the potential energy $U$.  A shows data for four temperature jumps to the same temperature ($T=0.60$ in units based on the pair-potential parameters).  B shows the normalized relaxation function, $R_U(t)$, of the thermal-equilibrium potential-energy time-autocorrelation function at this temperature (black full line) and the predictions based on this for
the temperature jumps (colored dashed curves). The data for the
normalized relaxation functions based on A are shown as colored
dots. The nonlinearity parameter was determined from the two smallest jumps in the same way as for the experimental data (see the Supplementary Materials).}\label{fig:sim}
\end{figure}

The results in \fig{fig:nonlin} demonstrate that nonlinear
physical-aging phenomena \note{in the intermediate regime} may be
predicted from a knowledge of the linear limit of aging. Previous
works have come close to this limit \cite{rek74,Niss2017}. 
While the linear limit is challenging to probe
experimentally, it is conceptually important. First, it
validates the central assumption of the material-time
formalism. Second, the linear response theory is well established via
the fluctuation-dissipation (FD) theorem that predicts the response
from thermal equilibrium fluctuations quantified via a 
time-correlation function \cite{reichl}. Our results therefore imply
that \note{intermediate} nonlinear physical aging can now, at least
in principle and for relatively small jumps, be predicted from
measurements of the equilibrium fluctuations, i.e., without perturbing
the system at all. We end the paper by illustrating this possibility by presenting results from a computer
simulation where thermal fluctuations are much easier to monitor than in experiments.

The system studied is the binary Lennard-Jones(LJ) mixture of Kob and
Andersen \cite{ka1}, which, for more than 20 years, has been the standard
model for computer simulations of glass-forming liquids. We simulated
a system of 8000 particles. The quantity monitored is the potential
energy $U$. Temperature-jump data were averaged over 1000 simulations
to reduce the noise. \Fig{fig:sim}A shows results for jumps from four
different temperatures to $T=0.60$ (in simulation units), plotted as
a function of the logarithm of the time passed after each
jump was initiated. The curves are quite different, showing that the
jumps are large enough to be notably nonlinear.

The FD theorem implies that the linear response to any small
temperature variation is uniquely determined by the thermal-equilibrium potential-energy
time-autocorrelation function $\langle U(0)U(t)\rangle$
\cite{nie96}. We evaluated this quantity at $T=0.60$. Using the 
single-parameter material-time formalism as above we then predict nonlinear
temperature-jump results (\fig{fig:sim}B). The only free parameter 
is the $\Lambda$ of \eq{eq:gamma2}, which is
determined from the two smallest jumps \cite{Hecksher2015a}. 
The colored dashed curves in \fig{fig:sim}B are the predictions for the
normalized relaxation functions based on the black line in the middle that gives
the thermal equilibrium normalized time autocorrelation function of the potential
energy; the full circles are the normalized data from \fig{fig:sim}A. 
Overall, the predictions work well, demonstrating that
\note{intermediate} nonlinear aging can be predicted from equilibrium
fluctuations. The minor deviations for the two largest jumps are not
unexpected, given that these involve temperature changes of more than
15\% for which the single-parameter ansatz in \eq{eq:gamma2} is likely not to be accurate.

\section*{DISCUSSION}

We have shown how physical aging \note{involving
temperature changes of a few percent} can be predicted from the linear
aging response, i.e., from the response to a very small temperature
variation. This validates the central assumption of the material-time
formalism. At the same time, it is clear that this formalism has
limitations. Thus, the largest up jump (3 K) is not well predicted
(Fig. 3A, bottom). This suggests that there are two regimes of
nonlinear aging: an intermediate regime where the
relaxation time variesfor, at most, a few decades and the material-time
concept describes the situation well, and a strongly nonlinear regime
where the formalism breaks down and a new theoretical approach is
needed. We speculate that even very large temperature down jumps may
fall into the intermediate regime because the system here
thermalizes gradually. This is in contrast to large up jumps, which are
known to result in heterogeneous states very far from equilibrium
\cite{Sepulveda2014}. \note{Aging far below the glass transition is also
likely to deviate from the predictions because processes faster than
the alpha relaxation may play a role here, particularly for
polymers \cite{Cangiaolosi2013,Monnier2021}}. Along this line of thinking, it is important to
note that the standard glass transition resulting from a continuous
cooling is likely to be described well by the material-time formalism,
i.e., is intermediately nonlinear because vitrification for a constant cooling rate takes
place over a narrow range of temperatures. 

In regard to the intermediate aging regime, the
implications of our findings are important both for the understanding
of aging in application and for the theoretical interpretation of the
aging dynamics. By reference to the FD theorem, the consequence is
that the properties governing the \note{intermediate} nonlinear
physical aging of a system far from equilibrium are embedded in the
thermal equilibrium fluctuations and can be predicted from these. This
means that there is no fundamental difference between the 
\note{intermediate} nonlinear and the linear aging responses. \note{
  Understanding physical aging is therefore intimately linked to
  characterizing and understanding the spectral shapes of linear responses
  and autocorrelation functions; a classical field where there has been
  important recent progress both experimentally
  \cite{Korber2020,Pabst2021} and theoretically \cite{Guiselin2021}.
} \note{The approach presented in this paper could prove useful for understanding the nonlinear response to electric fields. This is an active field \cite{Albert2016,Pyenongeun2016,Gabriel2021} in which concepts from physical aging have been used successfully \cite{Pyenongeun2016}. 
	
For future work it would also be interesting to see how far the description of physical aging in terms of linear response can be extended by including higher-order terms in the Taylor expansion of \eq{eq:gamma2}. This can hopefully lead to a complete picture of which samples and protocols exhibit aging governed by the same processes as those responsible of the linear alpha relaxation and which situations involve other processes and mechanisms  \cite{Sepulveda2014,Cangiaolosi2013,Monnier2021}.  }

\section*{MATERIALS AND METHODS}

The study involves the glass-forming liquids
VEC (99\% purity) from Sigma-Aldrich for
which data are shown in the main paper and
NMEC (96\% purity) from VWR for which
data are shown in the Supplementary Materials. Both liquids were stored
in a refrigerator at temperatures between 2 to 8~$^\circ$C and
used as received.

For each liquid a single sample was prepared for all the presented
experiments. The sample cell was a plane-plate capacitor with a
plate distance of 50~$\mu$m and a geometric capacitance of
$C_{geo}=16$~pF. The cell was filled under ambient conditions and
immediately mounted into a precooled cryostat. VEC was quenched to
$T_{cryo}=163$~K, and NMEC was quenched to $T_{cryo}=167$~K, at which the
samples were kept to equilibrate for a couple of days. The temperature
of the main cryostat was constant at $T_{cryo}=164$~K for VEC and
$T_{cryo}=167$~K for NMEC during the
experiments that lasted almost 1~year for each sample.

The temperature control of the experiments was obtained by a
microregulator integrated with the capacitor sample cell. The
regulation was achieved by controlling a Peltier element in contact
with a capacitor plate. Temperature was monitored with a negative temperature coefficient resistor
placed inside one of the capacitor plates. A figure showing the sample
cell with a microregulator can be found in the Supplementary Materials.
Further details on the microregulator and the main cryostat are given
in Ref. \onlinecite{Igarashi2008b}.  The microregulator can change
temperature by steps of a few millikelvin up to several Kelvin within
seconds and keep the temperature constant with variations of less than 1~mK
over weeks. All the temperature protocols shown, including the
sinusoidal protocol, were achieved by making jumps in temperature with
the microregulator.

The real and the imaginary part of the capacitance at 10~kHz was
monitored during the entire experiment with a sampling rate of
approximately one measurement per second. The measurements were
performed using an AH2700A Andeen Hagerling ultra precision
capacitance bridge. It is the combination of the fast and precise
temperature control with the high precision of the bridge that makes it
possible to measure aging in the linear limit.

The simulations used the Kob-Andersen 80/20 binary LJ mixture \cite{ka1}, which was simulated by means of standard $NVT$ Nosé-Hoover dynamics \cite{nose} using the GPU-optimized software RUMD \cite{RUMD}. A system of 8000 particles was simulated. In LJ units the time step was 0.0025. All pair potentials were cut and shifted at 2.5 times the length parameter $\sigma_{ij}$ of the relevant LJ pair potential ($i,j=A,B$).
At the reference temperature $T=0.60$ the potential-energy time-autocorrelation function was calculated as follows. First $10^7$ time steps of simulations were carried out for equilibration. After that, the time-autocorrelation function was calculated using the Fast Fourier Transform. The temperature jump simulations were carried out by the following procedure applied for all starting temperatures. First, $5\times 10^5$ time steps were spent on equilibration at the given starting temperature. After that a total of $5\times 10^8$ time steps were spent on the production runs from which 1000 independent configurations were selected to serve as starting configurations for a temperature jump to $T=0.60$. The Fig. 4 data represent averages over these 1000 jumps.

\section*{AUTHOR CONTRIBUTIONS}

T.H., J.C.D., and K.N. conceptualized the project. 
B.R., L.A.R., T.H., and K.N. initiated the experiments. 
B.R. and L.A.R. carried out the experiments. 
T.H. and K.N. supervised experiments and data treatment. 
B.R. carried out the data analysis.
T.S.I. and S.M. performed the computer simulations.
J.C.D. and K.N. wrote the manuscript.
B.R. wrote the Supplementary Materials with contributions from
J.C.D. and T.S.I.

\section*{ACKNOWLEDGMENTS}
This work was supported by the VILLUM Foundation's \textit{Matter}
grant (No. 16515). All data needed to evaluate the conclusions in the
paper are present in the paper and/or the Supplementary Materials. The
data presented in the paper are available at http://glass.ruc.dk/data/

%


\appendix
\clearpage
\beginsupplement

\section*{Supplementary Information}

This document provides supplementary information and figures complementing the data presented in the main publication. The data are organized into three sections. Sec.~\ref{sec:VPC} contains experimental data on 4-vinyl-1,3-dioxolan-2-one (VEC). Sec.~\ref{sec:NMEC} gives an overview on experimental data on N-methyl-$\epsilon$-caprolactam (NMEC). Sec.~\ref{CompSim} gives details on the computer simulations.

\section{VEC}\label{sec:VPC}
\subsection{Details on the data set measured on VEC}
This section contains 
\begin{itemize}
	\item[a)] Spectra of the storage and loss dielectric permittivity measured by dielectric spectroscopy
	\item[b)] Details on the temperature control
	\item[c)] Data set of the loss capacitance
	\item[d)] Comparisons of the experimental response with predictions for single and double-jump temperature protocols in the linear ($|\Delta T| \leq $\SI{100}{\milli\kelvin}) and nonlinear ($|\Delta T| > $\SI{100}{\milli\kelvin}) regime for the storage and partially also for the loss part of the capacitance.
\end{itemize}
\subsubsection{Spectra of storage and loss dielectric permittivity for VEC}

\begin{figure}[h!]
	\begin{minipage}[t]{0.475\columnwidth}
		\flushleft
		\textbf{a)}
	\end{minipage}	
	\begin{minipage}[t]{0.475\columnwidth}
		\flushleft
		\textbf{b)}
	\end{minipage}
	\includegraphics[width=\columnwidth]{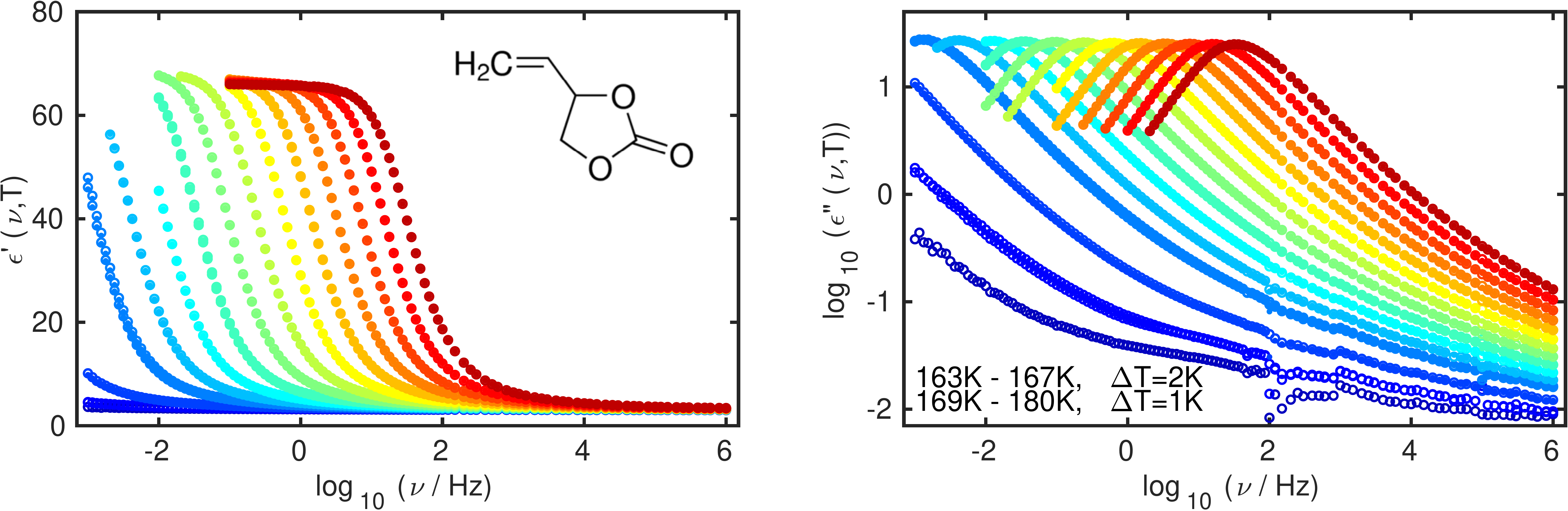}

	\caption{Dielectric spectra for VEC. Storage (a) and loss (b) dielectric permittivity, $\varepsilon'$ and $\varepsilon"$, as functions of frequency measured by a custom-built frequency generator and a commercial LCR meter in the temperature range \SI{163}{\kelvin}-\SI{180}{\kelvin}. The inset in panel a depicts the skeletal formula of VEC.}
	\label{Fig:SpecVPC}
\end{figure}

All data shown for VEC were measured on the same sample.The sample material was inserted into a parallel-plate capacitor with a plate distance of \SI{50}{\micro\meter} and a geometric capacitance of $C_{geo} = $\SI{16.2}{\pico\farad}. The sample was quenched to $T_{cryo} = $\SI{163}{\kelvin} by insertion into the pre-cooled cryostat and was held at that temperature for \SI{60}{\hour}. After that the spectra presented in Fig.~\ref{Fig:SpecVPC} were measured while tracking the temperature-specific voltage of the micro-regulator.

\subsubsection{Temperature control}
\begin{figure}[h!] 	
	\centering
	\includegraphics[width=.7\columnwidth]{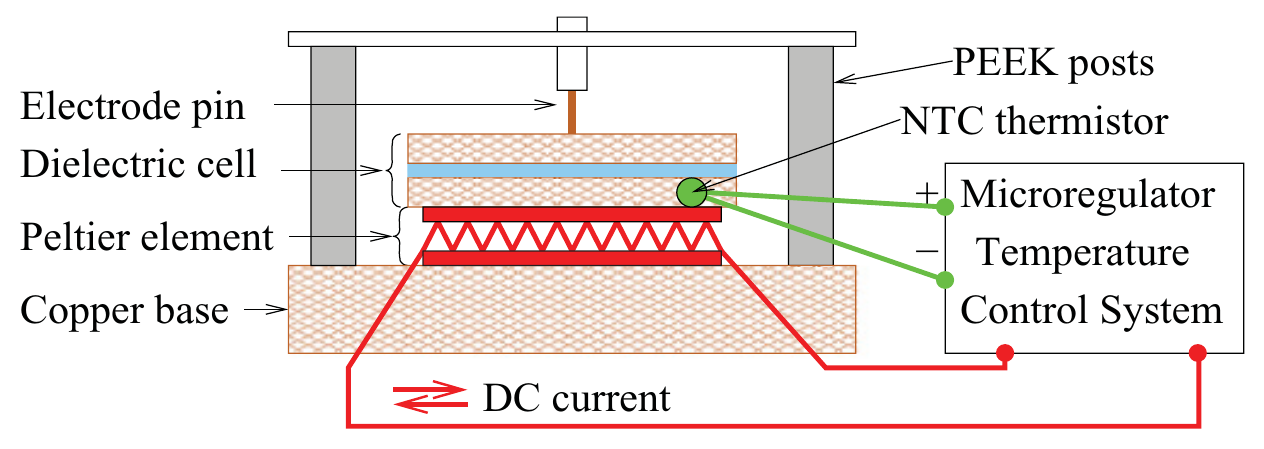}

	\caption{Schematic drawing of the capacitor (dielectric cell) and microregulator mounted in the main cryostat \cite{Hecksher2010}. The liquid is inserted into the \SI{50}{\micro\meter} gap between the disks of the dielectric cell (light blue). The Peltier element heats or cools the sample, depending on the direction of the electrical current powering the element. Details of the cryostat setup and the microregulator are given in Refs.~\onlinecite{Igarashi2008b} and \onlinecite{Igarashi2008x}.}
	\label{Fig:MR}
\end{figure}

The temperature-control of the microregulator was calibrated with the temperature-specific microregulator voltage tracked during the measurements of spectra in the temperature range between \SI{163}{\kelvin} and \SI{180}{\kelvin}. Then the cryostat-controlled temperature was adjusted to $T_{cryo} = $\SI{164}{\kelvin} and the micro-regulator was activated and set to $T = T_{MR} = $\SI{164.5}{\kelvin}. When the sample environment (cryostat) is held at $T_{cryo}$, the microregulator controls the exact sample temperature. Details on the microregulator setup are shown in Fig.~\ref{Fig:MR}. The microregulator can be tuned to ensure fast temperature equilibration, which is achieved after approximately \SI{4}{\second} for all temperature jumps performed on VEC. The microregulator ensures a long-time temperature stability better than \SI{1}{\milli\kelvin}.\\

\subsubsection{Overview of the loss-capacitance data}
The dielectric response was measured with an Andeen-Hagerling ultra-precision capacitance bridge (model 2700A). The capacitance, $C$, and the dielectric loss, $\tan \delta$, were tracked at the frequency $\nu = $\SI{10}{\hertz} with a time resolution of about \SI{1}{\second}.

The data in the main paper was reduced in its resolution by averaging the data over linear and logarithmic bins, with exeption of the initial data points of individual data sets. This averaging does not affect to visual appearance of the data and does thus not affect their evolution or smoothness, but only serves for decreasing the size of resulting image files. The number of data points for individual-jump data is in the order of \num{100} after reduction of the resolution, decreasing the number of data points by up to three orders of magnitude. The most intense zooms in Figs.~2 and 3 in the main paper show data at full resolution.\\

\begin{figure}[h!]
	\centering
	\includegraphics[width=0.9\columnwidth]{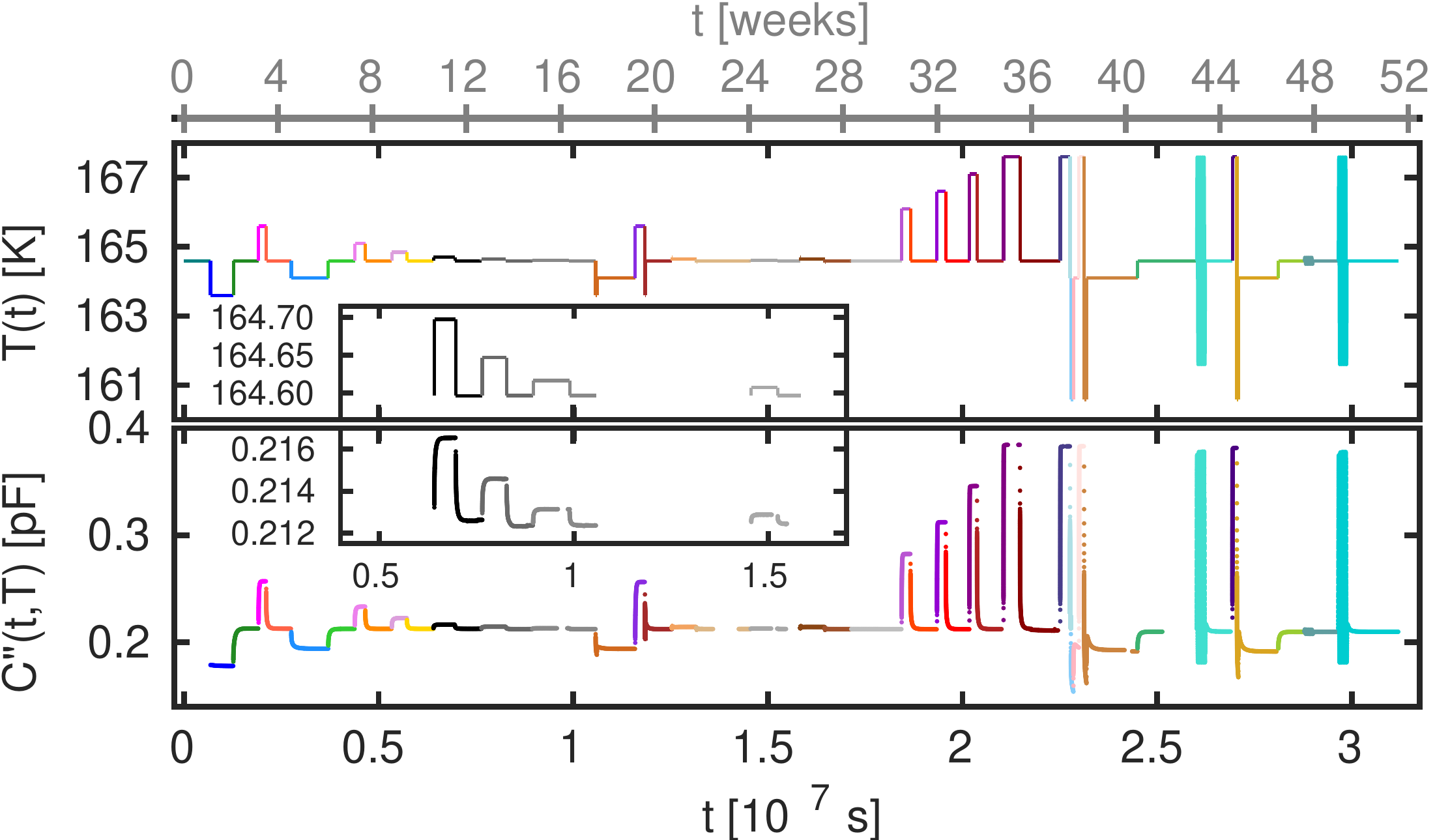}
	
	\caption{Overview of the temperature protocol of consecutive physical-aging experiments realized by temperature jumps facilitated by an NTC-thermistor-regulated Peltier element around $T_{ref} = $\SI{164.6}{\kelvin}, and of the loss capacitance, $C''(\nu = \text{\SI{10}{\kilo\hertz}})$, as functions of time (in seconds on the lower and weeks on the upper x-axis). Nonlinear aging data are colored while linear data are depicted on grey scales. The insets show single jumps with temperature amplitudes of \SI{100}{\milli\kelvin} or less with a zoom on the y-axes.}
	\label{Fig:TLossVPC}
\end{figure}

Individual temperature jumps with amplitudes ranging from $|\Delta T| = $\SI{2}{\milli\kelvin} to \SI{3.5}{\kelvin} were studied, as well as double-jump protocols with $\Delta T_1 = $-\SI{100}{\milli\kelvin}, +\SI{100}{\milli\kelvin}, +\SI{1}{\kelvin}, +\SI{2}{\kelvin} or +\SI{7}{\kelvin} and $\Delta T_2 = -\Delta T_1/2$, and multi-jump temperature protocols resembling an overall sinusoidal temperature variation with amplitudes of \SI{100}{\milli\kelvin} and \SI{3}{\kelvin}. The full temperature protocol is shown in Fig.~\ref{Fig:TLossVPC} together with the corresponding loss capacitance. The loss capacitance data are plotted with the original, as-measured resolution.\\

The main paper only considers the case in which the measured quantity $X$ is the real part of the capacitance at 10kHz, so there we denoted the normalized relaxation function by simply $R(t)$. The present document discusses also the case of $X$ being the imaginary part of the capacitance for which reason we define the general normalized relaxation function as follows

\be
R_X (t) = \frac{X(t) - X_{eq} (T_b)}{X_{eq}(T_i) - X_{eq}(T_b)} = \frac{\Delta X(t)}{\Delta X}\,.
\ee

\begin{figure}[h!]
	\begin{minipage}[t]{0.475\columnwidth}
		\flushleft
		\textbf{a)}
	\end{minipage}	
	\begin{minipage}[t]{0.475\columnwidth}
		\flushleft
		\textbf{b)}
	\end{minipage}
	\centering
	\includegraphics[width=\columnwidth]{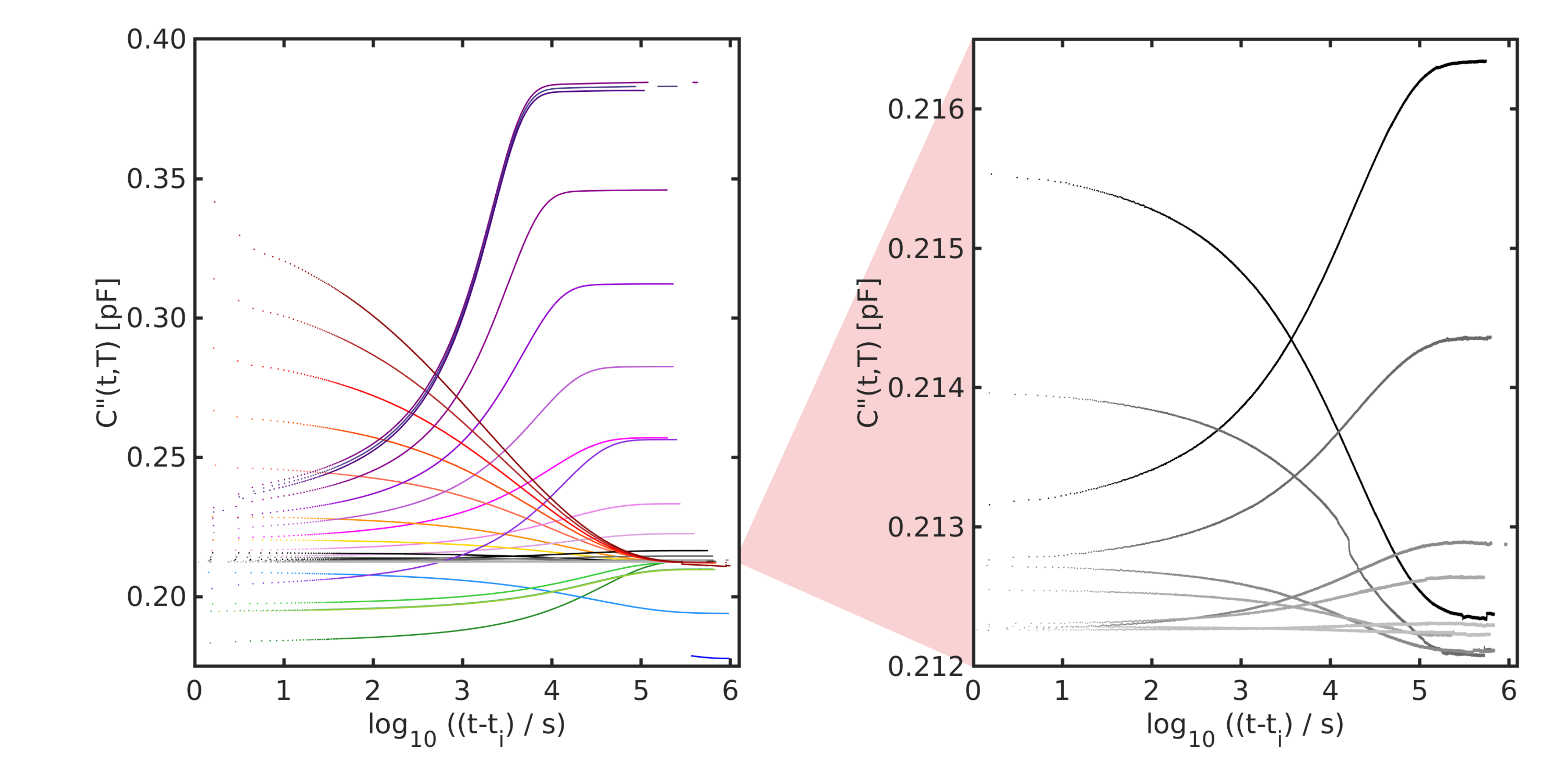}
	
	\caption{Loss capacitance data, $C"(\nu = $\SI{10}{\kilo\hertz}$)$, plotted as functions of the logarithm of the time that has passed since the initiation of a jump at $t=t_i$. Panel (a) depicts all single-temperature-jump data, panel (b) is a zoom on the jumps ranging from a \SI{2}{\milli\kelvin} to a \SI{100}{\milli\kelvin} temperature amplitude.}
	\label{Fig:LossJumpsVPC}
\end{figure}

\begin{figure}[h!]
	\centering
	\includegraphics[width=0.725\columnwidth]{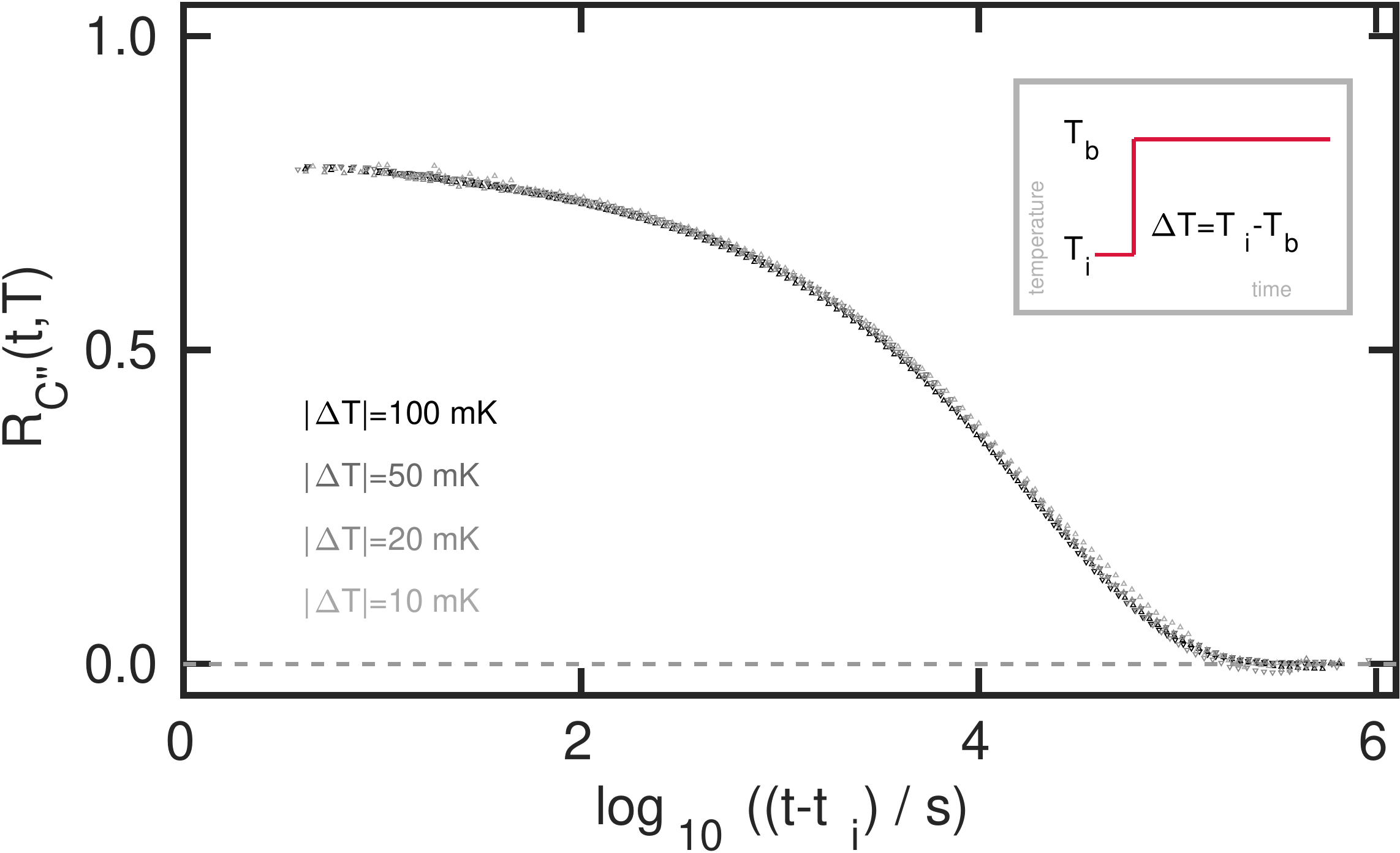}
	
	\caption{Normalized relaxation function for the loss part of the capacitance, $R_{C"}$, for the single temperature jumps with amplitudes between \SI{10}{\milli\kelvin} and \SI{100}{\milli\kelvin}, plotted as functions of the logarithm of the time that has elapsed since the initiation of a jump at $t=t_i$.}
	\label{Fig:LossRlinVPC}
\end{figure}

The single temperature jumps of the loss capacitance are plotted in Fig.~\ref{Fig:LossJumpsVPC} as functions of the logarithm of the time that has passed since the initiation of the jump, denoted by $t-t_i$. To visualize the linearity of the jumps from Fig~\ref{Fig:LossJumpsVPC}b, the normalized relaxation function is plotted in Fig.~\ref{Fig:LossRlinVPC}. The collapse of the data of all jumps with temperature amplitudes below \SI{100}{\milli\kelvin} demonstrates a linear-response regime (at the experimental accuracy). As jumps of \SI{2}{\milli\kelvin} show a considerably lower signal-to-noise ratio because the setup is here pushed closer to its limit in temperature control, only jumps with $|\Delta T| \geq $\SI{10}{\milli\kelvin} are depicted. Also the \SI{50}{\milli\kelvin} down-jump is omitted in this plot.

\clearpage

\subsubsection{Comparison of experimental data to predictions for VEC}
\begin{center}
	\textbf{Storage contribution of dielectric capacitance}\\
\end{center}

\begin{figure}[h!]	
	\begin{minipage}[t]{0.475\columnwidth}
		\flushleft
		\textbf{a)}
	\end{minipage}	
	\begin{minipage}[t]{0.475\columnwidth}
		\flushleft
		\textbf{b)}
	\end{minipage}	
	\includegraphics[width=.825\columnwidth]{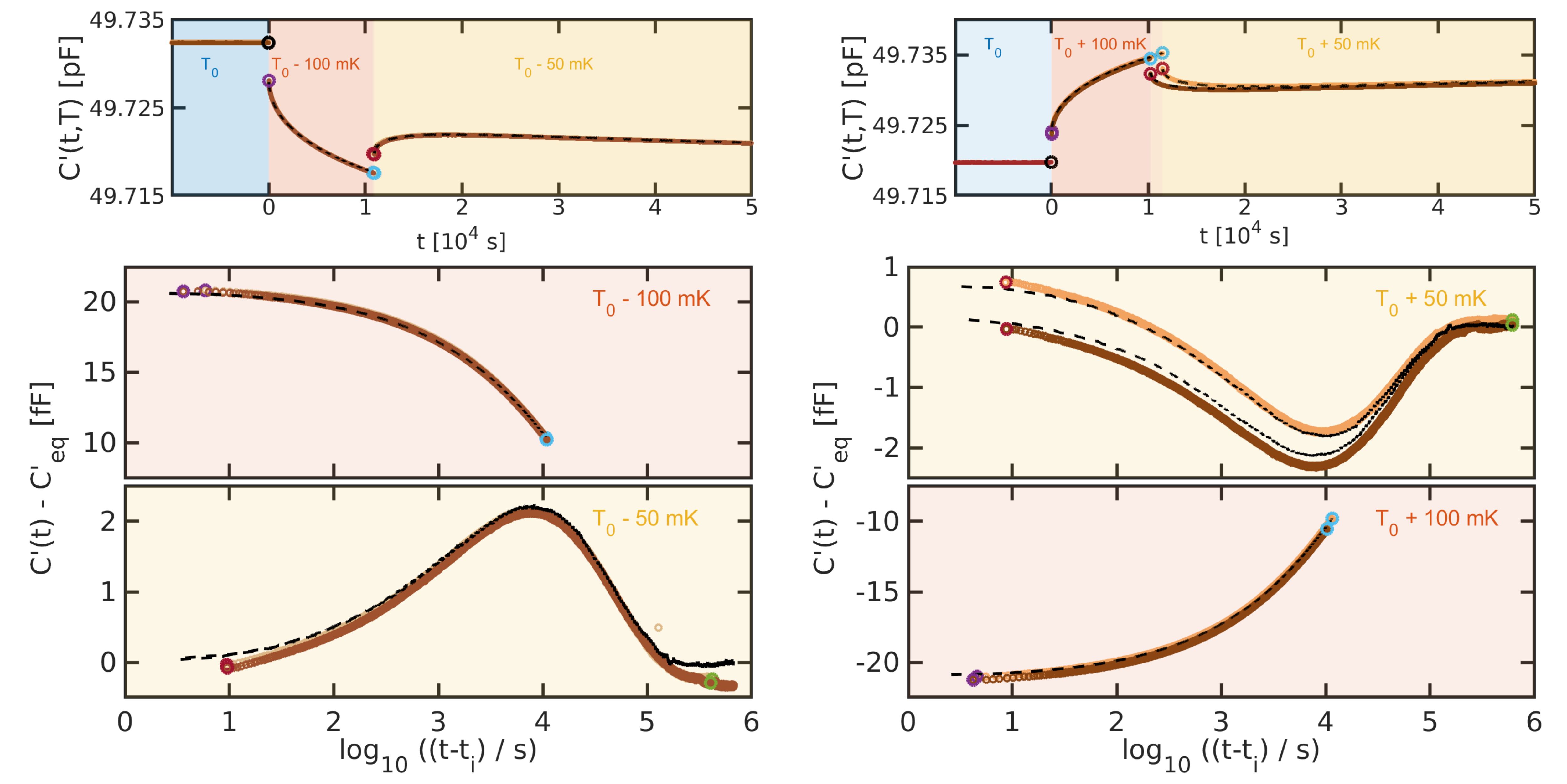}
	
	\caption{Experimental data and predictions of linear double jumps on VEC involving (a) temperature jumps of $\Delta T_1 = $+\SI{100}{\milli\kelvin} and $\Delta T_2 = $-\SI{50}{\milli\kelvin} and (b) $\Delta T_1 = $-\SI{100}{\milli\kelvin} and $\Delta T_2 = $+\SI{50}{\milli\kelvin}, each showing data for two realizations. Upper panels show data plotted against linear time, middle and lower panels depict data as functions of the logarithm of the time that has elapse since the initiation of a jump at $t=t_i$. No corrections were applied to yield the match between experimental data and prediction.}
	\label{Fig:LinDJsVPC}
\end{figure}

\begin{figure}[h!]
	\centering
	\includegraphics[width=.95\columnwidth]{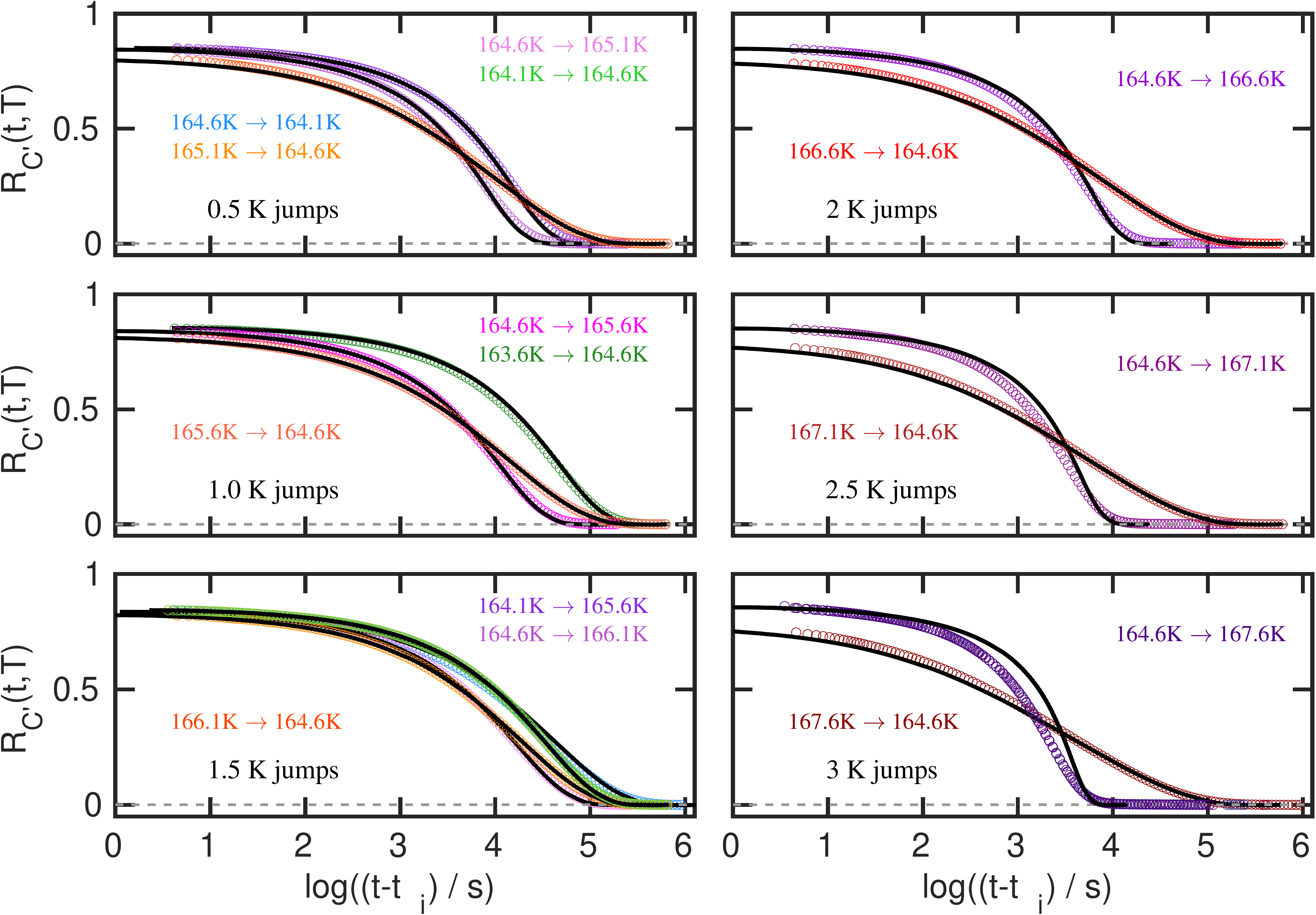}
	
	\caption{Normalized relaxation function based on the measured storage capacitance and predictions of nonlinear individual jumps on VEC involving temperature jumps between \SI{0.5}{\kelvin} and \SI{3}{\kelvin}. The full black lines give the theoretical prediction.}
	\label{Fig:NL_VPC}
\end{figure}

\begin{figure}[h!]
	
	\begin{minipage}[t]{0.475\columnwidth}
		\flushleft
		\textbf{a)}
	\end{minipage}	
	\begin{minipage}[t]{0.475\columnwidth}
		\flushleft
		\textbf{b)}
	\end{minipage}
	
	\includegraphics[width=.85\columnwidth]{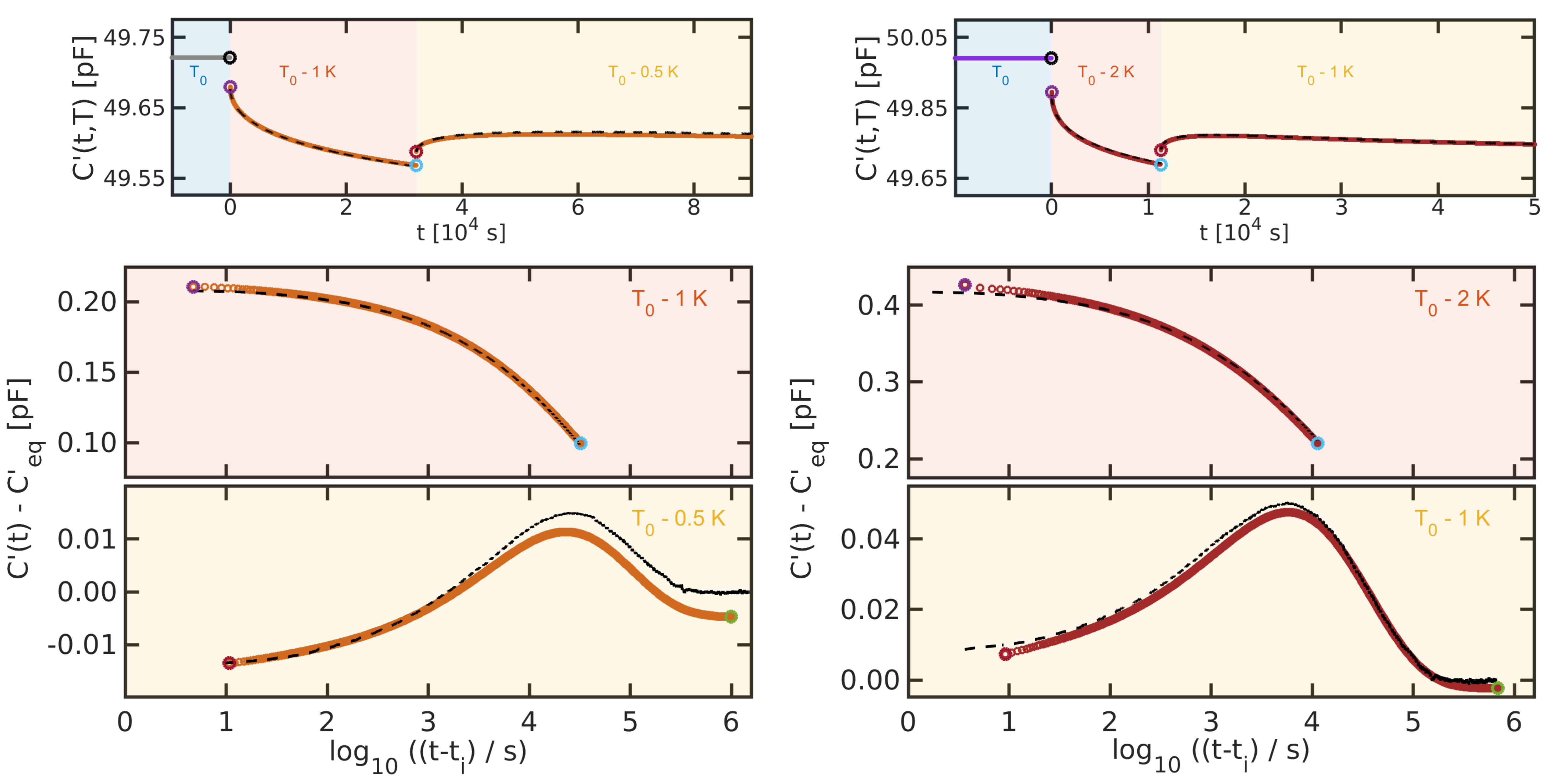}	
	
	\begin{minipage}[t]{.475\columnwidth}
		\flushleft
		\textbf{c)}
	\end{minipage}
	\begin{minipage}[t]{0.475\columnwidth}
		\flushleft
		\textbf{d)}
	\end{minipage}
	
	\begin{minipage}{0.425\columnwidth}
		\includegraphics[trim = 0cm 0cm 25.5cm 0cm, clip=true, width=\columnwidth]{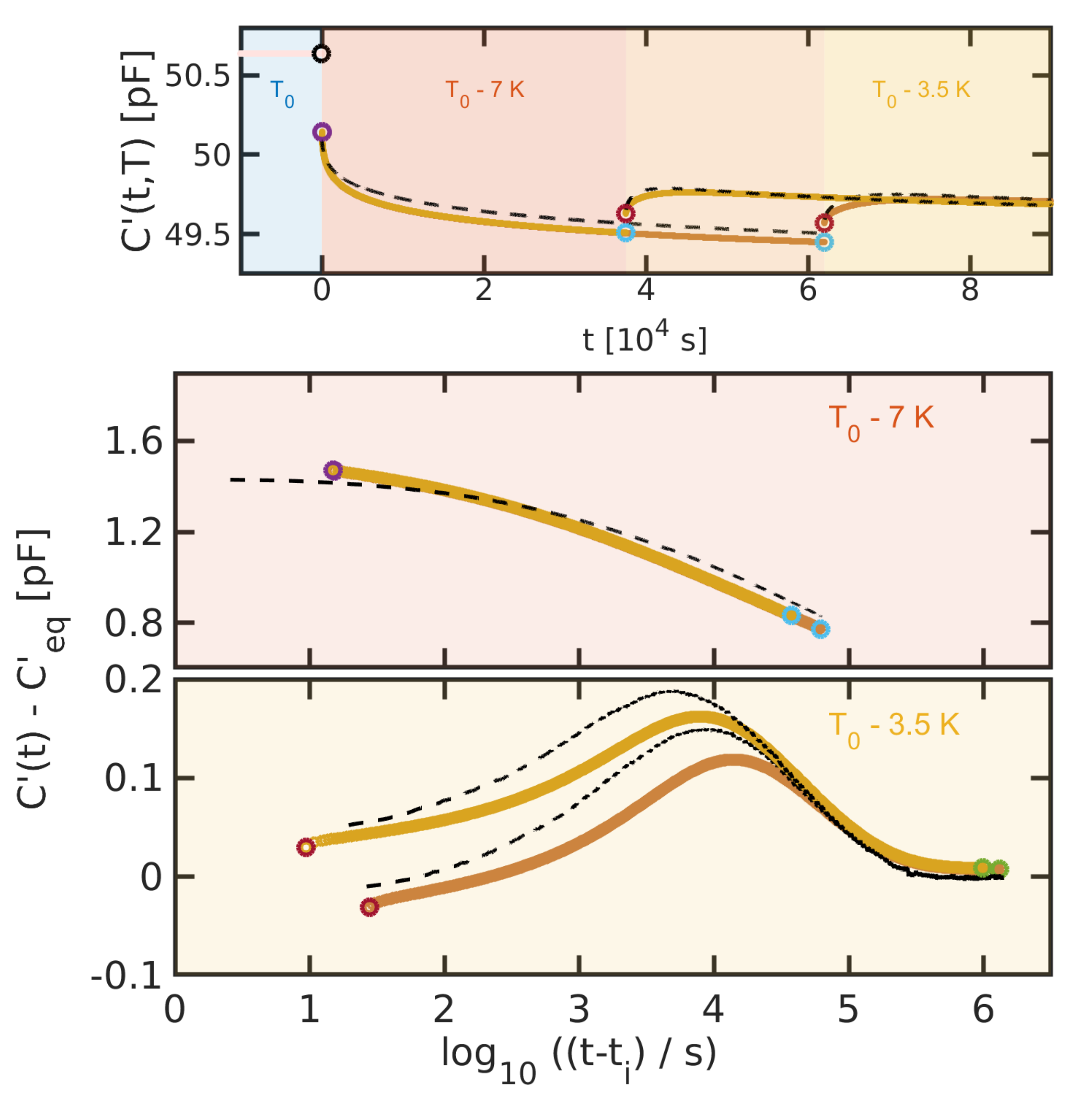}
	\end{minipage}
	\begin{minipage}{0.425\columnwidth}
		\includegraphics[trim = 0cm 0cm 25.5cm 0cm, clip=true, width=\columnwidth]{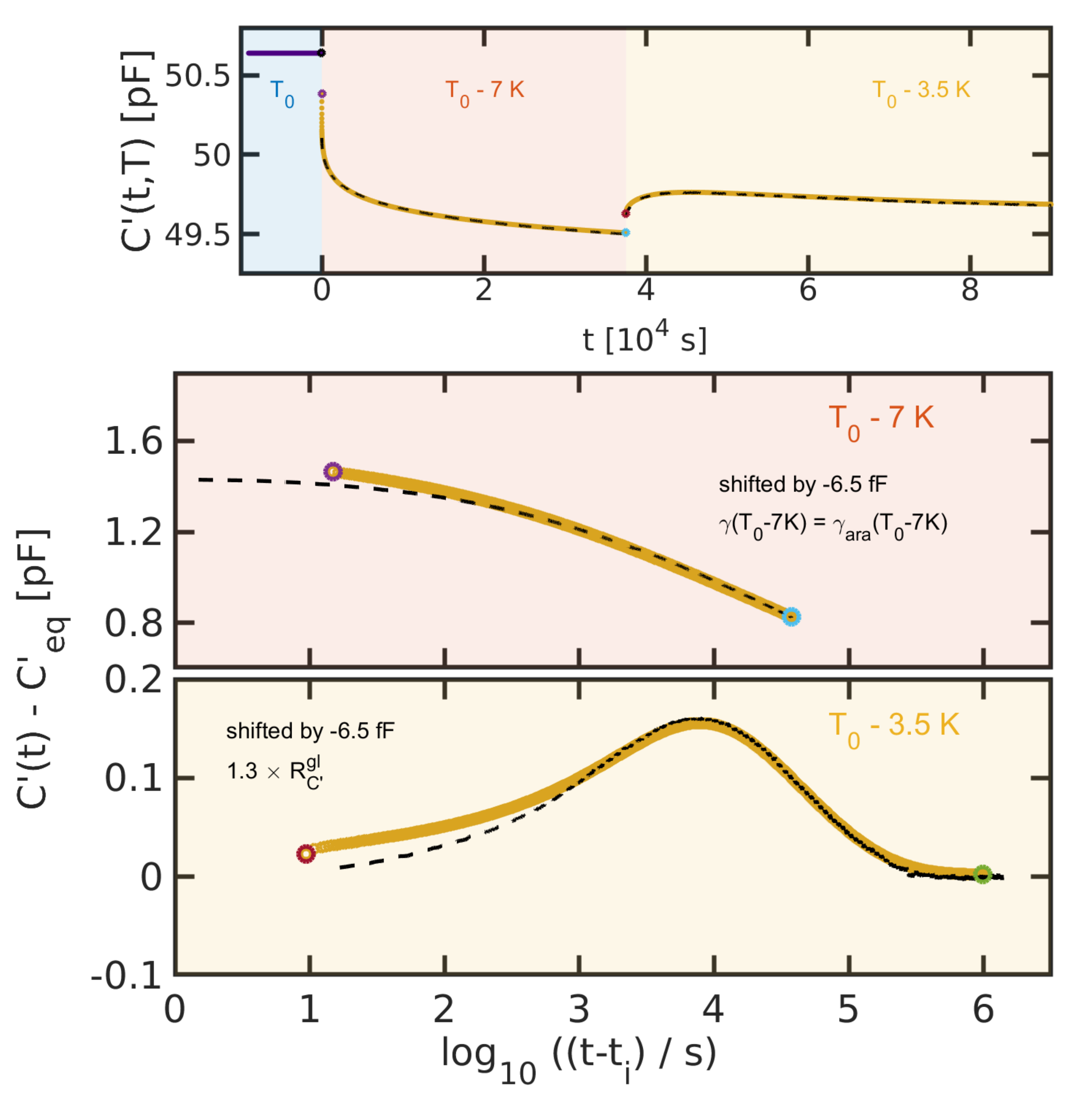}
	\end{minipage}
	
	\caption{Experimental data and predictions of nonlinear double jumps on VEC involving (a) temperature jumps of $\Delta T_1 = $+\SI{1}{\kelvin} and $\Delta T_2 = $-\SI{0.5}{\kelvin}, (b) $\Delta T_1 = $+\SI{2}{\kelvin} and $\Delta T_2 = $-\SI{1}{\kelvin}, (c) $\Delta T_1 = $+\SI{7}{\kelvin} and $\Delta T_2 = $-\SI{3.5}{\kelvin} for two realizations. The predictions are based on inter- and extrapolation of the glassy contribution  $X_{gl}$, the equilibrium response $X_{eq}$, $\Lambda$, and the equilibrium clock rate $\gamma_{eq}$. Note that for the predictions of the double jump with $\Delta T_1 = $+\SI{7}{\kelvin} shown in panel (c) the aforementioned quantities had to be extrapolated, while the predictions of double jumps in panels (a) and (b) are based on interpolations. For more details on how the predictions are derived, see section~\ref{sec:detDJpred}. No corrections were applied to yield the predictions shown in panels (a) to (c). In panel (d), the data were shifted by -\SI{6.5}{\femto\farad} along the y-axis to match to zero in the long-time limit of the second jump, and corrections for the clock rate and the glassy contribution of the second jump were applied.}
	\label{Fig:NLDJsVPC}
\end{figure}

Figures~\ref{Fig:LinDJsVPC} to~\ref{Fig:NLDJsVPC} show comparisons of experimental data in terms of the storage contribution of the dielectric capacitance, $C'$, and predictions derived from the storage contribution of the response to a linear \SI{50}{\milli\kelvin} down-jump.\\

\clearpage
In Fig.~\ref{Fig:LinDJsVPC}, experimental data and predictions are plotted following double-jump temperature protocols in the linear regime. One of the realizations in Fig.~\ref{Fig:LinDJsVPC}a is included in the main manuscript; the accurate match with the second realization confirms the achieved results. The double-jumps in Fig.~\ref{Fig:LinDJsVPC}b follow an inverted Ritland-Kovacs crossover protocol, starting with a temperature up jump followed by a down jump.

Note that the predictions capture the details of the responses with a high accuracy: While the second jump is initiated at precisely the same point for the two realizations in Fig.~\ref{Fig:LinDJsVPC}a, a difference in the initiation of the second jump of the two realizations shown in Fig.~\ref{Fig:LinDJsVPC}b is observed. This leads to different levels in the short-time response at $T = T_0 + $\SI{50}{\milli\kelvin}, which are captured with a high accuracy by the predictions.\\

The predictions and experimental responses to single temperature jumps with amplitudes larger than \SI{100}{\milli\kelvin}, i.e., nonlinear individual jumps, are plotted in Fig.~\ref{Fig:NL_VPC}. In addition to the data of the main manuscript, three additional jump amplitudes are presented here: $|\Delta T| = $\SI{0.5}{\kelvin}, $|\Delta T| = $\SI{1.5}{\kelvin}, and $|\Delta T| = $\SI{2.5}{\kelvin}. The details of the aging responses are captured accurately for all down jumps, and for up jumps with $|\Delta T| \leq $\SI{1}{\kelvin}, while the predictions show increasing deviations from the experimental data for larger up jumps. This signals a beginning deviation from the TN formalism that is known to work best for relatively small temperature variations. One possible scenario that qualitatively matches the observed behavior is the initiation of a heterogeneous growth process, as observed for very large temperature up-jumps for ultrastable vapor-deposited glasses.

In Fig.~\ref{Fig:NLDJsVPC}, experimental data and predictions are plotted following double-jump temperature protocols in the nonlinear regime. In addition to the nonlinear double-jump that is included in the main manuscript, two more nonlinear double-jumps are presented here with temperature amplitudes $\Delta T_1 =$+\SI{1}{\kelvin} in Fig.~\ref{Fig:NLDJsVPC}a  and $\Delta T_1 =$+\SI{7}{\kelvin} in Fig.~\ref{Fig:NLDJsVPC}c. Panels a to c show experimental data and predictions without any corrections.

From these three realizations of nonlinear double-jumps, only the \SI{7}{\kelvin} double-jump exceeds the temperature regime for which equilibrium-data can be interpolated; its predictions are based on extrapolations far (\SI{3}{\kelvin}) below the known equilibrium values. Thus it is not surprising that the prediction for the \SI{7}{\kelvin} double-jump deviates from the experimental data. To check under which parameter conditions predictions and experimental data coincide, we varied both the initial plateau of the second jump and the clock rate of the initial jump after making a minor adjustment of the overall data by shifting it by -\SI{6.5}{\femto\farad} along the y-axis to match to zero in the long-time limit of the second jump. The outcome is shown in Fig.~\ref{Fig:NLDJsVPC}d. The glassy contribution of the second jump was forced to match the data by multiplying the extrapolated parameter by \num{1.2} (visualized as grey diamond in Fig.~\ref{Fig:GlassR_VPC}). For the extrapolation of the clock rate the Avramov function was applied instead of the Vogel-Fulcher-Tammann (VFT) extrapolation, differing by a factor of \num{1.72} compared to the VFT-extrapolation (visualized as a black circle in Fig.~\ref{Fig:GammaN_VPC}). See  Sec.~\ref{sec:fittedParams} for details for inter- and extrapolations of response-related parameters.

\clearpage
\begin{center}
	\textbf{Loss contribution of dielectric capacitance}\\
\end{center}

\begin{figure}[h!]
	\centering
	\includegraphics[width=\columnwidth]{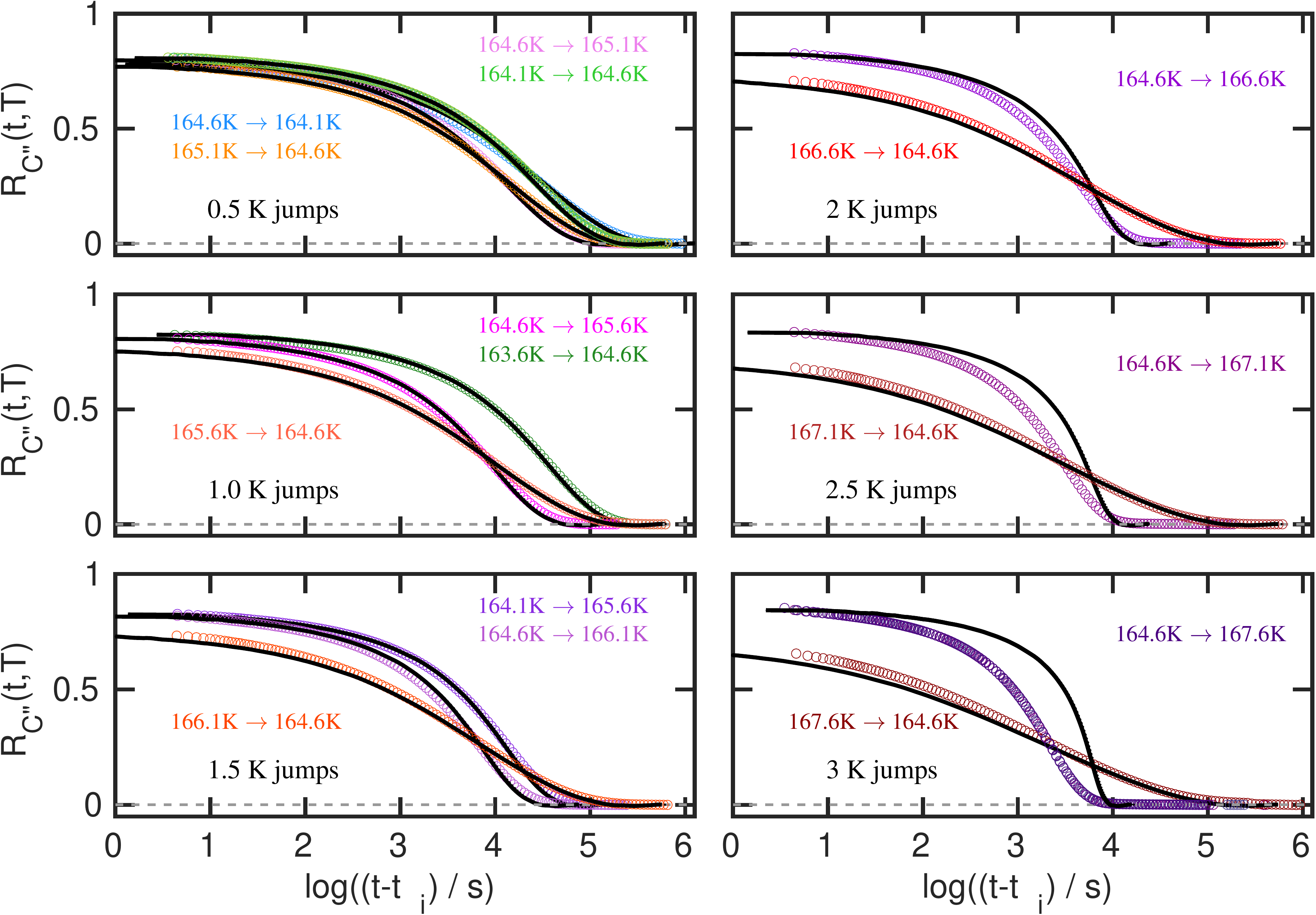}
	
	\caption{Normalized relaxation function based on the measured loss capacitance and predictions of nonlinear individual jumps on VEC involving temperature jumps between \SI{0.5}{\kelvin} and \SI{3}{\kelvin}. The full black lines give the prediction.}
	\label{Fig:NL_VPC_lo}
\end{figure}

\begin{figure}[h!]
	\centering
	\includegraphics[width=\columnwidth]{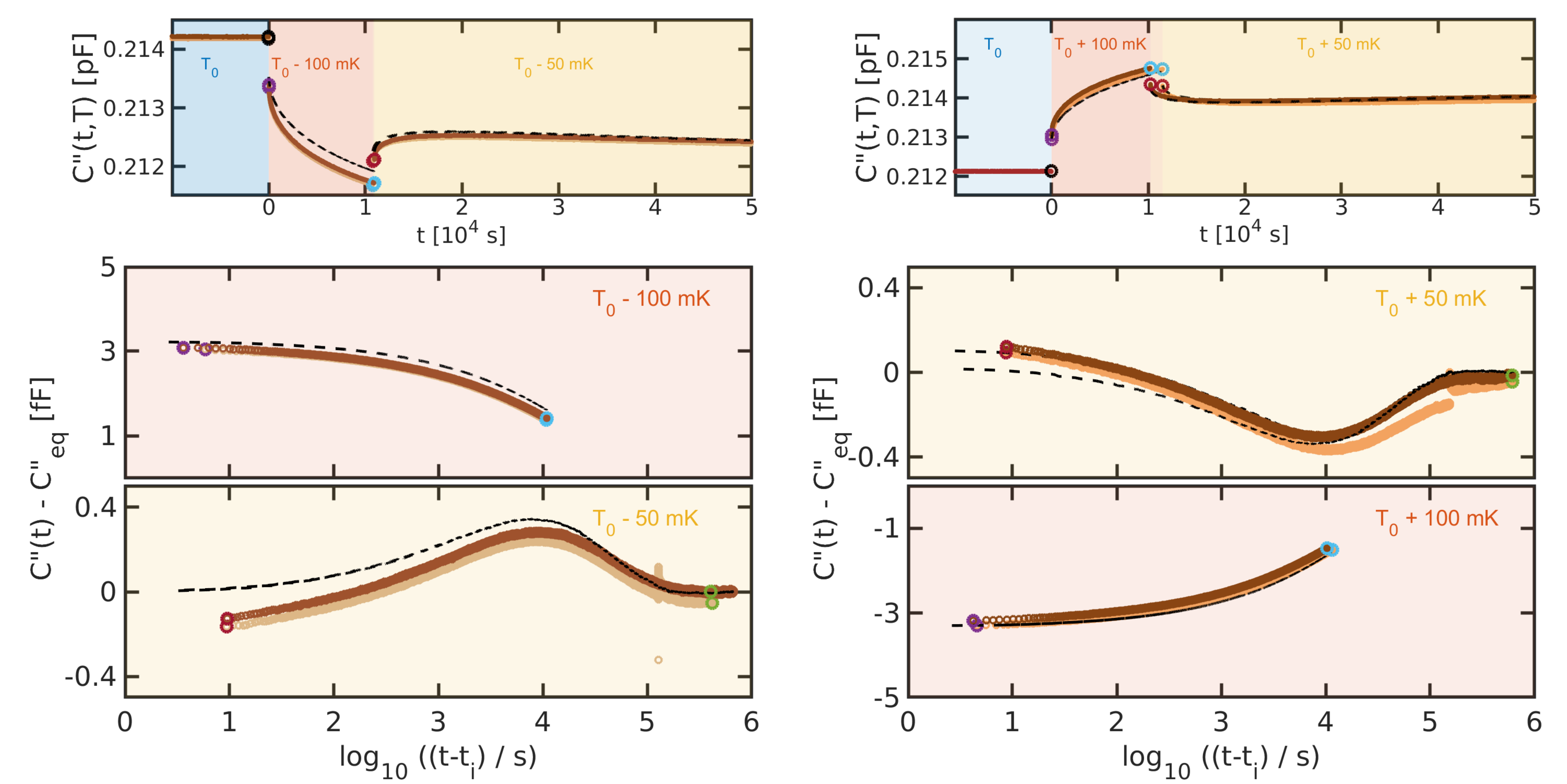}
	
	\caption{Experimental data and predictions of linear double jumps on VEC involving (a) temperature jumps of $\Delta T_1 = $+\SI{100}{\milli\kelvin} and $\Delta T_2 = $-\SI{50}{\milli\kelvin} and (b) $\Delta T_1 = $-\SI{100}{\milli\kelvin} and $\Delta T_2 = $+\SI{50}{\milli\kelvin}, each showing data for two realizations. Upper panels show data plotted against linear time, middle and lower panels depict data as functions of the logarithm of the time that has elapse since the initiation of a jump at $t=t_i$. No corrections were applied to yield the match between experimental data and prediction.}
	\label{Fig:LinDJ_VPC_lo}
\end{figure}

\begin{figure*}[h!]
	
	\begin{minipage}[t]{0.475\columnwidth}
		\flushleft
		\textbf{a)}
	\end{minipage}	
	\begin{minipage}[t]{0.475\columnwidth}
		\flushleft
		\textbf{b)}
	\end{minipage}
	
	\includegraphics[width=.9\columnwidth]{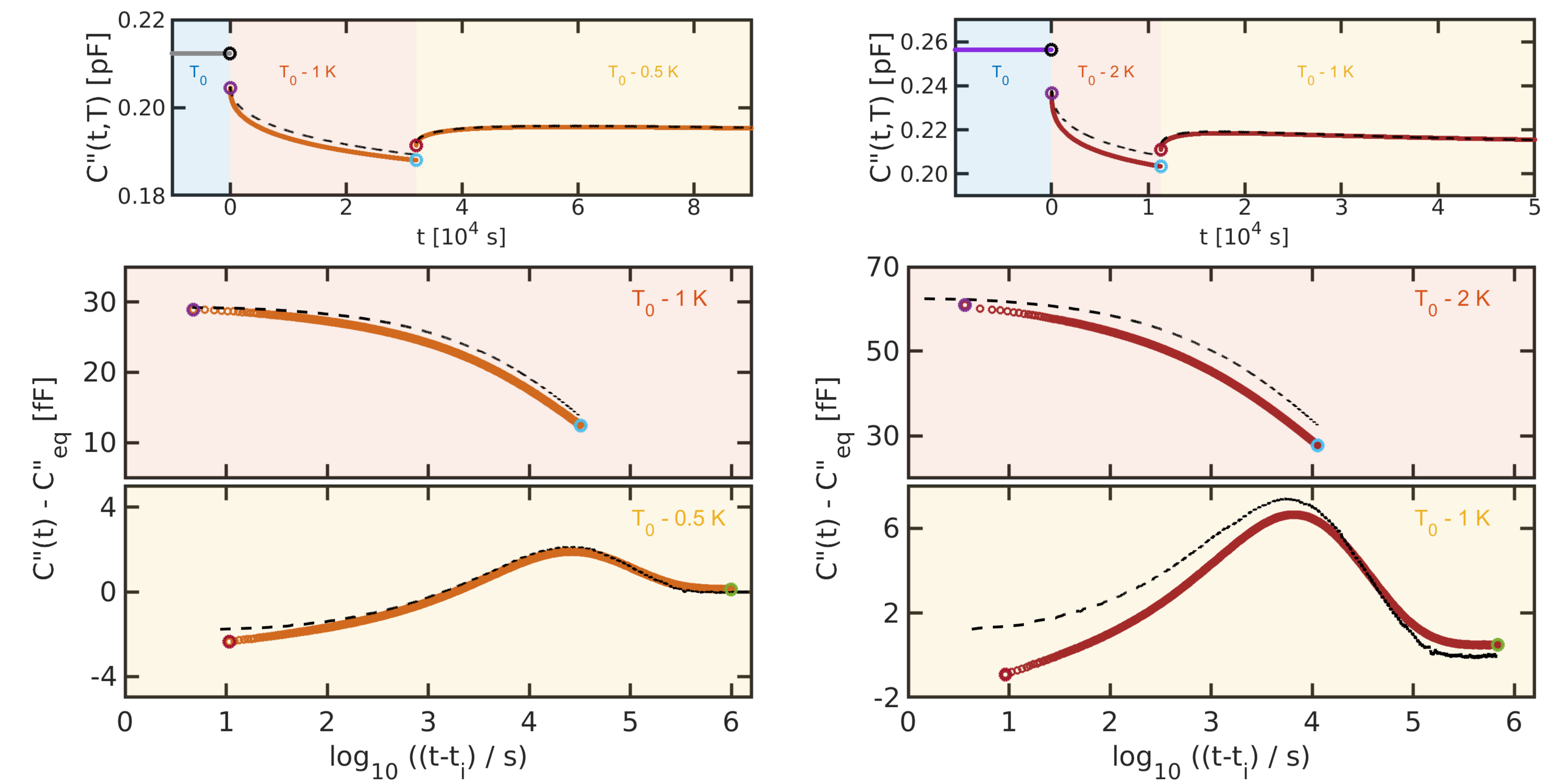}	

	\begin{minipage}[t]{0.25\columnwidth}
		\phantom{\textbf{c)}}
	\end{minipage}
	\begin{minipage}[t]{0.7\columnwidth}
		\flushleft
		\textbf{c)}
	\end{minipage}	

	\includegraphics[trim = 0cm 0cm 25.5cm 0cm, clip=true, width=.45\columnwidth]{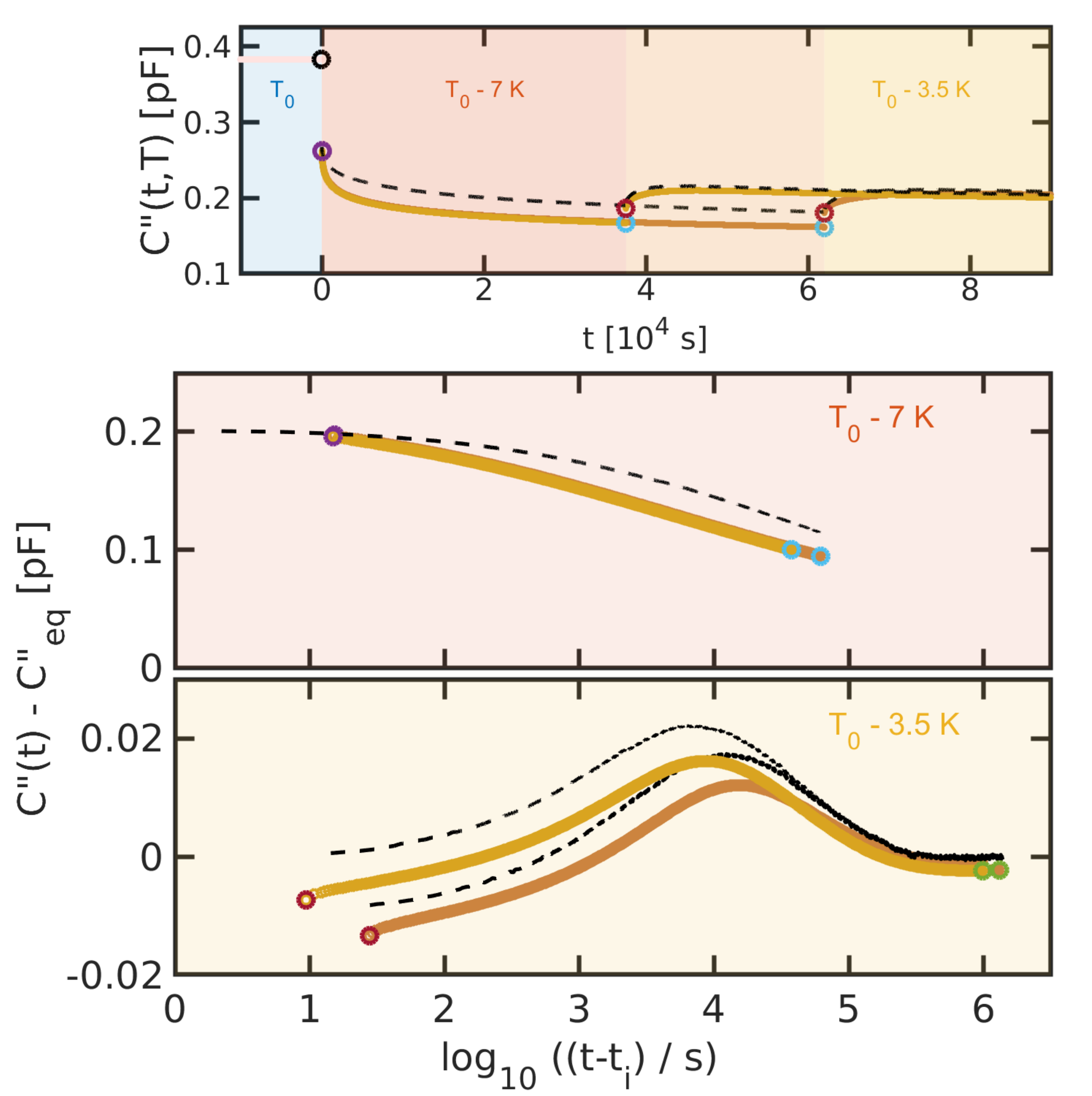}	
		
	\caption{Experimental data and predictions of nonlinear double jumps on VEC involving (a) temperature jumps of $\Delta T_1 = $+\SI{1}{\kelvin} and $\Delta T_2 = $-\SI{0.5}{\kelvin}, (b) $\Delta T_1 = $+\SI{2}{\kelvin} and $\Delta T_2 = $-\SI{1}{\kelvin}, (c) $\Delta T_1 = $+\SI{7}{\kelvin} and $\Delta T_2 = $-\SI{3.5}{\kelvin} for two realizations. The predictions are based on inter- and extrapolation of the glassy contribution  $X_{gl}$, the equilibrium response $X_{eq}$, $\Lambda$, and the equilibrium clock rate $\gamma_{eq}$. Note that for the predictions of the double jump with $\Delta T_1 = $+\SI{7}{\kelvin} shown in panel (c) the aforementioned quantities had to be extrapolated, while the predictions of double jumps in panels (a) and (b) are based on interpolations. For more details on how the predictions are derived, see section~\ref{sec:detDJpred}.}
	\label{Fig:NLDJsVPC_lo}
\end{figure*}

Figures~\ref{Fig:NL_VPC_lo} to~\ref{Fig:NLDJsVPC_lo} show comparisons of experimental data in terms of the loss contribution of the dielectric capacitance, $C"$, and predictions derived from the loss contribution of the response to a linear \SI{50}{\milli\kelvin} down-jump.\\

The predictions of the loss response to individual temperature jumps larger than \SI{100}{\milli\kelvin} collapse convincingly with experimental data for down-jumps and up-jumps of amplitudes up to \SI{1.5}{\kelvin} as shown in Fig.~\ref{Fig:NL_VPC_lo}. Predictions of up-jumps with higher amplitudes appear retarded when compared to the experimental data. This behavior is qualitatively comparable to observations made on the storage capacitance, but is more pronounced for the loss data.

\clearpage
\subsection{Evaluation of response-specific parameters}\label{sec:fittedParams}
This section covers:

\begin{itemize}
	\item[a)] Details on the drift correction
	\item[b)] Equilibrium values of the measured quantity, $X_{eq}$
	\item[c)] Glassy contribution to the normalized relaxation function, $R_X^{gl}$
	\item[d)] Equilibrium values of the relaxation rate, $\gamma_{eq}$
\end{itemize}
\subsubsection{Details on the drift correction}\label{sec:drift}
\begin{figure}[h!]
	
	\begin{minipage}[t]{\columnwidth}
		\flushleft
		\textbf{a)}
	\end{minipage}
	
	\begin{minipage}{\columnwidth}
		\centering
		\includegraphics[trim = 2cm 0cm 4.5cm 0cm, clip=true, width=\columnwidth]{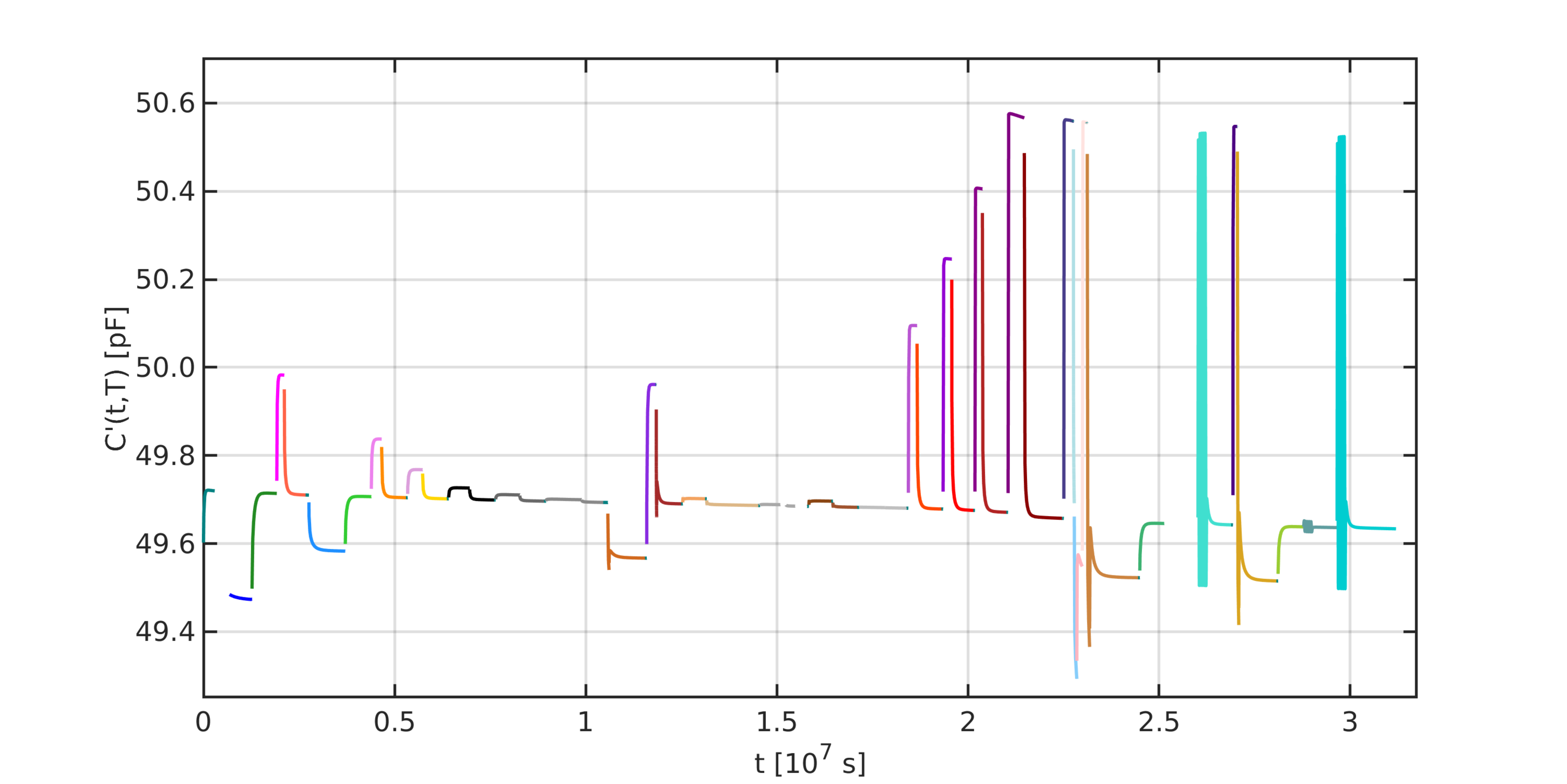}
	\end{minipage}
	
	\begin{minipage}[t]{\columnwidth}
		\flushleft
		\textbf{b)}
	\end{minipage}
	
	\begin{minipage}{\columnwidth}
		\centering
		\includegraphics[trim = 2cm 0cm 4.5cm 0cm, clip=true, width=\columnwidth]{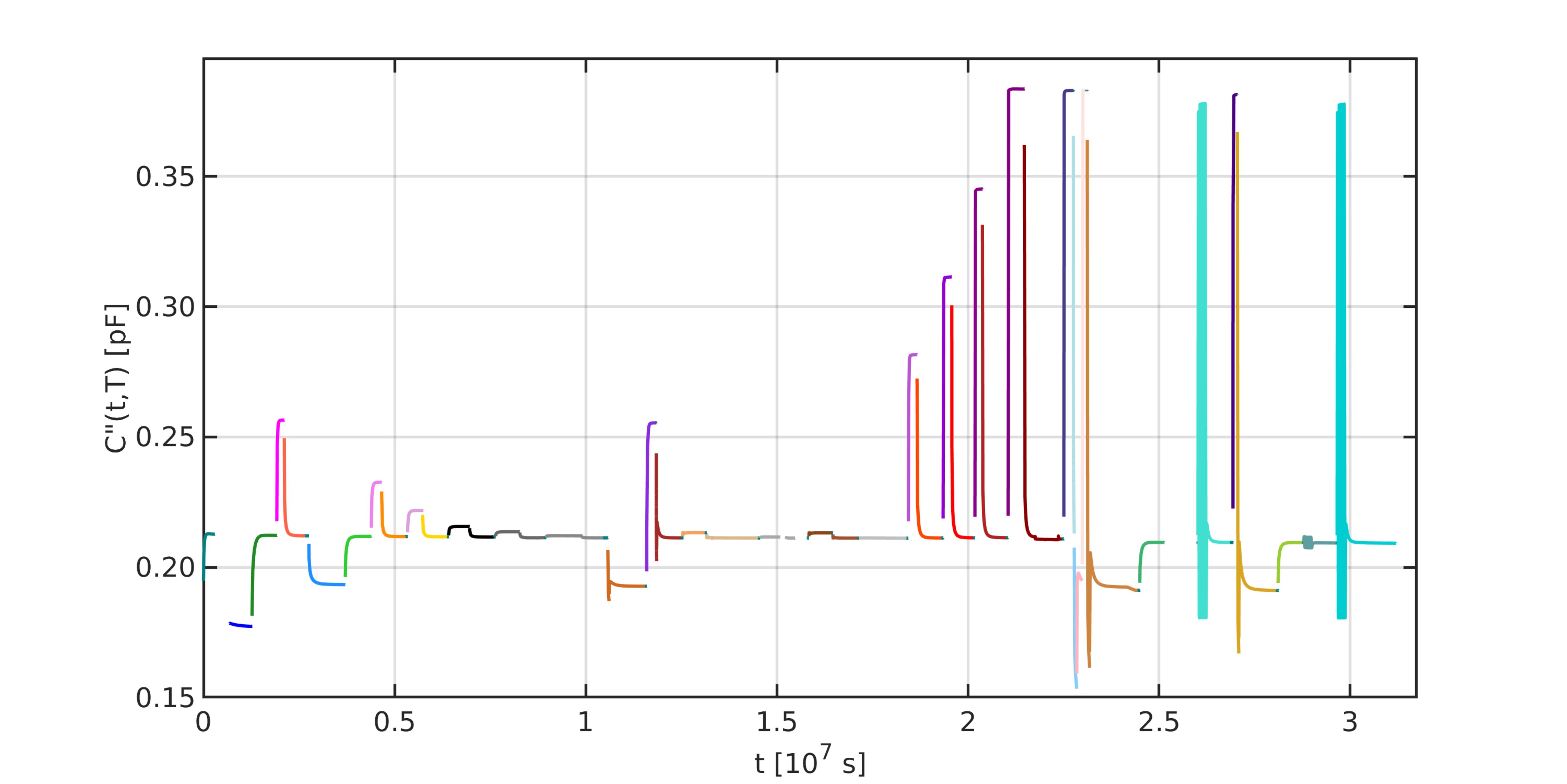}
	\end{minipage}
	
	\caption{Overview of the consecutive physical-aging experiments based on the temperature protocol presented in the main manuscript without drift correction for (a) the storage and (b) the loss contribution to the measured capacitance.}
	\label{Fig:OV_noCorr_VPC}
\end{figure}

The measured capacitance at constant temperature shows -- even when the sample is thermally equilibrated -- a small but steady decline (see Fig.~\ref{Fig:OV_noCorr_VPC}). This ``drift'' is described and analyzed in detail in this section.

The data presented in the main manuscript and the previous section are corrected for drift by subtraction of a linear drift, which is either determined on the basis of the individual jumps (``individual drift correction'') or sets of jumps (``loop-wise drift correction''), as described in the following.\\

\textit{Individual drift correction}\\

For each jump that ends in equilibrium (individual jumps or second parts of double jumps), the region of constant slope in the long-time regime of the data is determined manually, typically involving of the order $10^4$ data points. From a linear regression of these intervals according to 

\be
X(t) = 	r_X \times time + r_{X,0},
\ee
the drift is identified as the slope $r_X$ (for the storage and loss response, separately).\\

\textit{Loop-wise drift correction}\\

As for the individual drift correction, the region of constant slope in the long-time regime is determined manually for jumps that are assumed to have achieved equilibrium. For the loop-wise drift correction, subsequent jumps of same temperature amplitude $|\Delta T|$ are regarded as a set of jumps, called a \textit{loop}, typically consisting of four consecutive jumps (from or to $T_b$ by $\pm\Delta T$) or two consecutive jumps (from or to $T_b$ by $|\Delta T|$). The drift is determined from the slope $s_1$ of linear fits based on
\be
\begin{pmatrix}
	X_{j1}(t)\\
	X_{j2}(t)\\
	X_{j3}(t)\\
	X_{j4}(t)\\
\end{pmatrix}
= s_1 \times \begin{pmatrix}
	t_{j1} + s_2\\
	t_{j2} + s_3\\
	t_{j3} + s_4\\
	t_{j4} + s_5
\end{pmatrix}.
\ee
(for the response of the storage and loss component, separately). Thus we assume that all jumps of a loop may have different response ``offsets'' $s_2$ to $s_5$, but share the same ``drift'' slope $s_1$.

\begin{figure}[h!]	
	\begin{minipage}[t]{\columnwidth}
		\flushleft
		\textbf{a)}
	\end{minipage}
	
	\includegraphics[width=\columnwidth]{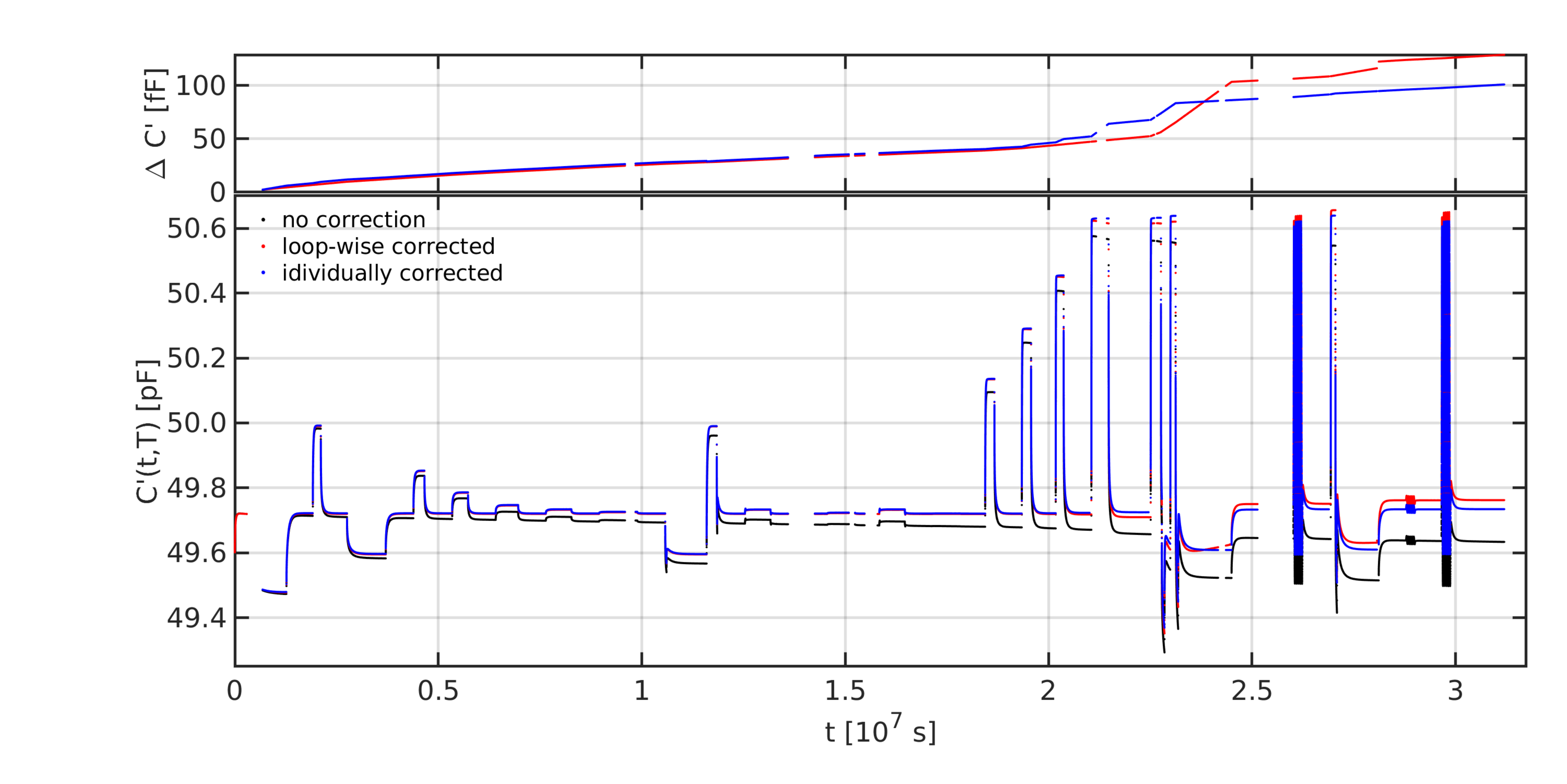}
	
	\begin{minipage}[t]{\columnwidth}
		\flushleft
		\textbf{b)}
	\end{minipage}
	
	\includegraphics[width=\columnwidth]{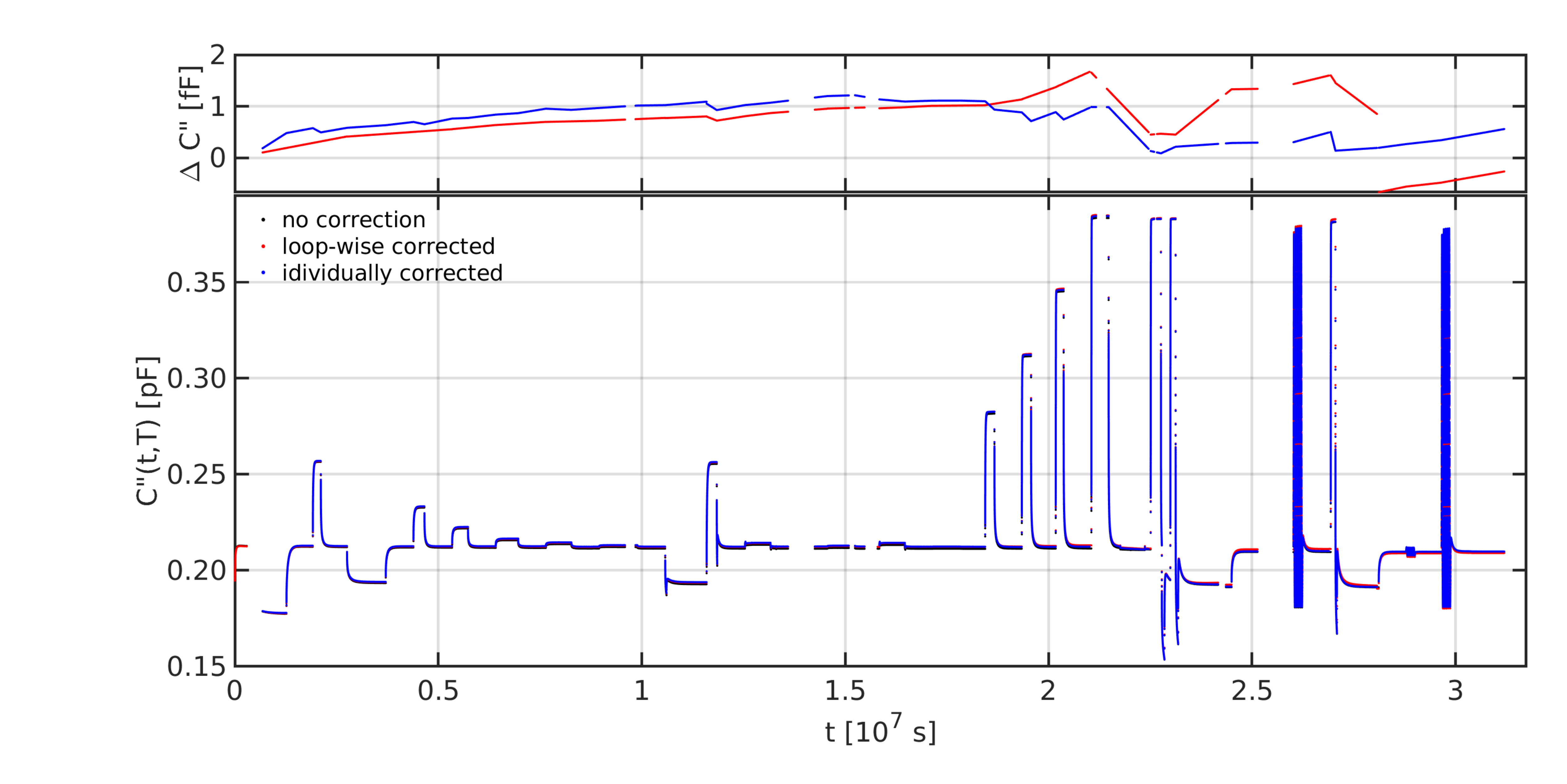}
	
	\caption{Comparisons of uncorrected, individually drift-corrected, and loop-wise dirft-corrected data of the storage (a) and loss (b) contribution to the measured capacitance. The upper panels in (a) and (b) show the difference between un-corrected data and individually corrected (blue) or loop-wise corrected data (red).}
	\label{Fig:CorrComp_VPC}
\end{figure}

Based on the individual correction and the loop-wise correction, the raw time-dependent response data are corrected by subtracting the cummulated drift-contribution since the initialization of the experiment. The resulting corrected data sets are shown in Fig.~\ref{Fig:CorrComp_VPC}a for the storage response and in Fig.~\ref{Fig:CorrComp_VPC}b for the loss response.

In the below sections \ref{Sec:equilR} to \ref{Sec:clora} it is shown that the difference in the data based on these two different drift-correction methods is minor. This demonstrates the validity of both approaches. Since the equilibrium level at $T_{ref}$ is more stable for loop-wise corrected data, this correction has been applied to all linear data. Nonlinear data are corrected with the individual drift-correction because the high-temperature equilibrium values are more steady in this case.

\begin{figure}[h!]	
	\begin{minipage}[t]{\columnwidth}
		\flushleft
		\textbf{a)}
	\end{minipage}
	
	\includegraphics[width=\columnwidth]{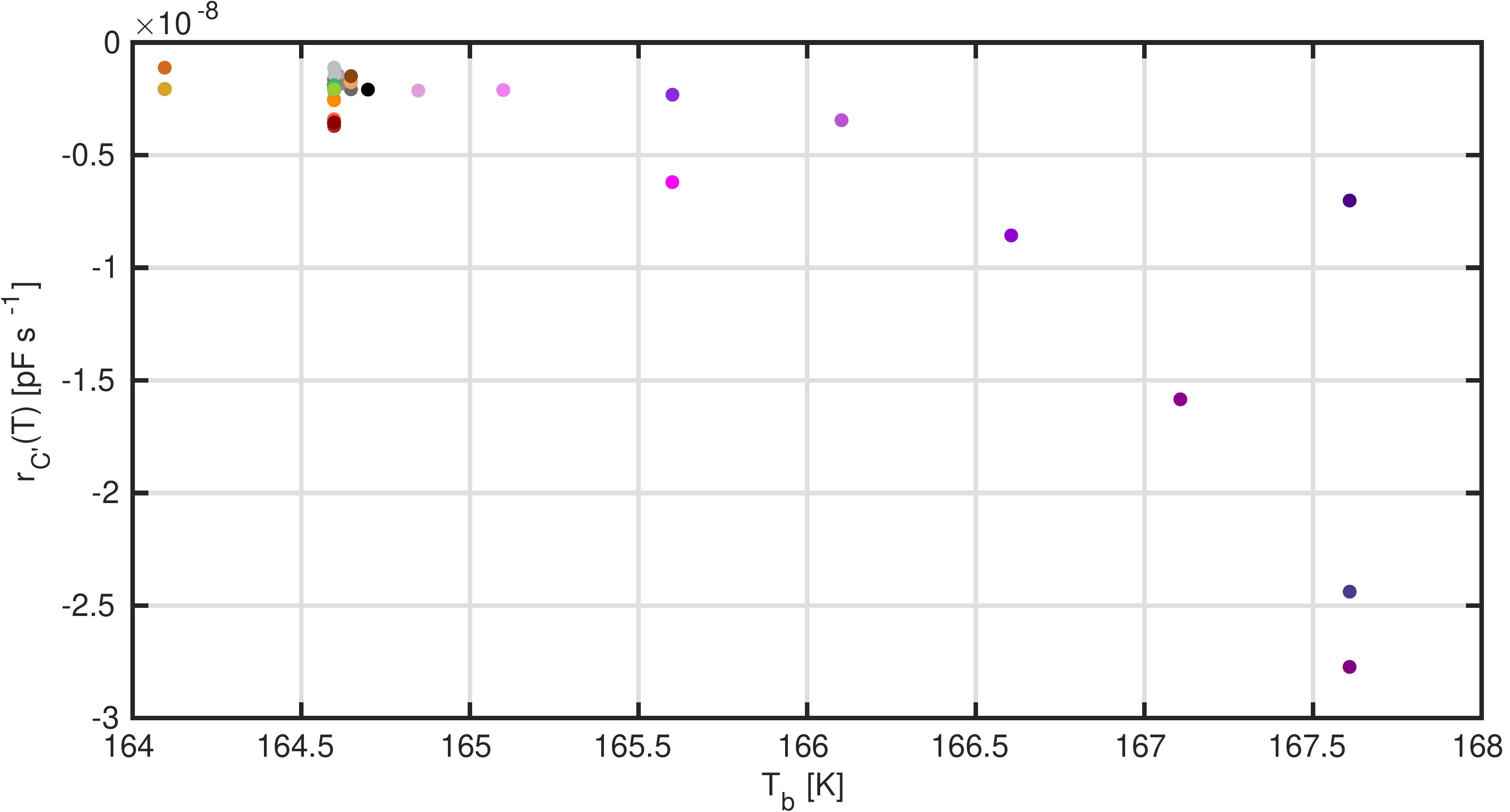}
	
	\begin{minipage}[t]{\columnwidth}
		\flushleft
		\textbf{b)}
	\end{minipage}
	
	\includegraphics[width=\columnwidth]{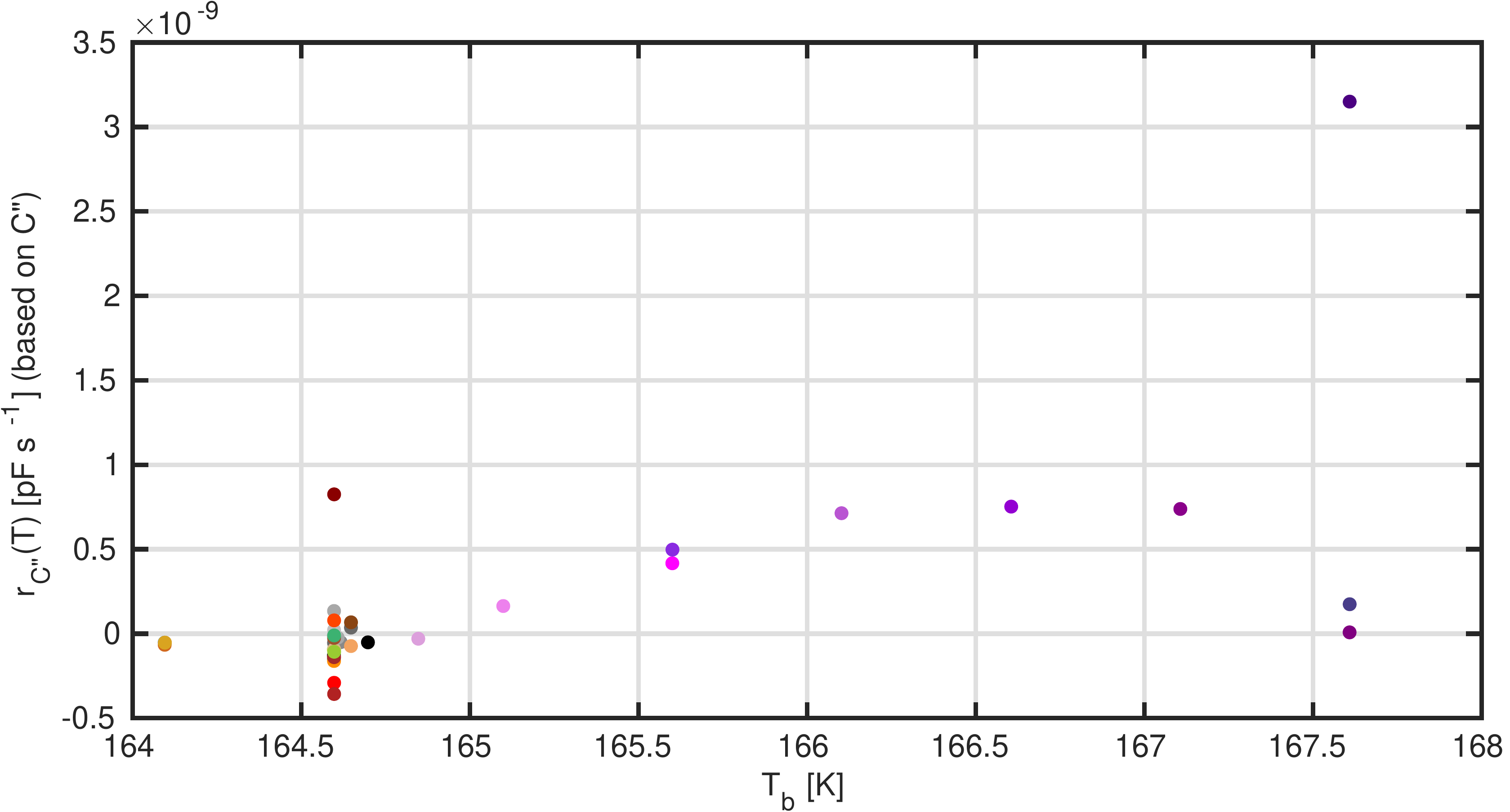}
	
	\caption{Temperature-dependence of the drift rate $r_1$ for (a) the storage and (b) the loss contributions to the measured capacitance based on the individual drift correction. The colors indicate the individual jumps as in Fig.~\ref{Fig:OV_noCorr_VPC}.}
	\label{Fig:DriftTemp_VPC}
\end{figure}

The drift rate $r_X$, determined from the individual drift correction, is plotted against the final temperature $T_b$ in Fig.~\ref{Fig:DriftTemp_VPC}. This figure shows temperature-dependent behavior for both the storage and the loss contribution; the evolution is more pronounced and systematic for the $C'$-data. Here, the drift rate is fairly constant for $T_b \leq T_{ref} + $\SI{1.5}{\kelvin} and increases with increasing temperature. For the loss data, $r_{C"}$ shows a weak increase towards positive rates where it becomes steady toward higher $T_b$. The cause of this drift may be a temperature-dependent geometric change of the setup or some components thereof, an effect that is estimated to be weak compared to the thermally caused changes in the dielectric permittivity of the investigated materials.

\clearpage
\subsubsection{Equilibrium values of the measured quantity $X_{eq}$}\label{Sec:equilR}

\begin{figure}[h!]	
	\begin{minipage}[t]{\columnwidth}
		\flushleft
		\textbf{a)}
	\end{minipage}
	
	\includegraphics[width=0.8\columnwidth]{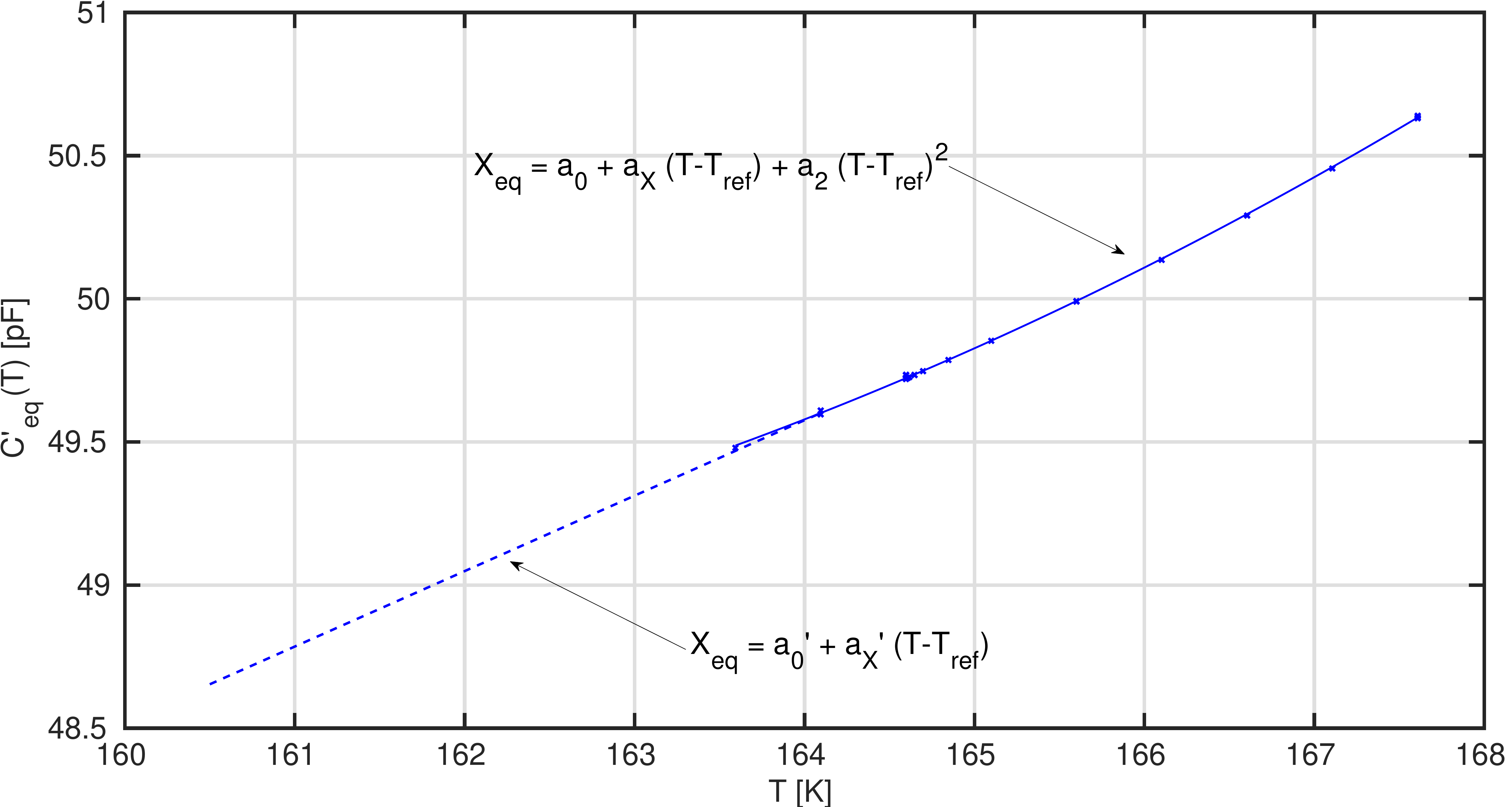}
	
	\begin{minipage}[t]{\columnwidth}
		\flushleft
		\textbf{b)}
	\end{minipage}
	
	\includegraphics[width=0.8\columnwidth]{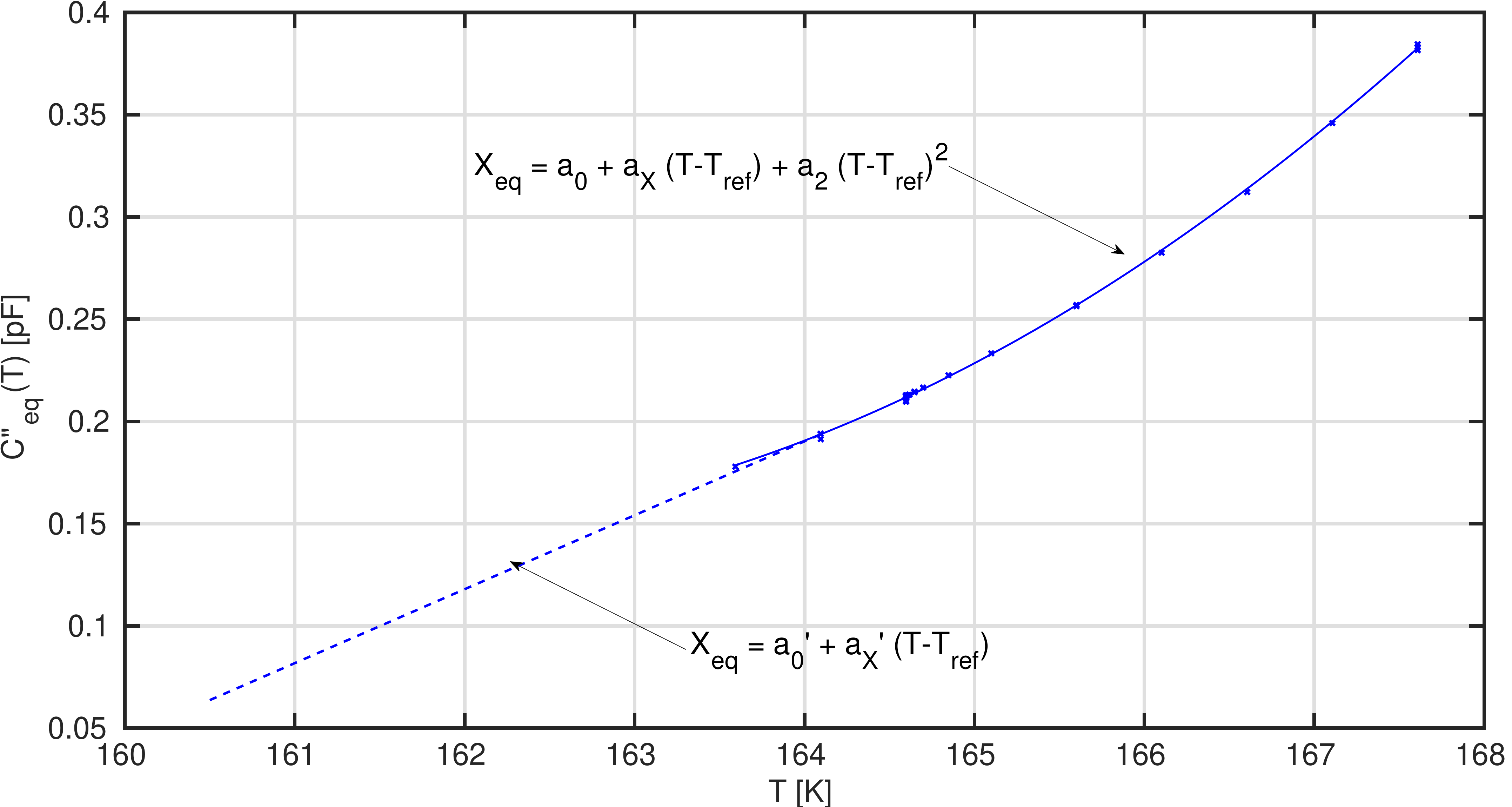}
	
	\caption{Equilibrium responses $X_{eq}$ in form of equilibrium values of (a) the storage and (b) the loss capacitance as functions of temperature for individually-corrected data. Full lines correspond to nonlinear interpolations of the data; dashed lines represent linear extrapolations based on the data between \SI{163.5}{\kelvin} and \SI{164.5}{\kelvin}.}
	\label{Fig:EquiTemp_VPC}
\end{figure}

A sample that is held at constant temperature after a temperature jump is regarded as in equilibrium when the response levels off to a constant value at long times. In that stage, the physical aging process is concluded and the material reflects the properties of the supercooled liquid. In the raw experimental data, this long-time limit is characterized by a drift, as described in the previous section. The equilibrium values of the storage and loss response at a given temperature $T_b$ are determined from the drift-corrected data, defined as the mean value of the corrected data in the interval used to determine the drift. The equilibrium data were plotted against temperature and interpolated by a non-linear polynomial function as shown in Fig.~\ref{Fig:EquiTemp_VPC}. These data depend only weakly on the applied drift-correction method, as is evident from the parameters in table~\ref{tab:EquiParm}. The individually drift-corrected data and fits were chosen to represent the temperature dependence of $X_{eq}$. On the basis of low-temperature data, a linear extrapolation was made that allows for estimating  $X_{eq}$ for $T\leq $\SI{163.5}{\kelvin}.

\begin{table}
	\begin{tabular}{l|c c|c c}
		&\multicolumn{2}{c}{$X = C'$ [pF]} 						& \multicolumn{2}{c}{$X = C"$ [pF]}	\\
		& indvl. corr. 				& loop-wise corr.   		& indvl. corr. 				& loop-wise corr. 	\\
		\hline
		
		$a_0$ [pF]	&\num{49.7230} 	& \num{49.7251} &\num{0.2118} 	& \num{0.2117} \\
		$a_X$ [pF K$^{-1}$]	&\num{0.2516} 	& \num{0.2493} &\num{0.0390} 	& \num{0.0390} \\
		$a_2$ [pF K$^{-2}$]	&\num{0.0168} 	& \num{0.0165} &\num{0.0059} 	& \num{0.0060} \\
		\hline
		$a'_0$ [pF]	&\num{49.7332} 	& \num{49.7366} &\num{0.2119} 	& \num{0.2119} \\
		$a'_X$ [pF K$^{-1}$]	&\num{0.2635} 	& \num{0.2640} &\num{0.0362} 	& \num{0.0363} \\
		
	\end{tabular}
	\caption{Parameters for nonlinear interpolation and linear extrapolation of individual and loop-wise corrected equilibrium data, $X_{eq}$, of the storage and the loss parts of the capacitance.}
	\label{tab:EquiParm}
\end{table}

\subsubsection{glassy contribution of the measured quantity}\label{Sec:glassyR}
\begin{figure}[h!]	
	\begin{minipage}[t]{\columnwidth}
		\flushleft
		\textbf{a)}
	\end{minipage}
	
	\includegraphics[width=0.8\columnwidth]{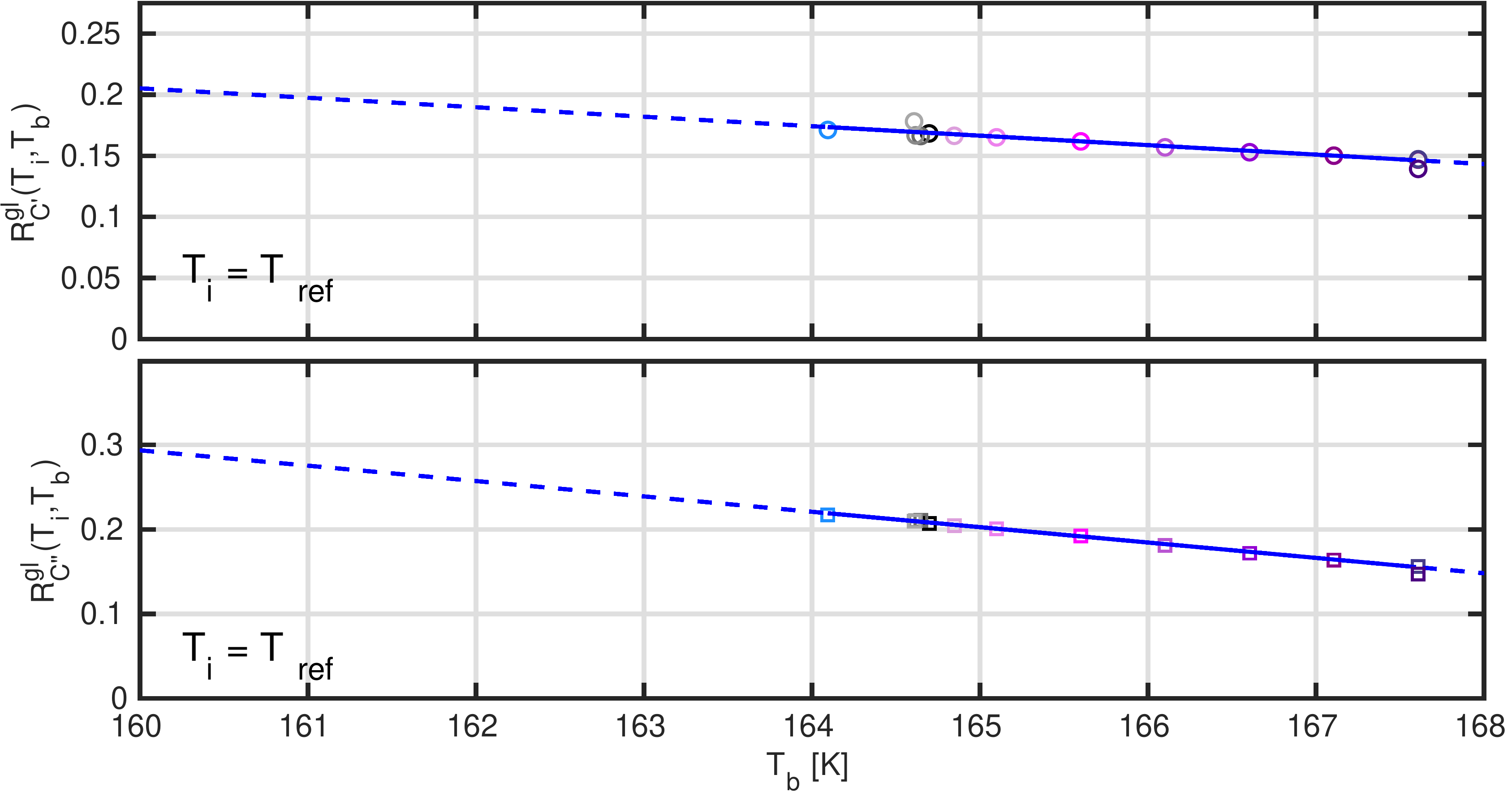}
	
	\begin{minipage}[t]{\columnwidth}
		\flushleft
		\textbf{b)}
	\end{minipage}
	
	\includegraphics[width=0.8\columnwidth]{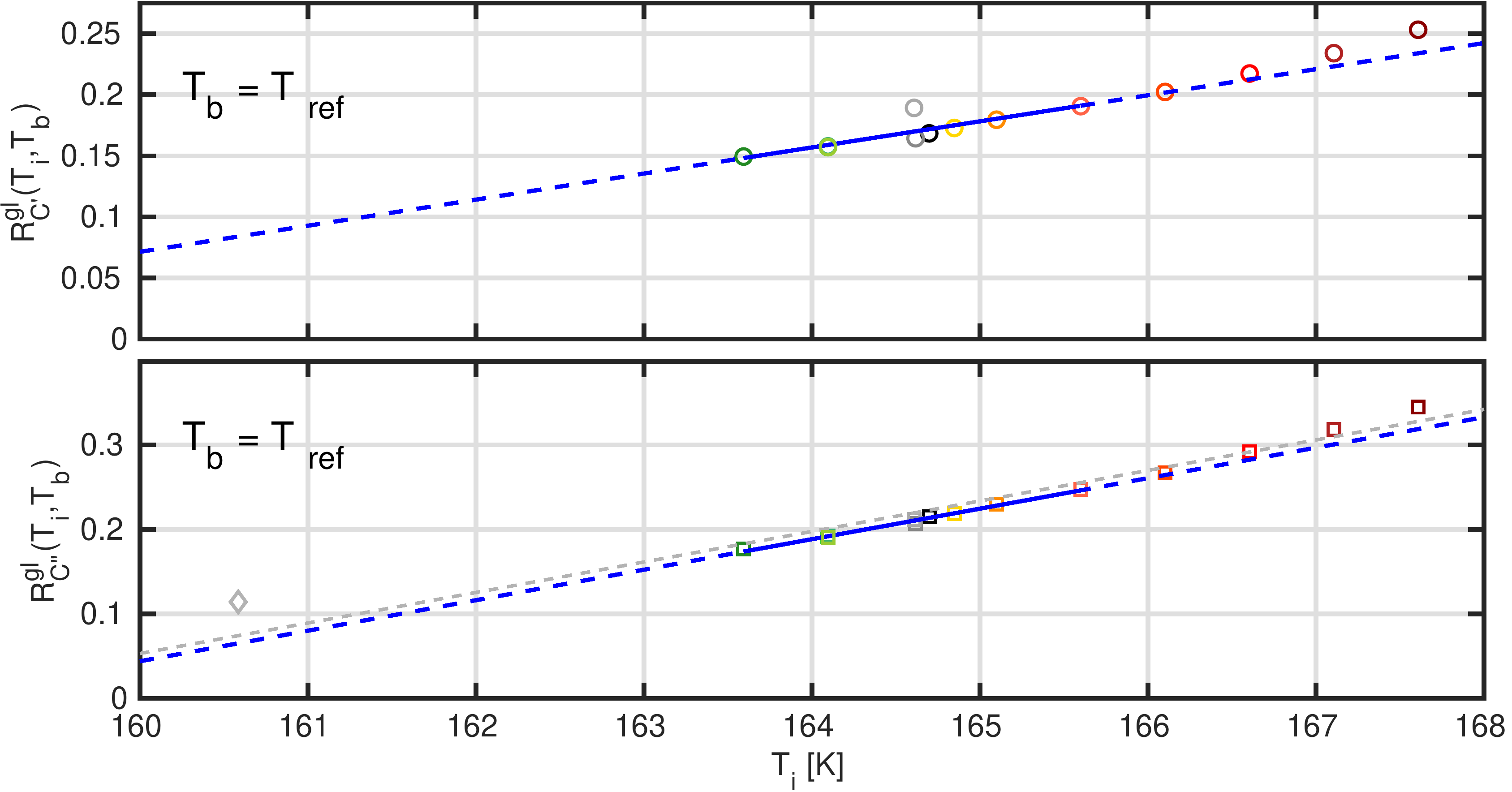}
	
	\caption{The glassy response contribution, $R^{gl}_X$, for jumps starting at $T_{ref}$ plotted as a function of final temperature, $T_b$ (panel ((a)), and for jumps ending at $T_{ref}$ plotted as a function of initial temperature, $T_i$ (panel (b)). The full lines represent the interval used for the fit, while the colored dashes lines represent the extrapolation over the full temperature range. The gray dashed line in the lower panel of (b) after a temperature jump represents the extrapolation of the glassy response contribution for jumps from temperatures $T_i$ to $T_b = $\SI{164.1}{\kelvin}, i.e., $R^{gl}_X(T_i,T_b=T_{ref}-$\SI{0.5}{\kelvin}). The gray diamond reflects the value of $R^{gl}_X$ that is applied for the second jump of the \SI{7}{\kelvin} double-jump protocol in Fig.~\ref{Fig:NLDJsVPC}d. All values shown in the lower part of panel (b) can be fitted by a second-order polynomial (not shown).}
	\label{Fig:GlassR_VPC}
\end{figure}

When plotting the normalized relaxation function, the level of the data just after the initiation of the jump, i.e. at small $t-t_i$, is significantly below unity. This is due to the elastic contribution of the material and, possibly, also to  secondary relaxation(s) occurring on time scales shorter than the time resolution of the experiment, i.e., at $t-t_i < $\SI{4}{\second}. This initial contribution is termed 'glassy contribution' in the following; it depends on both the initial temperature, $T_i$, and the final temperature,$T_b$, of a jump. The temperature dependence is fitted in the temperature ranges $T_{ref} - $\SI{1}{\kelvin} $\leq T_i \leq T_{ref} + $\SI{1}{\kelvin} and $T_{ref} - $\SI{0.5}{\kelvin} $\leq T_b \leq T_{ref} + $\SI{3}{\kelvin} by three parameters $t_1$, $t_2$ and $t_3$ according to $R^{gl}_X = t_1 + t_2 T_i  + t_3 T_b$  (see Table~\ref{tab:glR} for the parameters). Values connected to $T_i > T_{ref} + $\SI{1}{\kelvin} are excluded from the fit as the initial plateau is not sufficiently captured for jumps from elevated temperatures, resulting in an excess value of $R^{gl}_X$ as it includes a relaxational contribution. The deviations between fits of individually corrected and loop-wise corrected data are small; thus the individually corrected data represents the temperature dependence of $R^{gl}_X$ in Fig.~\ref{Fig:GlassR_VPC}.

\begin{table}
	\begin{tabular}{l|c c|c c}
		&\multicolumn{2}{c}{$X = C'$ [pF]} 						& \multicolumn{2}{c}{$X = C"$ [pF]}	\\
		& indvl. corr. 				& loop-wise corr.   		& indvl. corr. 				& loop-wise corr. 	\\
		\hline
		
		$t_1$ [-]		&\num{0.1697} 	& \num{0.1690} 	&\num{0.2102} 	& \num{0.2112} \\
		$t_2$ [K$^{-1}$]	&\num{0.0213} 	& \num{0.0212} 	&\num{0.0360} 	& \num{0.0362} \\
		$t_3$ [K$^{-1}$]	&\num{-0.0083} 	& \num{-0.0078} 	&\num{-0.0188} 	& \num{-0.0193} 		
		
	\end{tabular}
	\caption{Parameters for the extrapolation along $T_i$  and $T_b$ of individual and loop-wise corrected data for the glassy response contribution, $R^{gl}_X$, of the storage and the loss parts of the capacitance.}
	\label{tab:glR}
\end{table}

\clearpage

\subsubsection{Equilibrium values of the relaxation rate $\gamma_{eq}$}\label{Sec:clora}

\begin{figure}[h!]	
	
	\includegraphics[width=0.9\columnwidth]{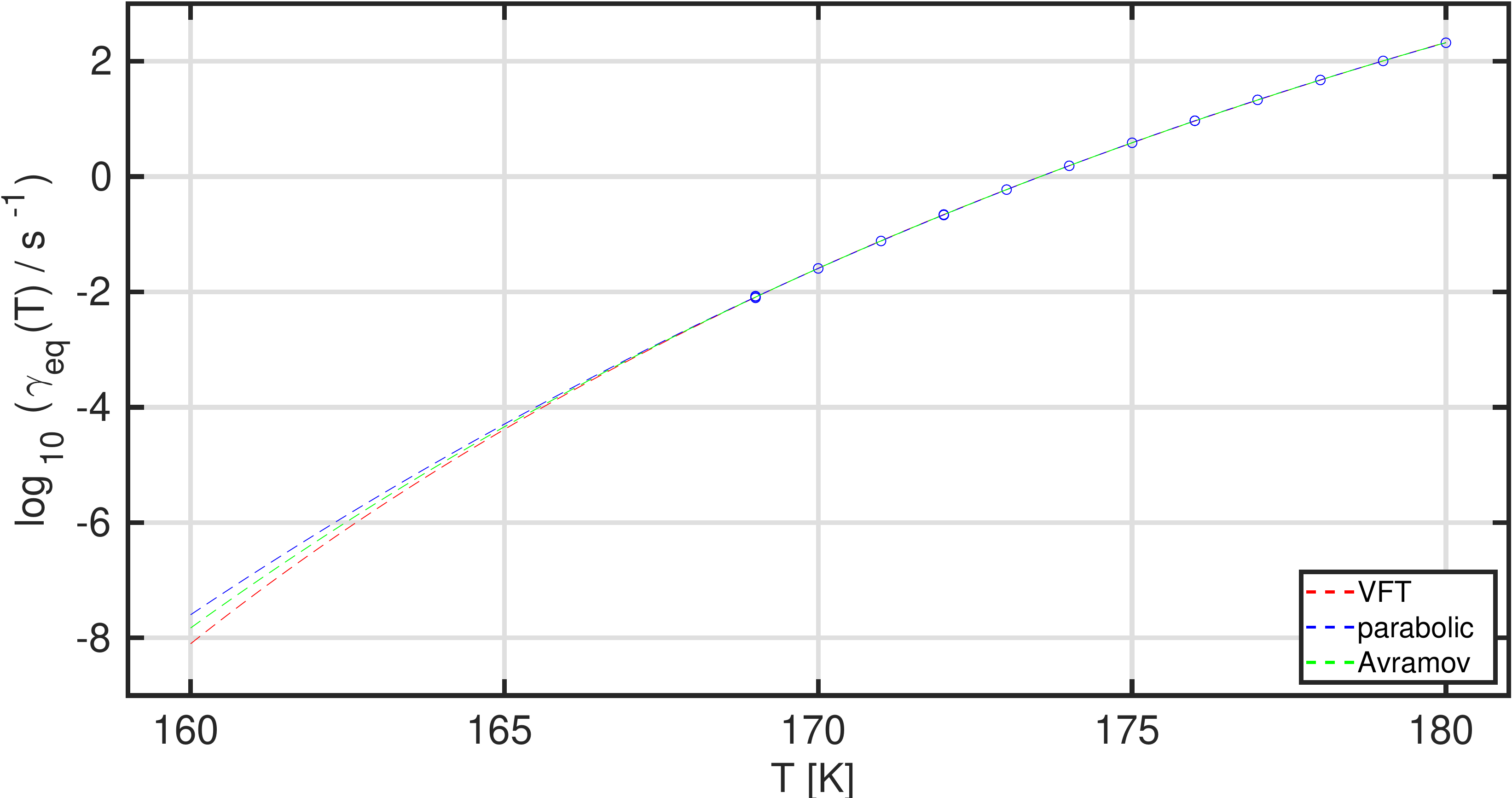}
	
	\caption{Temperature dependence of the inverse time scale derived from spectral loss-peak positions, $\gamma_{eq}$, and extrapolations based on the Vogel-Fulcher-Tammann (VFT), parabolic, and Avramov functions.}
	\label{Fig:SpecGamma_VPC}
\end{figure}

Fig.~\ref{Fig:SpecGamma_VPC} shows extrapolations by the Vogel-Fulcher-Tammann (VFT) expression, parabolic, and Avramov functions, based on the loss-peak frequencies of the spectra shown in Fig.~\ref{Fig:SpecVPC}.\\

VFT function: \hspace{30pt} $\log_{10} \gamma_{eq}(T) = u_1 - u_2 /( T - u_3)$ with $u_1 = $\num{18.1}, $u_2 = $\SI{791}{\kelvin}, and $u_3 = $\SI{130}{\kelvin}.\\

Parabolic function: \hspace{30pt} $\log_{10} \gamma_{eq}(T) = v_1 - v_2^2*(1/T - 1/v_3)^2$ with $v_1 = $\num{6.20}, $v_2 = $\SI{50}{\kelvin}, and $v_3 = $\SI{210}{\kelvin}.\\

Avramov function: \hspace{30pt} $\log_{10} \gamma_{eq}(T) =  w_1 - \frac{(w_2/T)^{w_3}}{\ln(10)}$ with $w_1 = $\num{10.4}, $w_2 = $\SI{275}{\kelvin}, and $w_3 = $\num{6.89}.\\

For the calculation of predictions, the value of $\gamma_{eq}$ was extrapolated by use of the VFT-function.

\clearpage
\subsection{Evaluation of prediction-related quantities and details on the double-jump temperature protocol}
This section covers:

\begin{itemize}
	\item[a)] The connection of time increments for two individual jumps in the material-time formalism
	\item[b)] The determination of $\Lambda$
	\item[c)] A check on the equilibrium clock rate $\gamma_{eq}$
\end{itemize}
\subsubsection{Connection of time increments for two individual jumps in the material-time formalism}
On the basis of the time-dependent aging rate,
\be
\gamma (t) = d\xi (t) / dt,
\ee
and the single-parameter aging ansatz according to which the aging rate is controlled by the measured quantity $X(t)$ as follows \cite{Hecksher2015a}
\be
\log (\gamma (t)) - \log (\gamma_{eq}(T)) = \Lambda \left( X(t) - X_{eq}(T) \right),
\ee
we get the following equation for the time-dependent aging rate
\be
\gamma (t) = d\xi (t) / dt =  \gamma_{eq}(T) \exp \left[ \Lambda \hspace{3pt} \Delta X(t) \right].
\ee
Thus the incremental change in material time is given by
\be
d\xi (t)  =  \gamma_{eq}(T) \exp \left[ \Lambda \hspace{3pt} \Delta X(t) \right] dt\,.
\ee
Consider two jumps, A and B, at times $t^*_A$ and $t^*_B$ corresponding to the same value of the normalized relaxation function, i.e., $R_A(t^*_A) = R_B(t^*_B)$, and impose the condition $d\xi (t^*_A) = d\xi (t^*_B)$. On this basis we get

\be\label{eq:dtSPA}
dt^*_B = \frac{\gamma_{eq}(T_{b,~A})}{\gamma_{eq}(T_{b,~B})} \exp \left[- \Lambda \left( \Delta X_B - \Delta X_A \right) R_A(t^*_A)\right] dt^*_A\,.
\ee
Eq.~(\ref{eq:dtSPA}) determines how a specific time step $dt^*_A$ from jump A can be transformed to the corresponding time step $dt^*_B$ of a different jump B, based on 1) the input of the equilibrium clock rate for each jump, 2) the overall change in the equilibrium response for each jump, 3) the constant $\Lambda$, and 4) the normalized response of jump A as a function of time.

\subsubsection{Determination of $\Lambda$}\label{Sec:LambdaVEC}

\begin{figure}[h!]		
	\includegraphics[width=0.8\columnwidth]{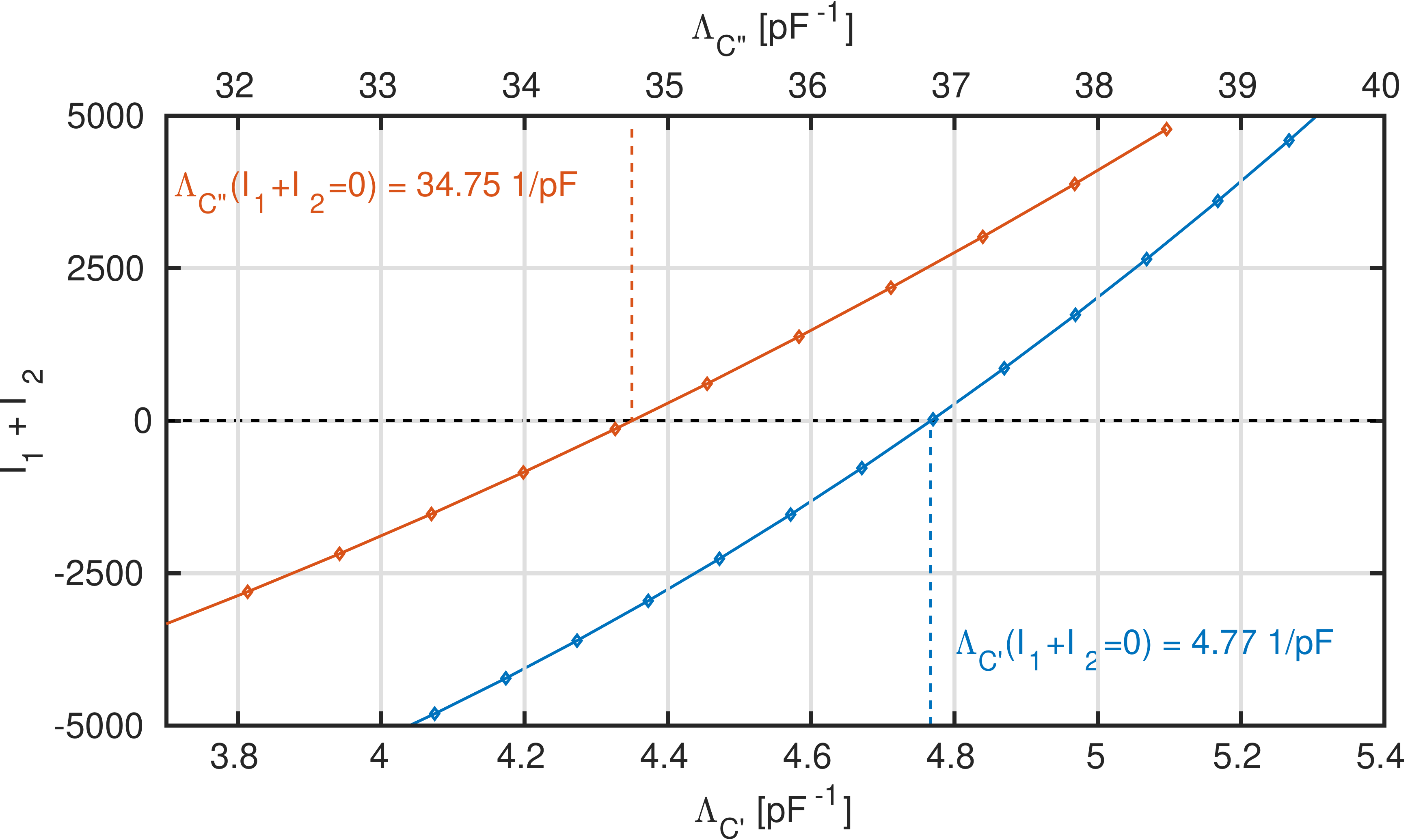}
	
	\caption{Determination of $\Lambda$ for storage and loss contribution of the data. The sum of integrals, $I_1 + I_2$, plotted as a function of $\Lambda$ derived from individually corrected data of a pair of temperature jumps of \SI{1}{\kelvin} amplitude towards the reference temperature, $T_{ref}$.}
	\label{Fig:Xconst_VPC}
\end{figure}

The nonlinearity parameter $\Lambda$ was determined by means of the integral criterion developed in Ref. \onlinecite{Hecksher2015a}: For two jumps $A$ and $B$ towards the same final temperature the equilibrium clock rates are the same, i.e., $\gamma_{eq}(T_{b,~A}) = \gamma_{eq}(T_{b,~B})$. In this case, eq.~\ref{eq:dtSPA} simplifies to

\begin{equation}\label{eq:Xconst}
	dt^*_B = \exp \left[- \Lambda \left(\Delta X_B - \Delta X_A\right) R_A(t^*_A)\right] dt^*_A\,.
\end{equation}
The integral over the expression in eq.~\ref{eq:Xconst} allows us to write the difference between times $t_A$ and $t_B$ as follows:

\begin{equation}\label{eq:Integral_part1}
	t^*_B - t^*_A = \int_{0}^{t_B(R)} dt^*_B - \int_{0}^{t_A(R)} dt^*_A = \int_{0}^{t_A(R)} \left( \exp \left[- \Lambda \left(\Delta X_B - \Delta X_A\right) R_A(t^*_A)\right] -1 \right) dt^*_A\ = I_1
\end{equation}
A similar expression can be written for $t^*_A - t^*_B = I_2$. This means that $\Lambda$ can be determined as the value where $I_1 + I_2$ is equal to zero. This procedure is applied to the storage and loss contribution of jumps separately, resulting in one parameter for each contribution, which is used for all predictions. Fig.~\ref{Fig:Xconst_VPC} shows the dependence of $I_1 + I_2$ on $\Lambda_{C''}$ and $\Lambda_{C"}$. For the storage contribution we get $\Lambda_{C''}(I_1 + I_2=0) = $\SI{4.77}{\per\pico\F} based on the individually corrected storage data of the two jumps of magnitude \SI{1}{\kelvin} to the reference temperature. For the loss contribution the result is $\Lambda_{C"}(I_1 + I_2=0) = $\SI{34.75}{\per\pico\F} based on individually corrected loss data of the same set of jumps as used for the storage data. In the analysis the term $\left(\Delta X_B - \Delta X_A \right)$ is written as $-2 \Delta T_A a_x$, where $a_x$ describes the linear part of the relation between the equilibrium quantity and temperature as in Fig.~\ref{Fig:EquiTemp_VPC} and $\Delta T_A$ is the temperature change of jump A.

\subsubsection{Check on the equilibrium clock rate}
\begin{figure}[h!]		
	\includegraphics[width=0.8\columnwidth]{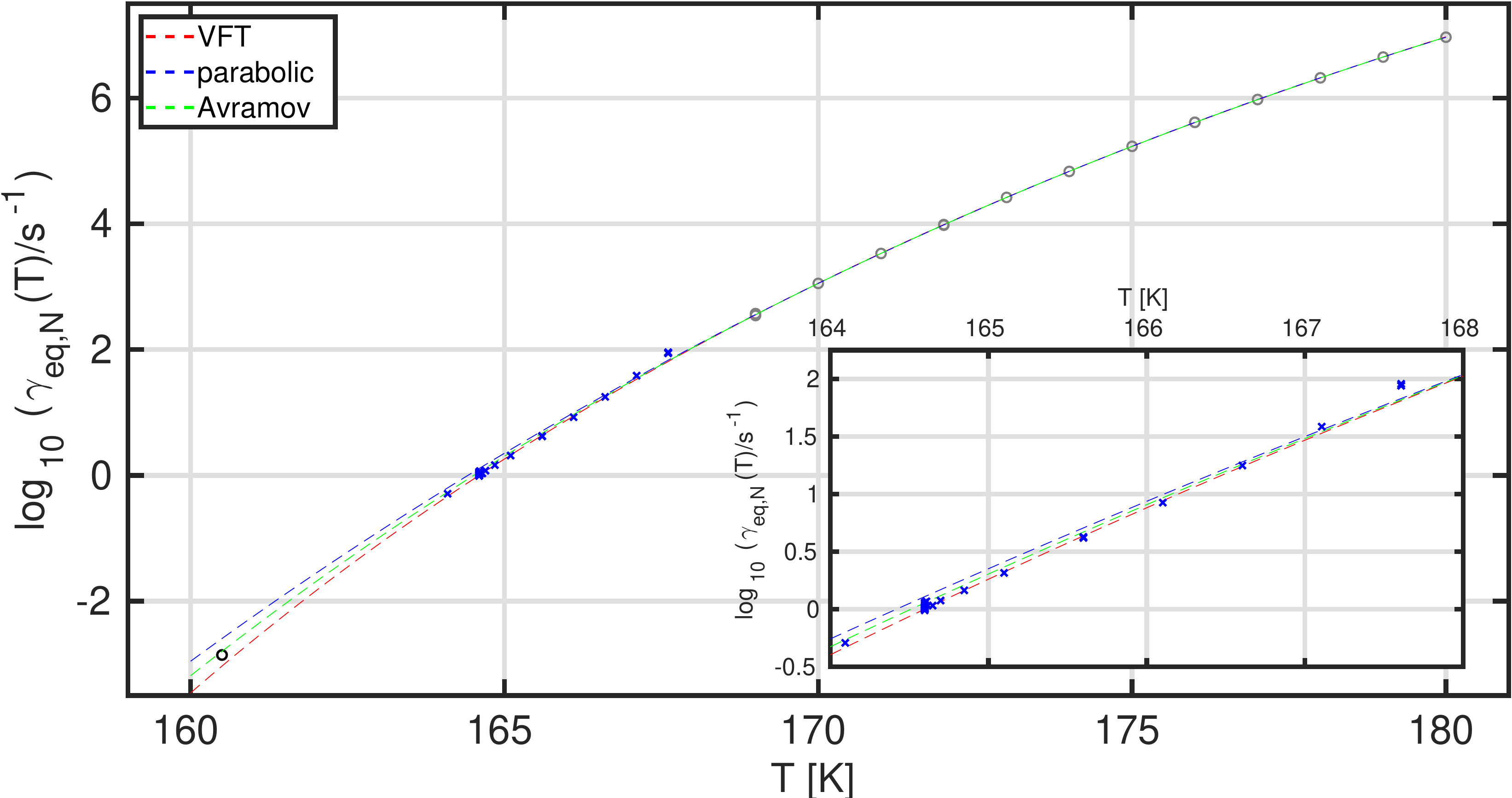}
	
	\caption{Logarithmic normalized equilibrium clock rate, $\gamma_{eq,N}$ as a function of the temperature. Dashed lines correspond to extrapolations based on the clock rates determined from spectral data (grey cirles), crosses correspond to clock rates from  fitting the storage contribution of individual jumps. The inset shows a zoom of $\gamma_{eq,N}$ derived from the individual jumps and extrapolations from the spectral data.}
	\label{Fig:GammaN_VPC}
\end{figure}

For two individual jumps, $A$ and $B$, the relation between their incremental time steps, $dt^*_A$ and $dt^*_B$, is described by Eq.~(\ref{eq:dtSPA}). If one of these jumps ends at the reference temperature $T_{ref}$, the normalized equilibrium clock rate can be written as $\gamma_{eq,N}(T_{b,~B}) = {\gamma_{eq}(T_{b,~B})}/{\gamma_{eq}(T_{b,~A} = T_{ref})}$. Using the extrapolation of gamma values based on spectral data, we can check how well a fit of $\gamma_{eq,N}(T_{b,~B})$ relates to it. In this case the normalized equilibrium clock rate $\gamma_{eq,N}(T_B)$ is the only fitting parameter. The normalized equilibrium clock rate that is derived from the fit of storage data of individual jumps is plotted in Fig.~\ref{Fig:GammaN_VPC} together with the spectra-based clock rates shown in Fig.~\ref{Fig:SpecGamma_VPC}, and which are normalized by the extrapolated value of $\gamma_{eq}$ at $T_{ref}$.\\

\clearpage
\subsection{Details on calculating the predictions}
This section covers the calculation of predictions for:
\begin{itemize}
	\item[a)] Single temperature jumps
	\item[b)] Double temperature jumps
	\item[c)] Multiple temperature jumps
\end{itemize}

In order to calculate the prediction for a single temperature jump, the normalized response, $R_X(t)$, to a different single temperature jump is used as  basis. This $R_X(t)$ is transformed back to its response amplitude in the duration of its time intervals to match the to-be-predicted data. This transformation utilizes the material-time aging formalism involving inter- and extrapolations of $\gamma_{eq}$, $\Lambda$, and $X_{eq}$, as well as the inter- or extrapolation of the glassy contribution $R^{gl}_X$.

The basis data set for all predictions of this work is a linear down-jump of \SI{50}{\milli\kelvin} amplitude toward $T_{ref}$. Note, however, that one is not restricted to linear jumps; the formalism allows for making predictions on the basis of any jump, linear or nonlinear. For VEC, only predictions for the storage data are presented.

In the following, details of the calculation of the various predictions are given. To clarify which parameters or data are connected to the linear data sets that serve as a basis for the predictions, the notation $R_{lin}$ is used for the normalized relaxation function for the linear jump of \SI{50}{\milli\kelvin} amplitude that serves as a basis for the prediction, while $R_X$ is the normalized relaxation function for the to-be-predicted data set. As the basis data is a linear data set, its time can be referred to as the material time $\xi$. The individual data points of the data set are addressed by the index $k$. After the transformation of the material time to match the to-be-predicted data, the transformed time-data is referred to as the predicted time, $t^{pred}$. The time data connected to the to-be-predicted data are referred to as laboratory time, $t^{lab}$.

\subsubsection{Predictions: Single temperature jumps}\label{sec:idvlJ}
The following procedure describes the calculation of a prediction for an individual jump from $T_i$ to $T_b$. The response amplitude of the prediction, $\Delta X(\xi_k)$, is calculated from the normalized relaxation function $R_{lin}(\xi_k)$ by
\be\label{eq:DXij}
\Delta X(\xi_k) = R_{lin}(\xi_k) \Delta X \left( \frac{1-R^{gl}_X(T_i,T_b)}{1-R^{gl}_{lin}}\right),
\ee
with the overall response amplitude $\Delta X = X_{eq}(T_i) - X_{eq}(T_b)$.\\

The time interval of the prediction, $dt_k$, is calculated according to the material-time aging formalism by
\be\label{eq:dtk}
dt_k = \frac{\gamma_{eq}(T_b)}{\gamma^{lin}_{eq}(T_b^{lin})} \exp \left[- \Lambda \hspace{3pt} \Delta X(\xi_k,\xi_{k+1}) \right] d\xi_k,
\ee
with $d\xi_k = \xi_{k+1} - \xi_k$ and $\Delta X(\xi_k,\xi_{k+1}) = \frac{\Delta X(\xi_k) + \Delta X(\xi_{k+1})}{2}$. The predicted time is given by
\be
t_k^{pred} = \sum_{k=1}^{N} dt_k.
\ee

\subsubsection{Predictions: Double temperature jumps}\label{sec:detDJpred}
The prediction of the response to a temperature double jump can be understood as the prediction of an individual jump for the initial jump of the double-jump protocol, and a superposition of two individual jumps made in material time for the second jump of the protocol. For a visualization of this procedure, see Fig.~\ref{Fig:NMECLinKov}.

The calculation of each of the double-jump predictions can be broken down into the two steps described above, i.e., the calculation of the response amplitude and the calculation of the time steps. Starting with the initial jump, the response amplitude of the prediction can be calculated as in Eq.~(\ref{eq:DXij}). Then, the time steps are transformed from material time to the time of the experiment as described in Eq. (\ref{eq:dtk}). However, this is not done for the complete jump, but is interrupted at the point where the second jump is initiated. This is when the predicted time, $t^{pred}$, is equal to the last point in laboratory time of the initial jump, $t^{lab}$, of the to-be-predicted data set, just before the second jump is initialized.

Next the prediction for the second jump can be calculated. Since the point in time of the jump is known ($t^{obs}$), it is also known at which material time $\xi$ the jump occurs, which is denoted by $\xi_{1}$. As the prediction of the second jump is calculated from the superposition of the predictions for the individual jumps in material time, the response of the total prediction is given by
\be\label{eq:DXtot}
\Delta X_{tot}(\xi_k) = \Delta X_{1}(\xi_k-\xi_1) +  \Delta X_{2}(\xi_k).
\ee

The transformation from material time to $t^{pred}$ is analogous to the approach described for the individual jumps in Eq.~(\ref{eq:dtk}), where $\Delta X$ is replaced by $\Delta X_{tot}$ from Eq.~(\ref{eq:DXtot}).

\subsubsection{Predictions: Multiple temperature jumps}
The prediction of a response that involves multiple jumps is very similar to the prediction of a double jump: the approach iteratively determines the response amplitude of a prediction as a superposition of the individual responses in material time. This is again followed by the transformation of the time step from material time to $t^{pred}$ in order to match the laboratory time of the to-be-predicted experiment. The only difference is that the superposition in material time involves several individual predictions:
\be\label{eq:DXtotMulti}
\Delta X_{tot}(\xi_k) = \sum_{i=1}^{N} \Delta X_{i}(\xi_k-\xi_i) \hspace{5pt} \text{for} \hspace{5pt} \xi_k > \xi_N
\ee
\clearpage
\section{Overview on the data set measured on NMEC}\label{sec:NMEC}
This section contains 
\begin{itemize}
	\item[a)] Spectra of the storage and loss dielectric permittivity
	\item[b)] Temperature-dependent data set of the storage and loss capacitance at $\nu = $\SI{10}{\hertz}
	\item[c)] Comparisons of experiments to predictions for single and double-jump temperature protocols in the linear ($|\Delta T| \leq $\SI{100}{\milli\kelvin}) and nonlinear ($|\Delta T| > $\SI{100}{\milli\kelvin}) regimes
\end{itemize}
\subsubsection{Spectra of storage and loss dielectric permittivity for NMEC}

\begin{figure}[h!]
	
	\begin{minipage}[t]{0.475\columnwidth}
		\flushleft
		\textbf{a)}
	\end{minipage}	
	\begin{minipage}[t]{0.475\columnwidth}
		\flushleft
		\textbf{b)}
	\end{minipage}
	
	\begin{minipage}{0.475\columnwidth}
		\centering
		\includegraphics[width=\columnwidth]{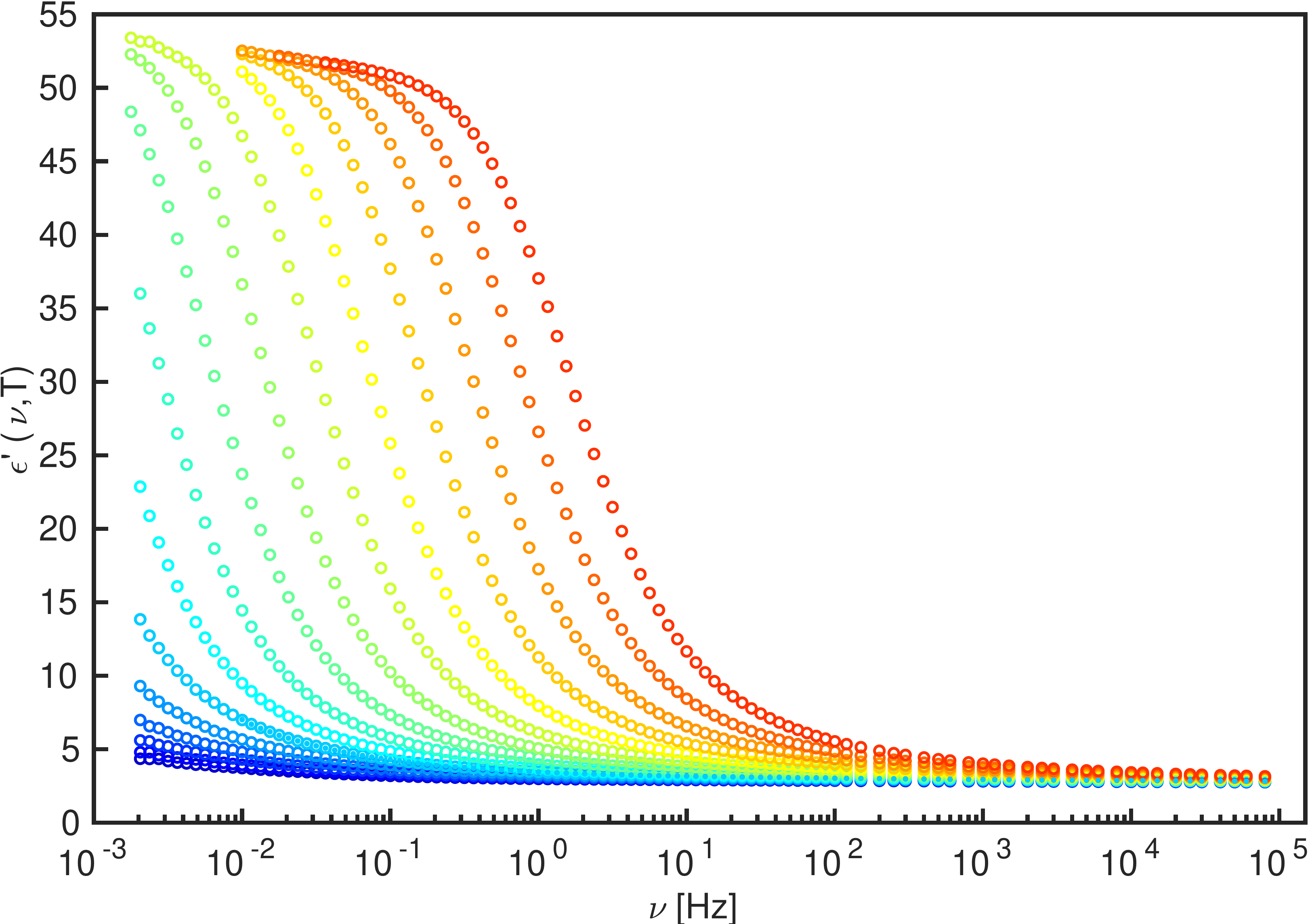}
	\end{minipage}
	\begin{minipage}{0.475\columnwidth}
		\centering
		\includegraphics[width=\columnwidth]{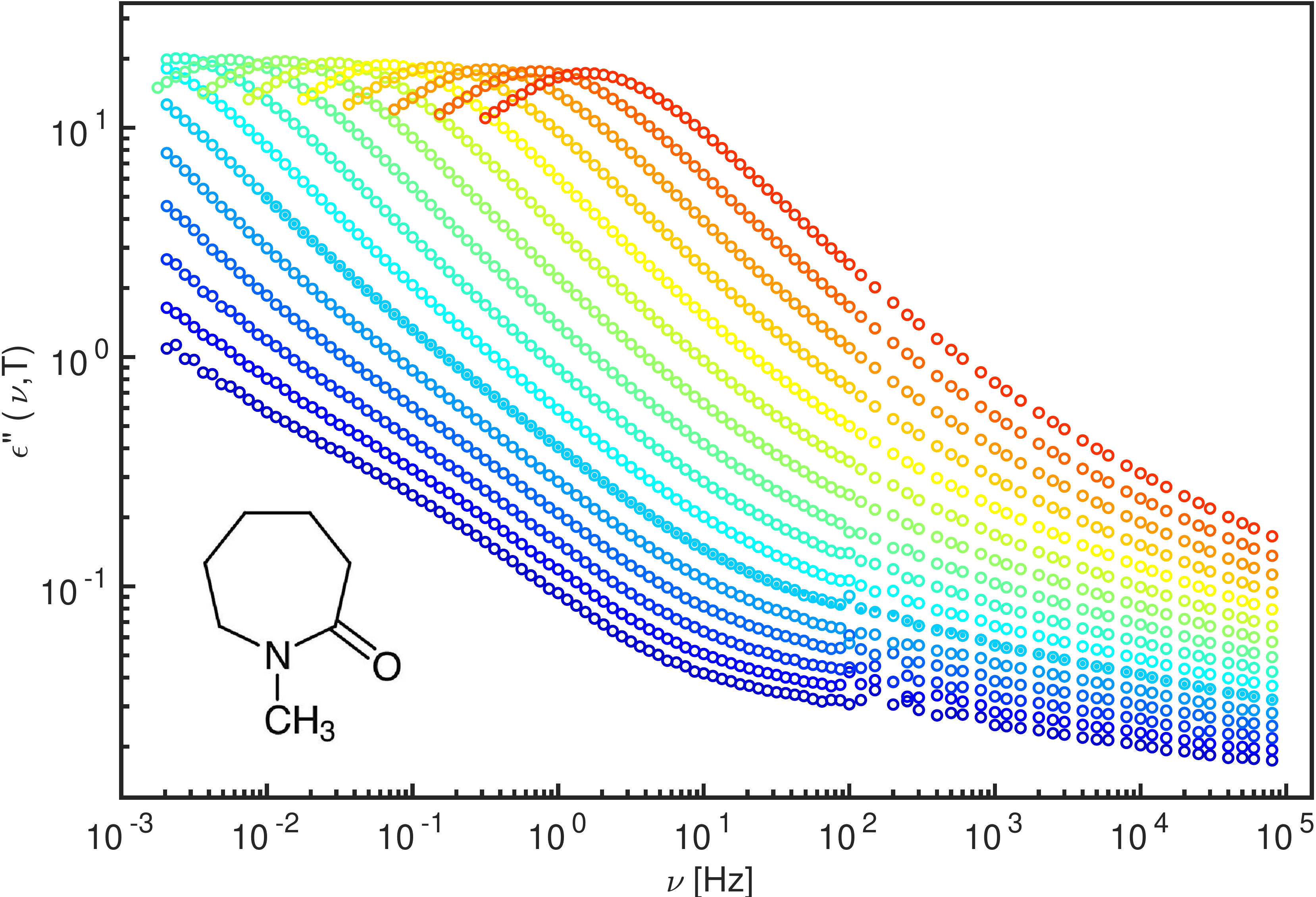}
	\end{minipage}
	\caption{Dielectric spectra for NMEC. Storage (a) and loss (b) capacitance as functions of frequency, measured by means of a custom-built frequency generator and a commercial LCR meter in the temperature range from \SI{165}{\kelvin} to \SI{180}{\kelvin} in \SI{1}{\kelvin} steps. The inset in panel (b) depicts the skeletal formula of NMEC.}
	\label{Fig:SpecNMEC}
\end{figure}

All data shown for NMEC were measured on a single sample measured by a capacitor with geometric capacitance $C_{geo} = $\SI{15.7}{\pico\farad}. The sample was initially quenched to $T_{cryo} = $\SI{167}{\kelvin} by insertion into the pre-cooled cryostat and was held at that temperature to equilibrate. Subsequently, the spectra presented in Fig.~\ref{Fig:SpecNMEC} were measured while tracking the temperature-specific voltage of the micro-regulator, which was then used to calibrate the temperature control of the micro-regulator.

\clearpage
\subsubsection{Storage and loss capacitance data for NMEC}
\begin{figure}[h!]	
	\centering
	\includegraphics[width=.95\columnwidth]{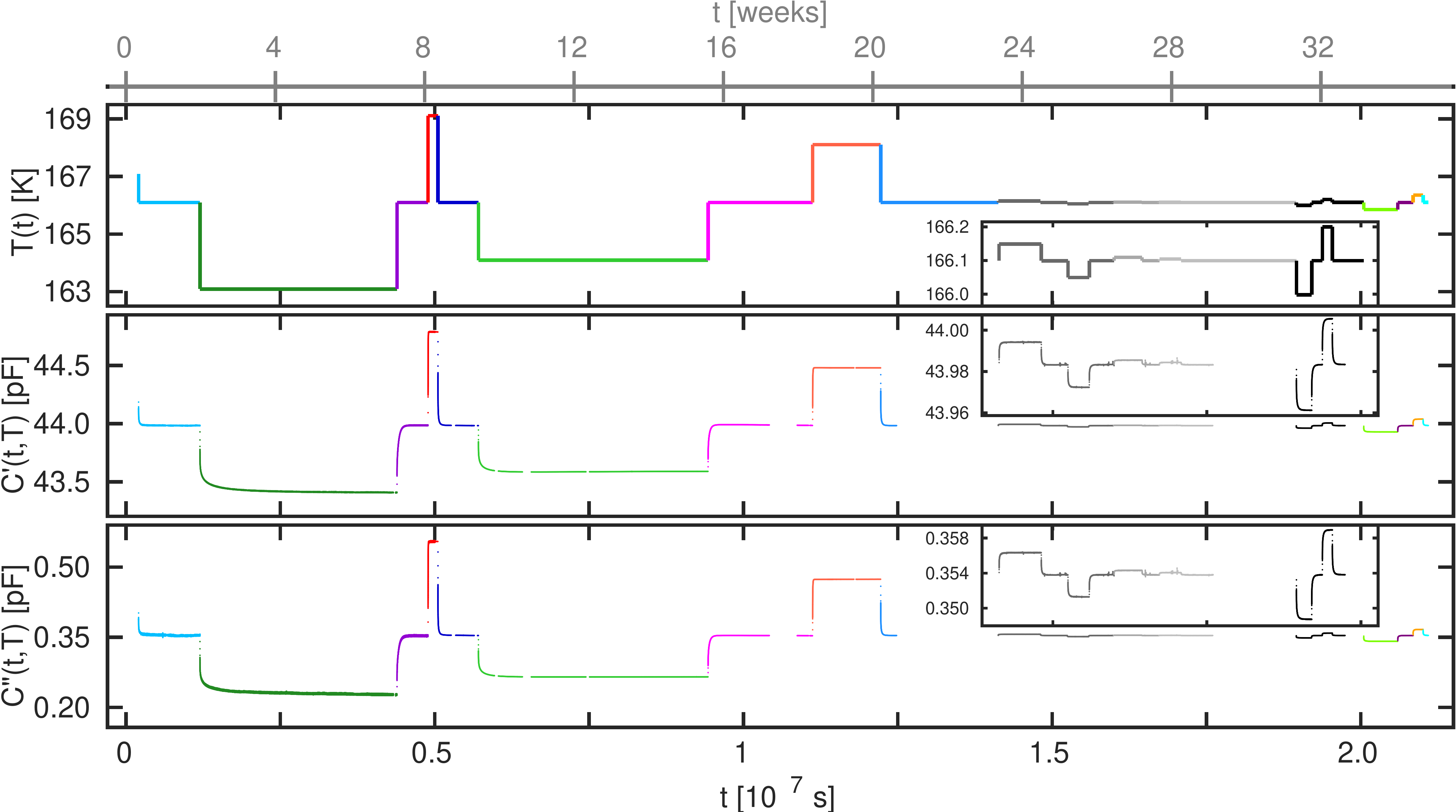}
	
	\caption{Overview of the temperature protocol and raw data of the full experiment on NMEC after a loop-wise drift correction. The temperature protocol of experiments realized by modulations around $T_{ref} = $ \SI{166.1}{\kelvin} is plotted in the upper panel together with the real and loss part of the measured capacitance $C (\nu = $\SI{10}{\hertz}$)$ in the lower panels, each as functions of time on a linear scale. Jumps larger than \SI{100}{\milli\kelvin} are colored while jumps of magnitude \SI{100}{\milli\kelvin} or less are depicted on a gray scale. The insets show details of jumps with $|\Delta T| \leq $\SI{100}{\milli\kelvin} and share the same time-axis as the main panels.}
	\label{Fig:NMEC_OV}
\end{figure}

After the measurements of the above spectra, the microregulator was activated and set to $T = T_{ref} = $\SI{166.1}{\kelvin}. By means of the Andeen-Hagerling capacitance bridge the capacitance, $C$, and the dielectric loss, $\tan \delta$, were tracked at the frequency $\nu = $\SI{10}{\hertz}. Individual temperature jumps with amplitudes from $|\Delta T| = $\SI{5}{\milli\kelvin} up to \SI{3}{\kelvin} were initiated, as well as double jumps with $\Delta T_1 = $+\SI{50}{\milli\kelvin} and $\Delta T_2 = -\Delta T_1/2$. An overview of the temperature protocol is shown in Fig.~\ref{Fig:NMEC_OV} together with the corresponding storage and loss capacitance. The first four jumps were measured with the dielectric setup used for the measurements of the spectra in Fig.~\ref{Fig:SpecNMEC}. These data emphasize the significantly higher accuracy of measurements with the Andeen-Hagerling AH2700 used for obtaining the aging data. After approximately \num{35} weeks of consecutive measurements, a failure in the cooling unit of the the cryostat unfortunately terminated the measurement sequence.

The analysis of the capacitance data for NMEC mainly follows the procedure described for VEC in the previous section. However, for NMEC only a loop-wise drift correction was performed, individually corrected data was not generated.

Note the difference in signal-to-noise ratio for the conventional dielectric setup and the Andeen-Hagerling instrument that becomes obvious in the fourth jump of the data set ($T_{ref}$ to $T_{ref} + $\SI{3}{\kelvin}, red color), where the measurement setups were switched.

\begin{figure}[h!]
	
	\begin{minipage}[t]{0.475\columnwidth}
		\flushleft
		\textbf{a)}
	\end{minipage}	
	\begin{minipage}[t]{0.475\columnwidth}
		\flushleft
		\textbf{b)}
	\end{minipage}
	
	\centering
	\includegraphics[width=.95\columnwidth]{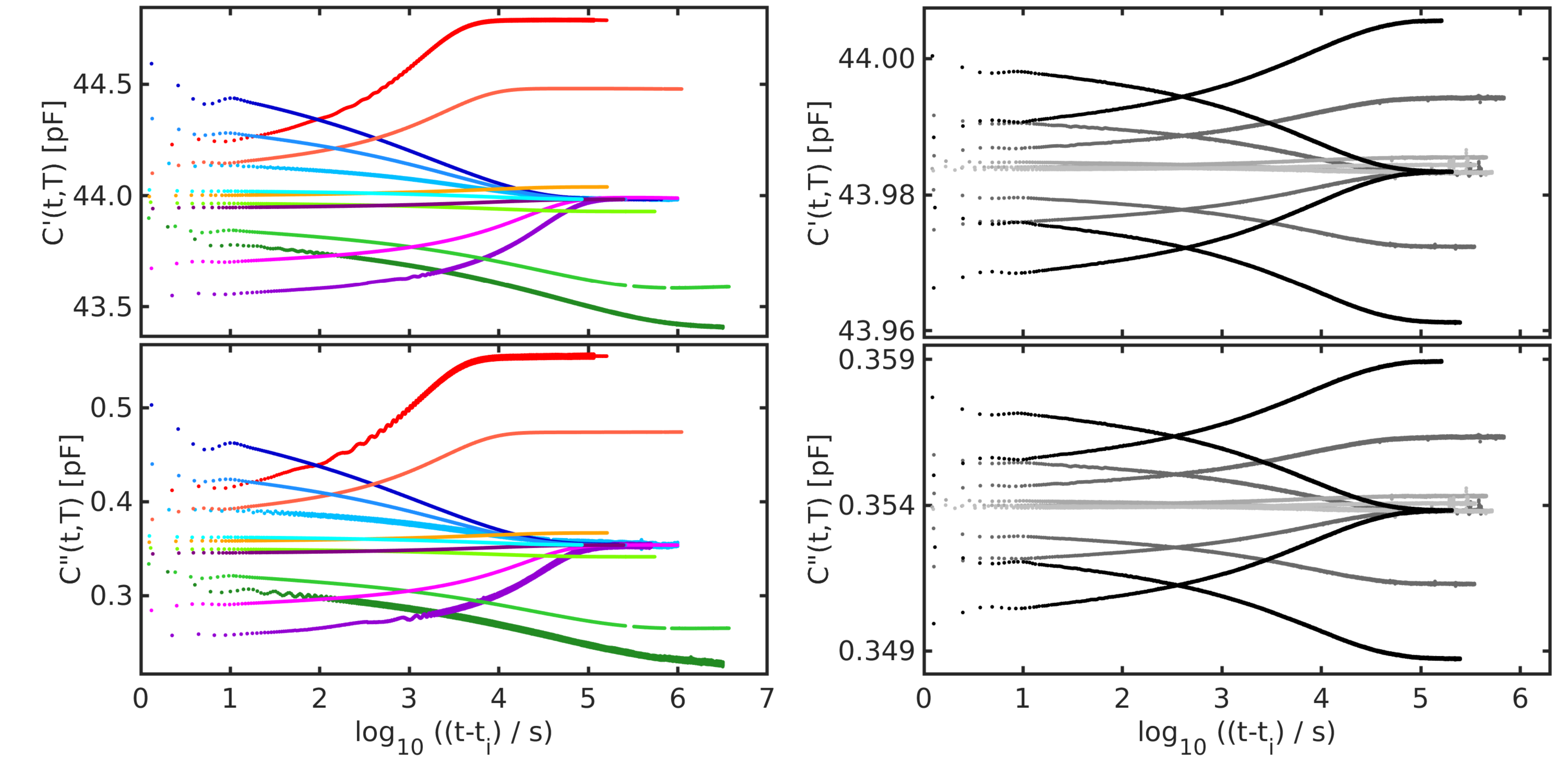}
	
	\caption{The storage and loss capacitance data, $C'(\nu = $\SI{10}{\kilo\hertz}$)$ and $C"(\nu = $\SI{10}{\kilo\hertz}$)$, plotted as functions of the logarithm of the time that has passed after the initiation of each jump at $t_i$. Panel (a) depicts jumps with $|\Delta T| > $\SI{100}{\milli\kelvin} and panel (b) is a zoom on jumps of magnitude \SI{100}{\milli\kelvin} or less.}
	\label{Fig:NMECjumps}
\end{figure}

\clearpage
The response to individual temperature jumps is plotted against logarithmic time in Fig.~\ref{Fig:NMECjumps}. The initial \SI{10}{\second} of each jump are clearly still influenced by the temperature regulation and were excluded from the analysis. After this thermalization, the initial plateau is only captured for jumps from $T < T_{ref}$ to higher temperatures (purple colors in Fig.~\ref{Fig:NMECjumps}a); for all other jumps a significant amount of relaxation has taken place, implying that the data set on NMEC does not allow for an analysis of the glassy contribution. Thus, this analysis was omitted in case of NMEC.

Fig.~\ref{Fig:NMECjumps}b depicts the individual temperature jumps with temperature amplitudes between  \SI{5}{\milli\kelvin} and  \SI{100}{\milli\kelvin} that correspond to the data magnified in the insets of Fig.~\ref{Fig:NMEC_OV}.

\begin{figure}[h!]	
	\centering
	\includegraphics[width=.8\columnwidth]{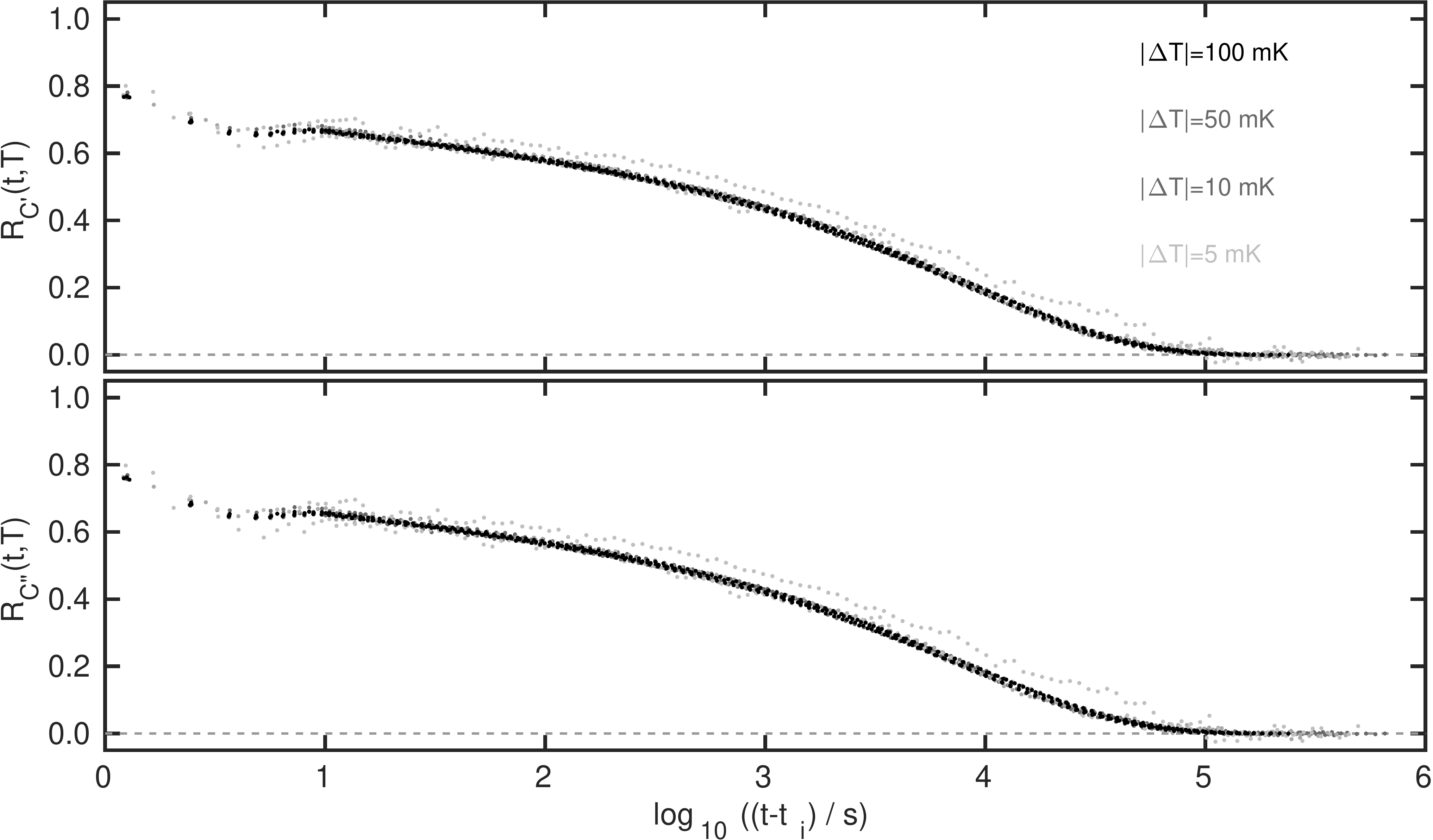}
	
	\caption{Normalized relaxation function for the real and loss parts of the capacitance data, $R_{C'}$ and $R_{C"}$, of single temperature jumps with amplitudes between \SI{5}{\milli\kelvin} and \SI{100}{\milli\kelvin}.}
	\label{Fig:NMEC_Rlin}
\end{figure}

\clearpage
In Fig.~\ref{Fig:NMEC_Rlin} the linear aging data are plotted in the normalized representation. Data with amplitudes down to \SI{10}{\milli\kelvin} collapse as expected for linear relaxation. The responses to \SI{5}{\milli\kelvin}-jumps show deviations from the master curve, even though the data still follow the evolution of the collapsed data. This deviation is clearly influenced by thermal fluctuation.

\clearpage
\subsubsection{Comparison of experimental data to predictions for NMEC}

\begin{figure*}[h!]
	
	\begin{minipage}[t]{0.6\columnwidth}
		\flushleft
		\textbf{a)}
	\end{minipage}	
	\begin{minipage}[t]{0.375\columnwidth}
		\flushleft
		\textbf{b)}
	\end{minipage}
	
	\begin{minipage}{.6\textwidth}
		\centering
		\includegraphics[trim = 2cm 0cm 15cm 10cm, clip=true, width=.85\textwidth]{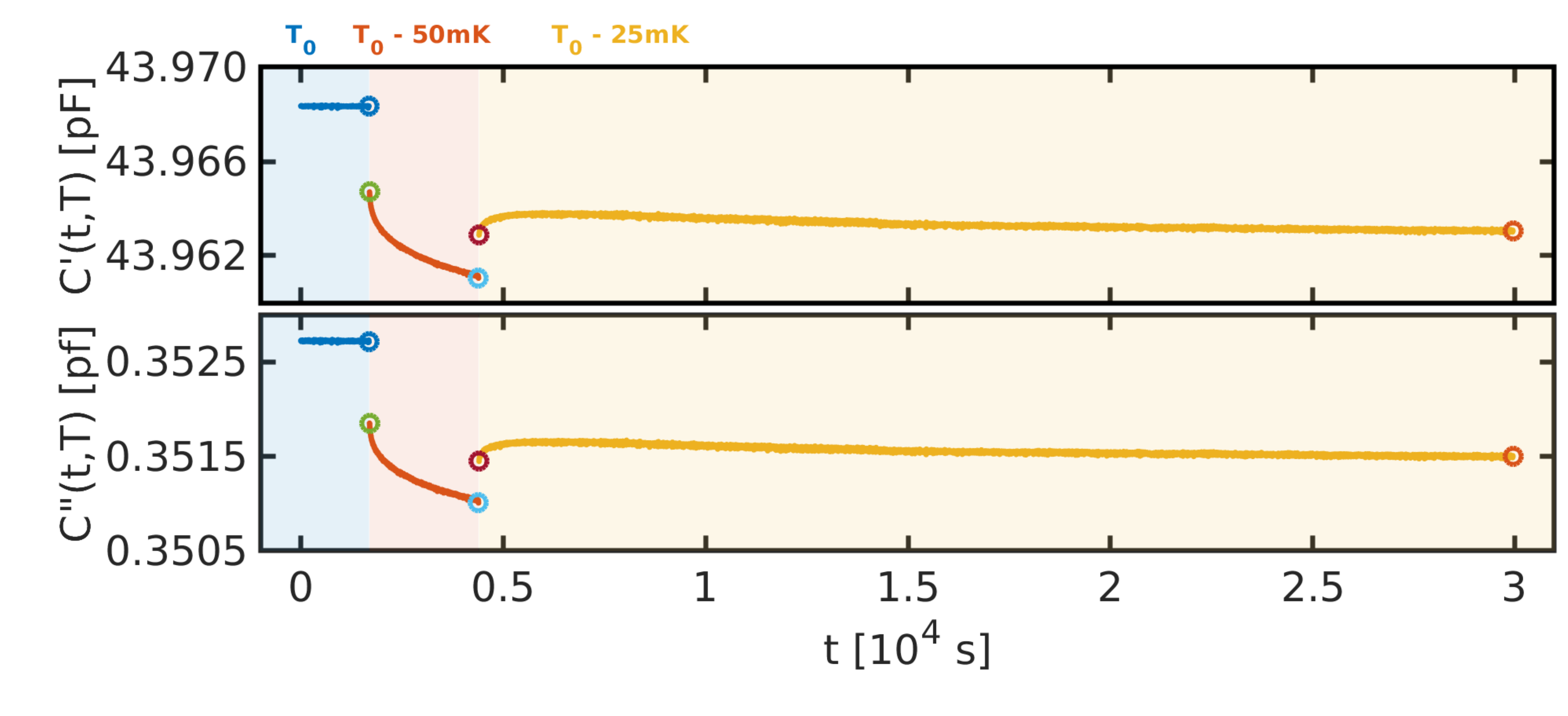}
	\end{minipage}
	\begin{minipage}{0.3\textwidth}
		\centering
		\includegraphics[width=0.85\textwidth]{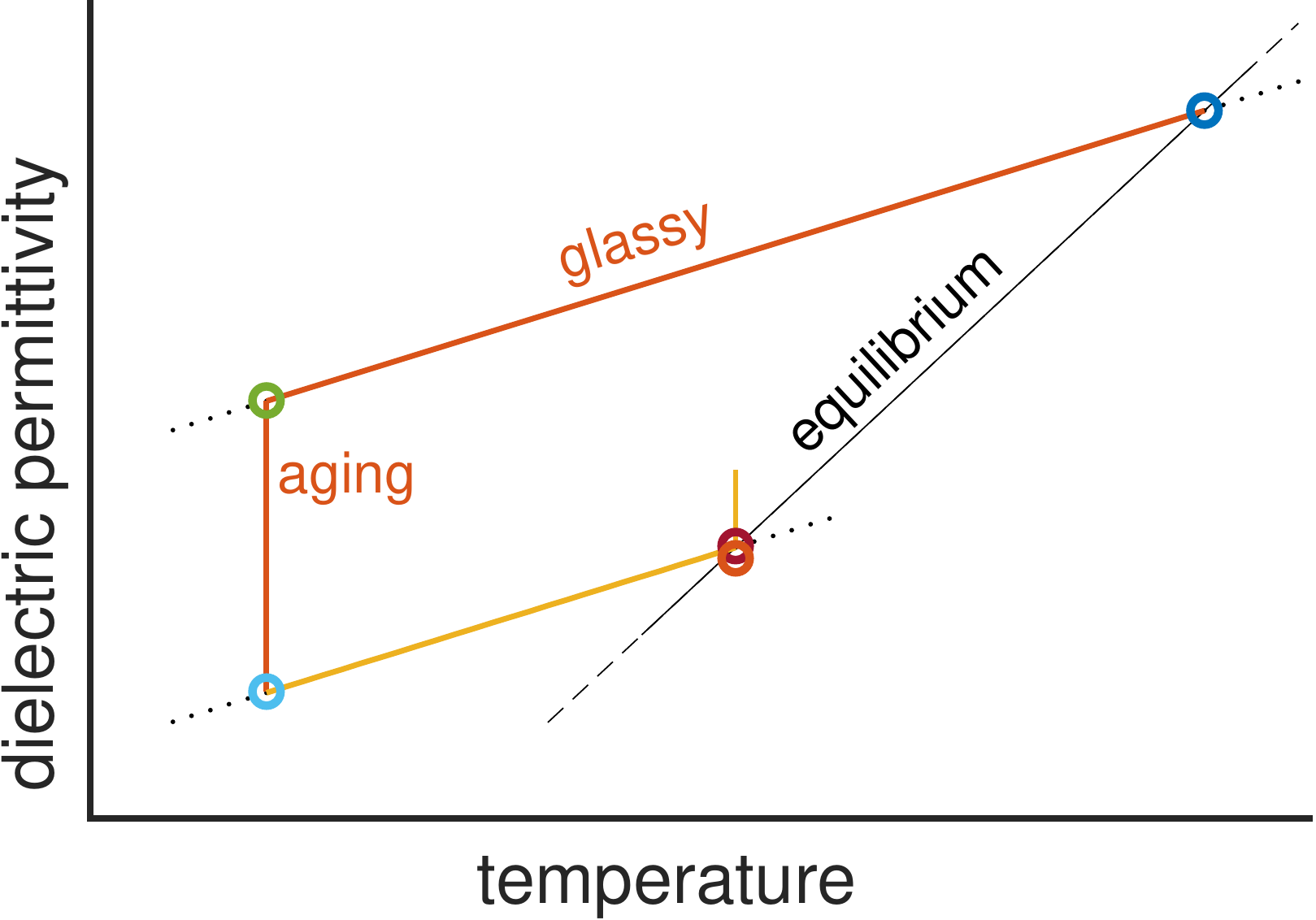}
	\end{minipage}
	
	\begin{minipage}[t]{0.6\columnwidth}
		\flushleft
		\textbf{c)}
	\end{minipage}	
	\begin{minipage}[t]{0.375\columnwidth}
		\flushleft
		\textbf{d)}
	\end{minipage}
	
	\begin{minipage}{.6\textwidth}
		\centering
		\includegraphics[trim = 2.5cm 0cm 9.75cm 1cm, clip=true, width=0.95\textwidth]{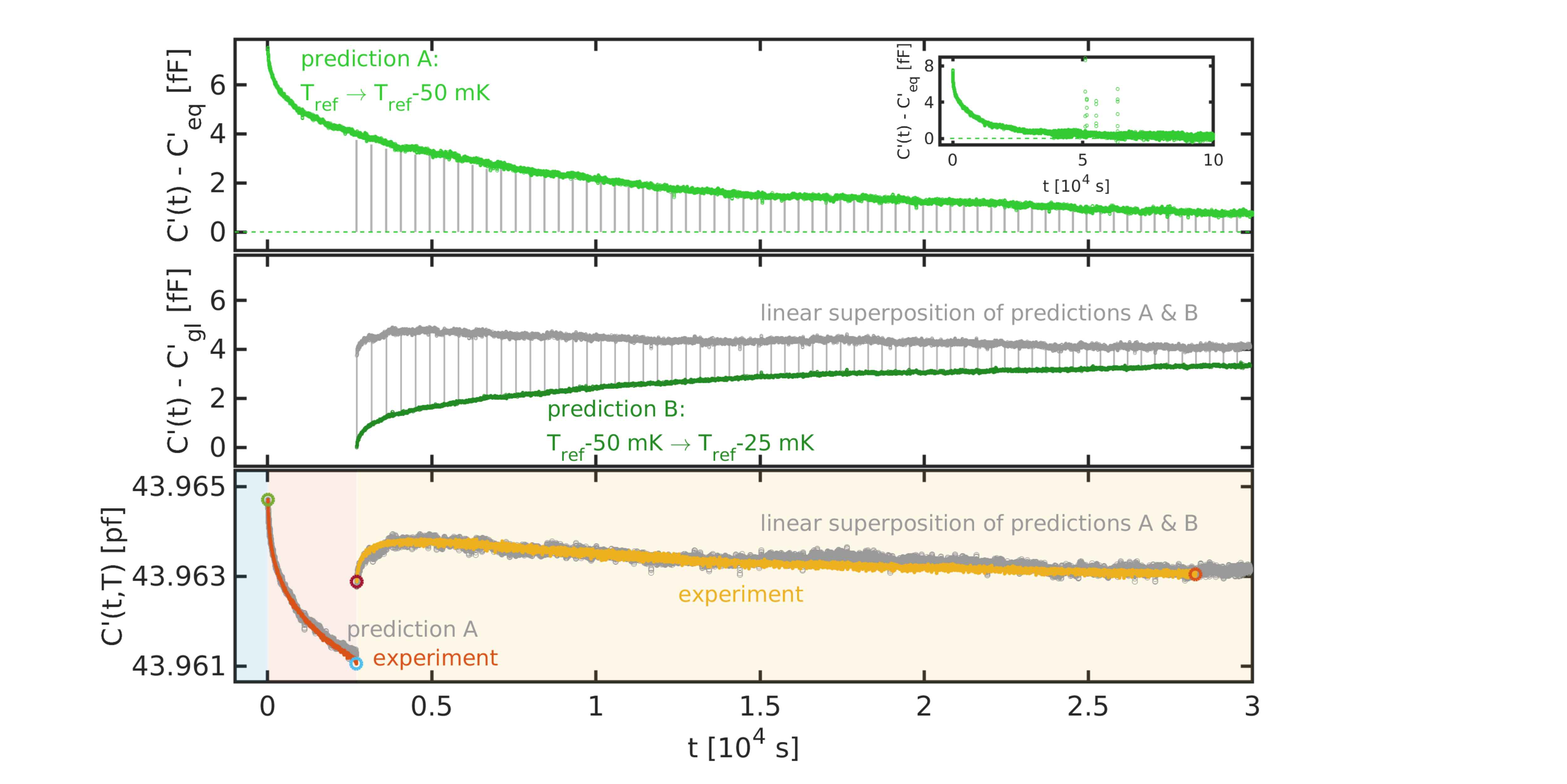}
	\end{minipage}
	\begin{minipage}{.375\textwidth}
		\centering
		\includegraphics[trim = 3cm 0cm 27.5cm 0cm, clip=true, width=.95\textwidth]{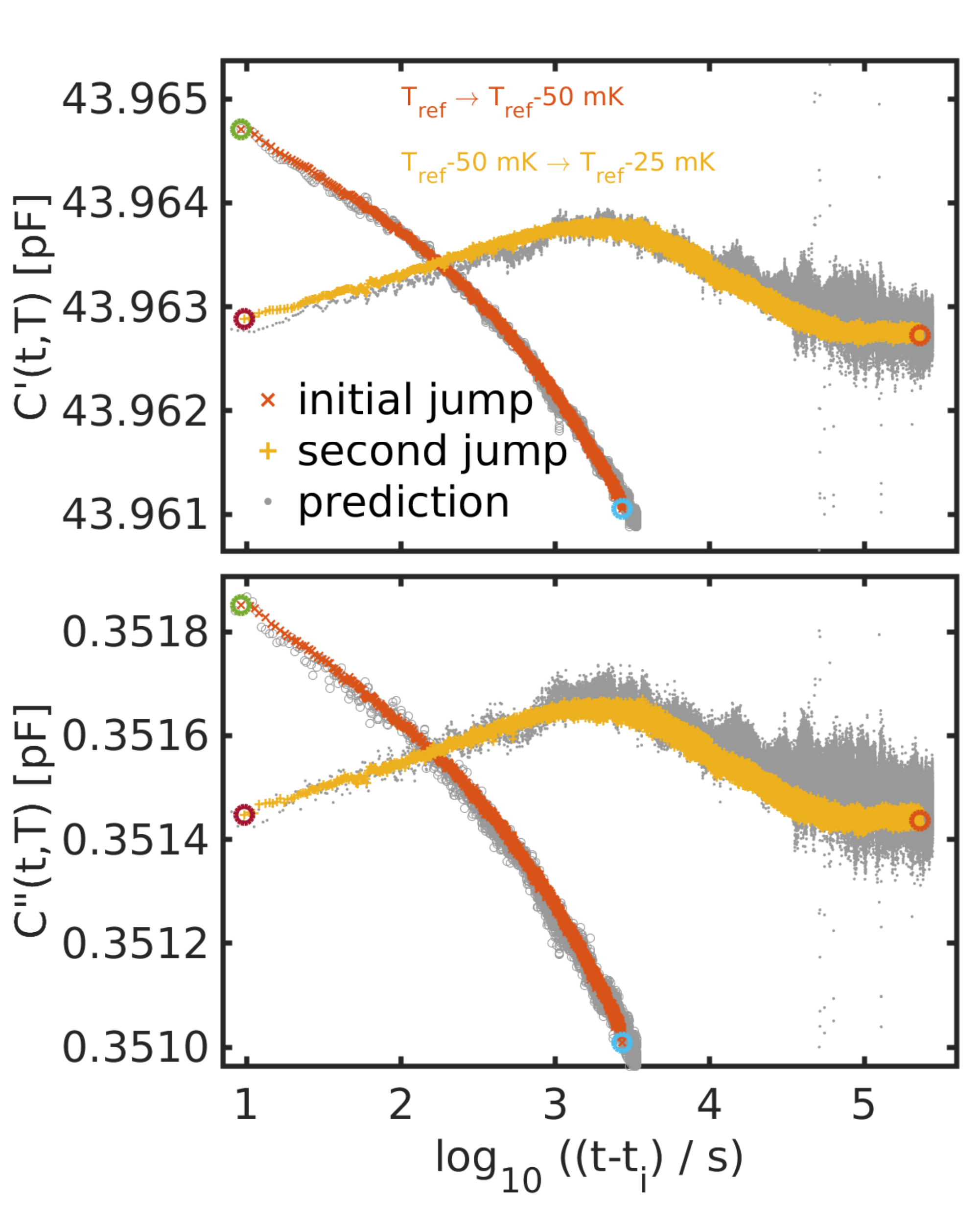}
	\end{minipage}
	
	\caption{Experimental data and prediction of a linear double jump on NMEC with $\Delta T_1 = $+\SI{50}{\milli\kelvin} and $\Delta T_2 = $-\SI{25}{\milli\kelvin}. (a) The data given as a function of linear time. (b) Schematic representation of a crossover experiment visualizing the glassy and aging contributions to the capacitance for two subsequent temperature jumps. Colored circles mark identical experimental situations for the different representations of the data (throughout panel (a) to panel (d). (c) Prediction of the experimental response based on an individual, independently measured linear temperature jump. Top panel: Prediction of the dielectric aging response to a temperature jump from $T_{ref}$ to $T_{ref} - $\SI{50}{\milli\kelvin} as a function of logarithmic time based on an individual jump from $T_{ref} + $ \SI{10}{\milli\kelvin} to $T_{ref}$, after subtraction of the equilibrium value $X_{eq}$. Inset top panel: Prediction of the aging response as a function of time, depicting as in the main panel the complete response towards equilibrium. Middle panel: Prediction of the dielectric aging response to a temperature jump from $T_{ref} - $\SI{50}{\milli\kelvin} to $T_{ref} - $\SI{25}{\milli\kelvin} as a function of time based on the same data as in the prediction of the top panel, after subtraction of the glassy contribution to the jump. Grey vertical lines reflect the same $\Delta C'(t)$ values at a given point in time from the top panel to visualize the linear superposition of the two predicted curves that yield the prediction of the dielectric response of the second jump (dark grey data points). Bottom panel: Experimental response (storage data) plotted together with the prediction from the middle panel. (d) Collapse of predictions and experimental data for both storage and loss contributions, illustrating the applicability of linear superposition in the linear-response limit.}	
	\label{Fig:NMECLinKov}
\end{figure*}

Fig.~\ref{Fig:NMECLinKov} visualizes the superposition of individually predicted jumps to yield the overall prediction for the double-jump temperature protocol. This example covers a double jump in the linear regime with $\Delta T_1 = $+\SI{50}{\milli\kelvin} and $\Delta T_2 = $-\SI{25}{\milli\kelvin}. In (a) the storage and loss capacitance are plotted against linear time. (b) reflects distinct points in the thermal protocol, such as the initial equilibrium, the glassy response contribution, the initiation of the second jump, and the final equilibrium, in a plot of the dielectric response as a function of temperature. Storage data from (a) are presented in the lower panel of (c) in comparison to the calculated prediction. (d) shows a comparison of storage and loss capacitance from (a) to predictions on a logarithmic time scale. More details are given in the figure caption.

\clearpage
In order to predict the response to temperature jumps larger than \SI{100}{\milli\kelvin} not only the equilibrium reponse value $X_{eq}$ and the response to a linear jump (\SI{50}{\milli\kelvin} up-jump to $T_{ref}$) is used, but also the single parameter $\Lambda$ and the equilibrium clock rate $\gamma_{eq}$. As for VEC and as used for the prediciton of linear double jumps, $X_{eq}$ was determined for the storage and loss capacitance analogous to the procedure described for VEC in section~\ref{Sec:equilR}.\\

\begin{figure}[h!]	
	\centering
	\includegraphics[width=.8\columnwidth]{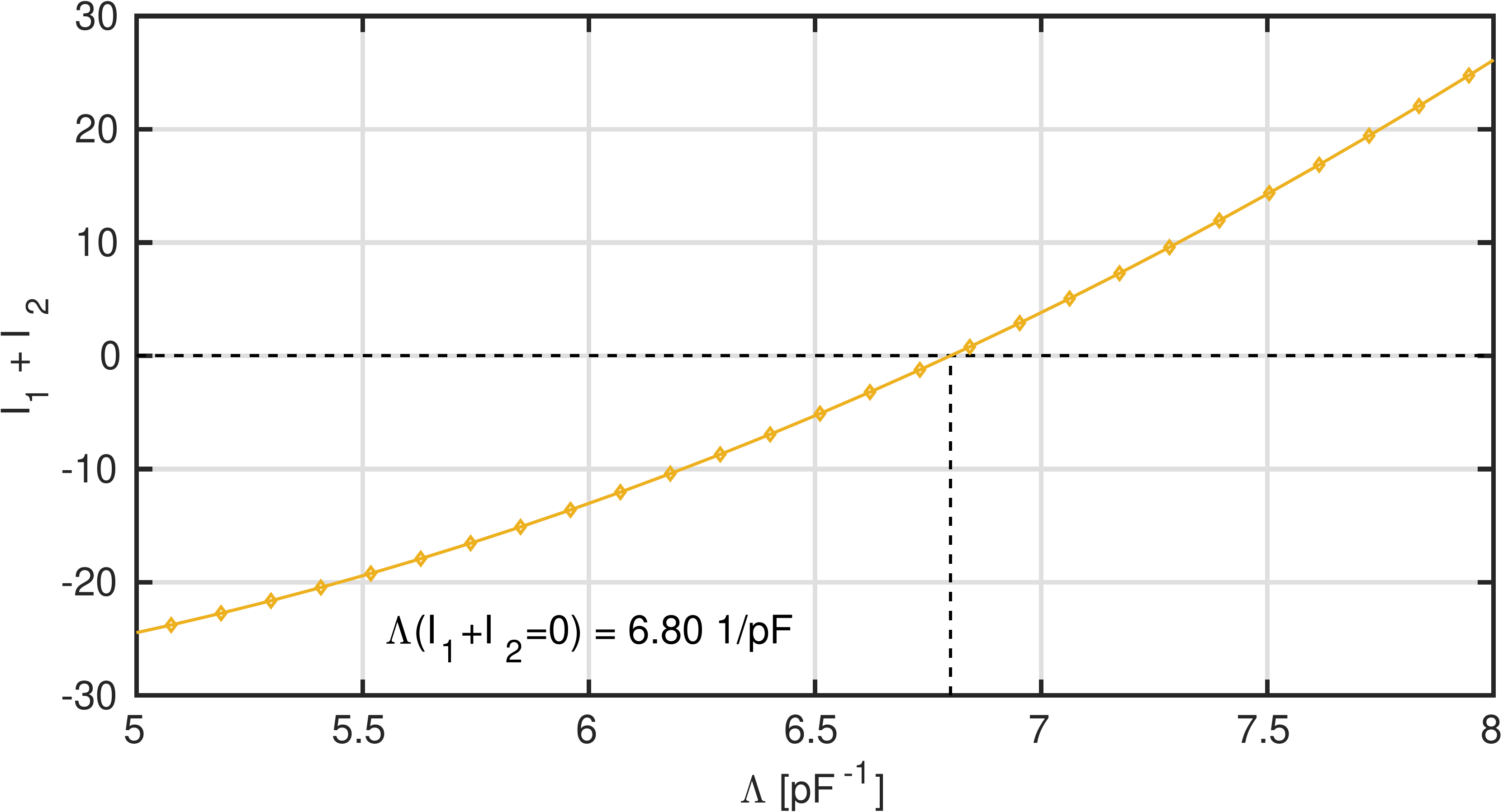}
	
	\caption{Determination of $\Lambda$. The sum of integrals, $I_1 + I_2$, plotted as a function of $\Lambda$ derived from loop-wise corrected data of a pair of temperature jumps of \SI{100}{\milli\kelvin} amplitude towards the reference temperature, $T_{ref}$.}
	\label{Fig:Xconst_NMEC}
\end{figure}

The parameter $\Lambda$ for NMEC was determined by the integral criterion as described in section~\ref{Sec:LambdaVEC} for VEC. Fig.~\ref{Fig:Xconst_NMEC} visualizes the minimization of the sum of integrals $I_1+I_2$ for the case of storage data, resulting in a value of $\Lambda_{C'} =$~\SI{6.80}{\per\pico\F}.

\begin{figure}[h!]	
	\centering
	\includegraphics[width=.8\columnwidth]{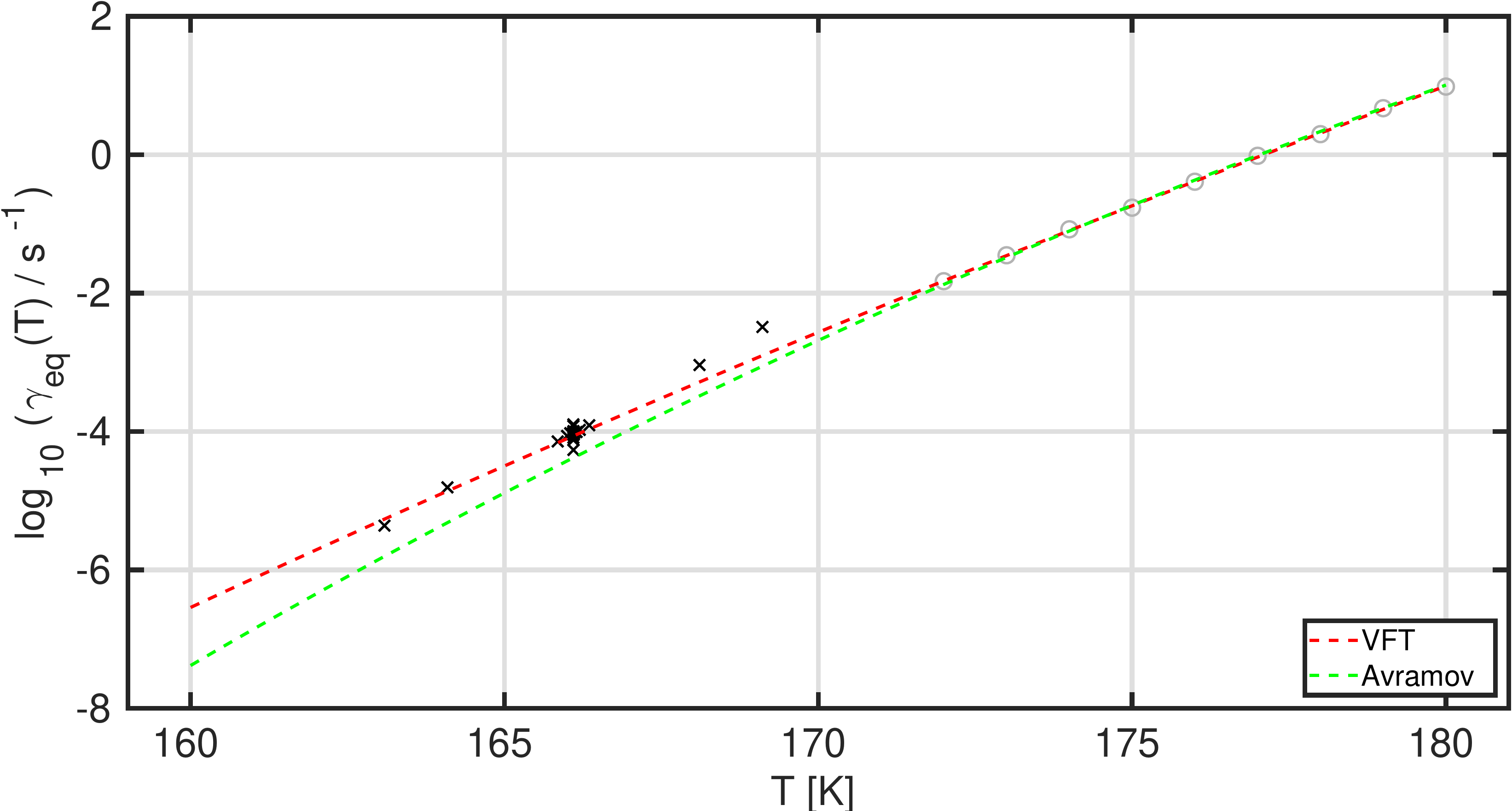}
	
	\caption{Logarithmic normalized equilibrium clock rate, $\gamma_{eq,N}$, as a function of temperature. Dashed lines correspond to extrapolations based on the clock rates determined from spectral data (grey circles), crosses correspond to clock rates from fits to the storage contribution of individual jumps.}
	\label{Fig:Gamma_NMEC}
\end{figure}


The equilibrium clock rate is based on an extrapolation from the fit of relaxation times derived from dielectric spectra to a Vogel-Fulcher-Tammann function, analogous to the description given in Sec.~\ref{Sec:clora} for VEC. However, clock rates determined from fits to the storage response of individual jumps at the highest investigated temperatures do not collapse with the VFT-extrapolation based on clock rates derived from spectra, as visualized in Fig.~\ref{Fig:Gamma_NMEC}.\\

\begin{figure*}[h!]
	\begin{minipage}{.495\textwidth}
		\flushleft
		\textbf{a)}
	\end{minipage}
	\begin{minipage}{.495\textwidth}
		\flushleft
		\textbf{b)}
	\end{minipage}	
	\begin{minipage}{.495\textwidth}
		\centering
		\includegraphics[width=.9\columnwidth]{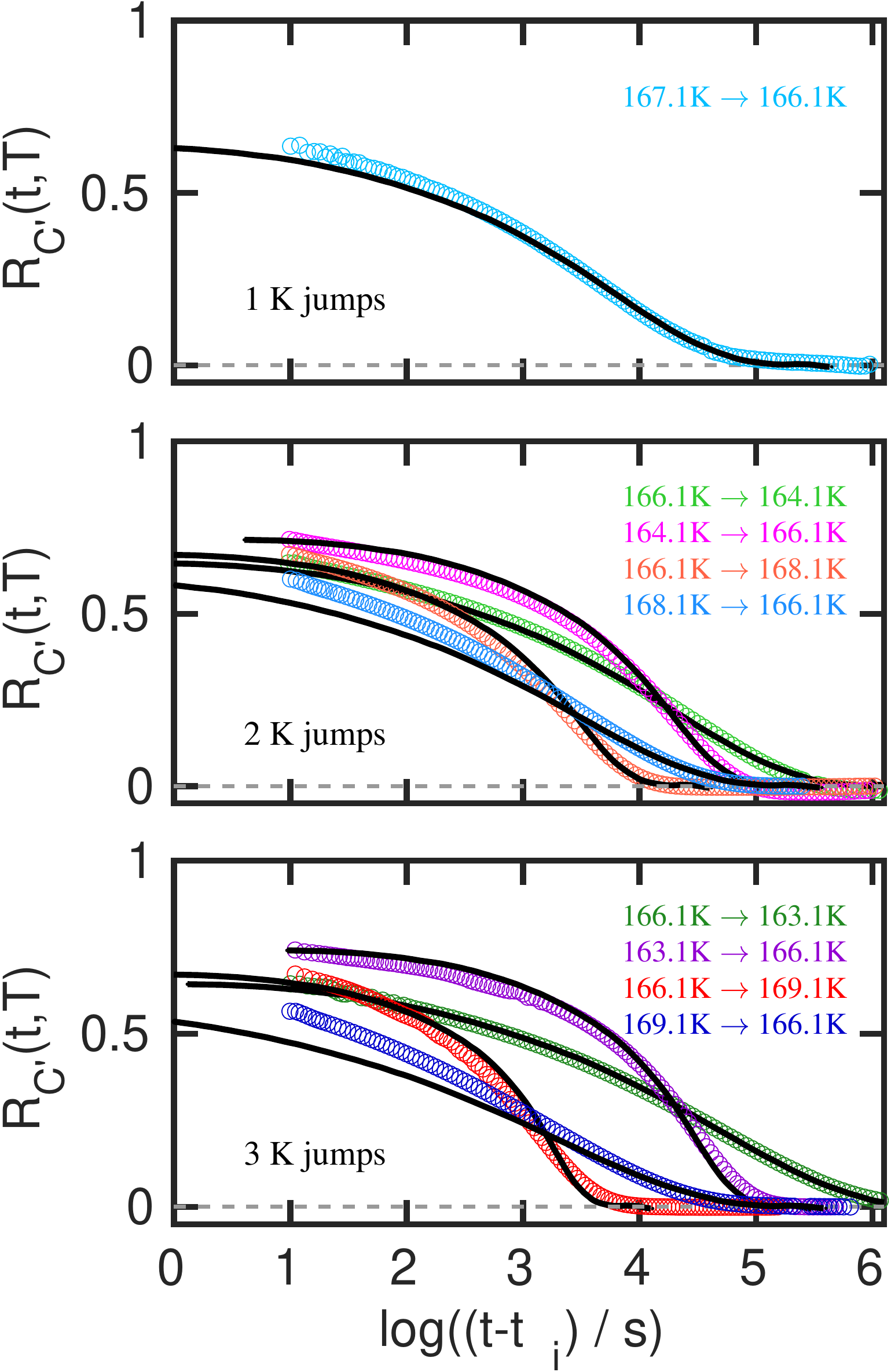}
	\end{minipage}
	\begin{minipage}{.495\textwidth}
		\centering
		\includegraphics[width=.9\columnwidth]{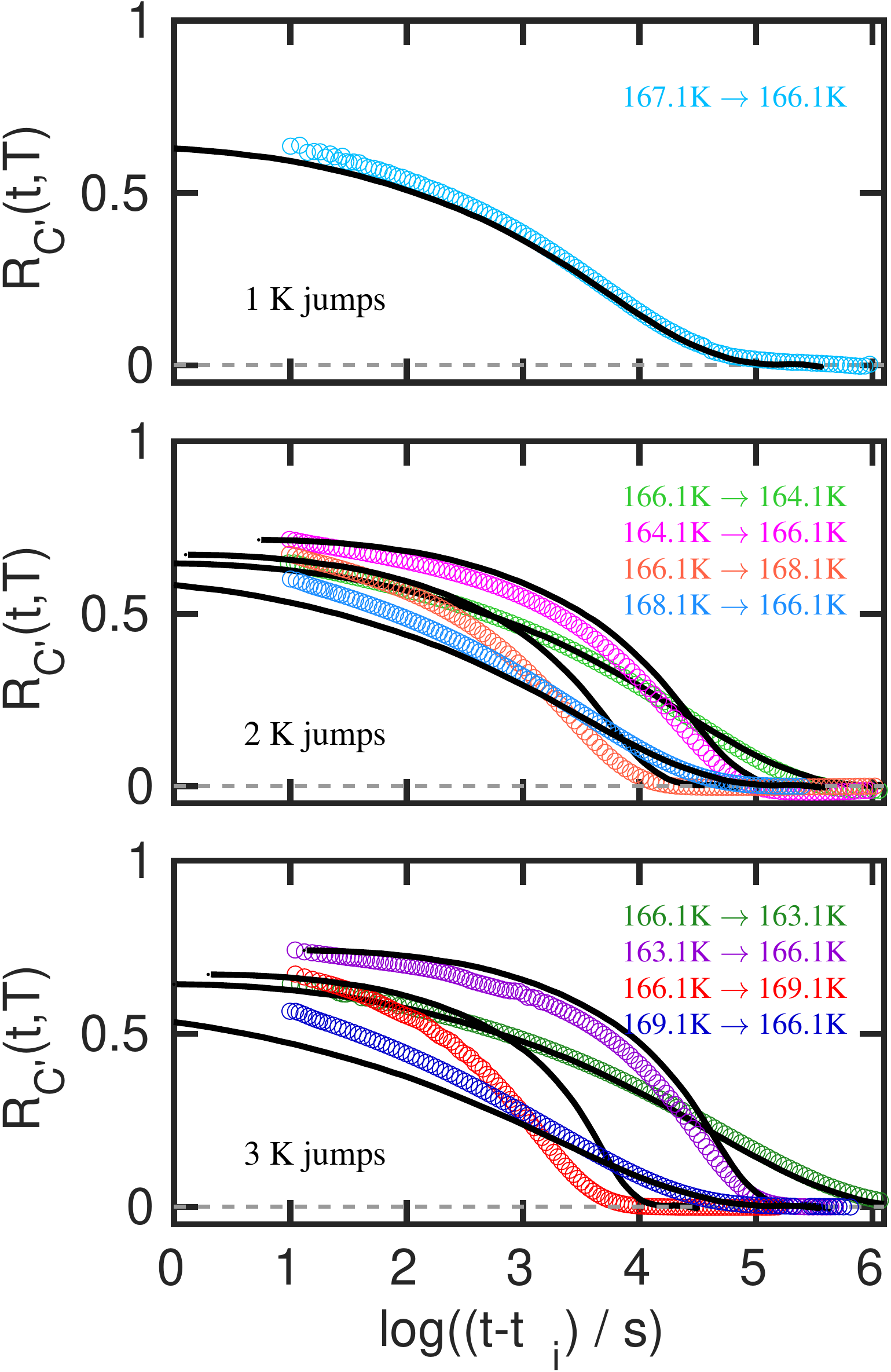}
	\end{minipage}

	\caption{Experimental data of nonlinear individual jumps on NMEC involving temperature jumps between \SI{1}{\kelvin} and \SI{3}{\kelvin}. The predictions (black lines) are based on different parameter sets: a) predictions are based on the glassy contribution as observed for the experimental data for individual jumps and with clock rates from fits to the experimental data for individual jumps. b) predictions are based onthe glassy contribution as observed for the experimental data for individual jumps as in a), but clock rates are extrapolation by the Vogel-Fulcher-Tammann function from loss-peak positions determined from spectral data.}
	\label{Fig:NMEC_NLpred}
\end{figure*}

The storage contribution of the capacitance for individual jumps with amplitudes larger than \SI{100}{\milli\kelvin}, i.e., nonlinear individual jumps, are plotted in Fig.~\ref{Fig:NMEC_NLpred} together with predictions. As described before, predictions are based on a linear response data set $R_{lin}$, equilibrium response values, $C'_{eq}$, the equilibrium clock rate, and the single parameter, $\Lambda_{C'}$. Again, predictions are based on a \SI{50}{\milli\kelvin} up-jump as used for the linear case as in Fig.~\ref{Fig:NMECLinKov}. Equilibrium response values are interpolated analogous to the procedure described for VEC, and also the single parameter $\Lambda$ is derived in analogy as shown in Fig.~\ref{Fig:Xconst_NMEC}. In order to take varying levels of the glassy response contribution into account, the as-measured levels were applied instead of making an interpolation, as the initial plateau value at short times is not well resolved for most individual jumps for NMEC. As shown in Fig.~\ref{Fig:Gamma_NMEC}, the values of the clock rate that are derived from fits of the individual jumps do not follow the tested extrapolations as nicely as the for VEC. In Fig.~\ref{Fig:NMEC_NLpred} the outcome for predictions based on fitted (a) and extrapolated clock rates (b) are visualized. While the fitted clock rates yield a better collapse between predictions and experimental data, the extrapolation of clock rates by the VFT-function yields qualitatively good results but more deviations than VEC-predictions, with strongest differences for jumps from the reference temperature upwards, similar to the trend observed for VEC.

\clearpage

\section{Computer simulations}\label{CompSim}

\subsubsection{Model}

The simulations studied the 80/20 Kob-Andersen binary Lennard-Jones (KABLJ) mixture \cite{ka1}, which was simulated with $NVT$ Nose-Hoover dynamics \cite{nose} using the GPU-optimized software RUMD \cite{RUMD}. The potential of KABLJ is $v_{ij}(r) = [(\sigma_{ij}/r)^{12} - (\sigma_{ij}/r)^{6}]$ ($i,j=A,B$) with $\sigma_{AA} = 1$, $\sigma_{AB} = 0.80$, $\sigma_{BB} = 0.88$, and
$\epsilon_{AA} = 1$, $\epsilon_{AB} = 1.5$, $\epsilon_{BB} = 0.5$. The masses are all unity and a system of $N$ = 8000 particles was simulated. In LJ units the time step was $\Delta t$ = 0.0025. All pair potentials were cut and shifted at $r_{\rm c}$ = 2.5$\sigma_{ij}$.\\

\subsubsection{Auto-correlation function}

At the reference temperature $T=0.60$ the potential-energy time-autocorrelation function was calculated as follows. First $10^7$ time steps  were carried out for equilibration where equilibration was confirmed from two consecutive runs comparing the self-part of the intermediate scattering function. After that a run of $5\times 10^{6}$ time steps was carried out dumping the potential energy every 32 time steps. The auto correlaton function was calculated using the Fast Fourier Transformation method as implemented in RUMD (see also Ref. TILDE for more information).\\

\subsubsection{Aging}

The temperature jump simulations were carried out by the following procedure applied for all starting temperatures. First, $5 \times 10^8$ time steps were spent on equilibration at the given starting temperature. After that 1000 configurations were generated from a $5 \times 10^{8}$ simulation dumping configurations every $2^{19}$ time steps. This ensures that the configurations are statistically independent at the lowest temperature $T$ = 0.50 studied. For each of these 1000 configurations an aging simulation of $10^{6}$ time steps was performed. During the aging, the potential energy was dumped every 8 time steps. The curves shown in the main text are averages over these 1000 aging simulations.\\

\subsubsection{Prediction}

The fluctuation-dissipation theorem in the particular version for the dynamic specific heat \cite{nie96} was used to calculate the aging response to a very small temperature perturbation $\delta T$ according to

\begin{equation}
\delta U(t) = \frac{\delta T}{k_{\rm B}T_{0}^{2}}\Big(\langle (\Delta U)^{2}\rangle_{0} - \langle \Delta U(0)\Delta U(t)\rangle_{\rm 0} \Big),
\end{equation}
where the subscript denotes the initial temperature $T_{0}$, $\Delta U(t) = U(t) - U_{\rm 0}$, and $\delta U$ gives the change in potential energy from the initial state point; the first term corresponds to integrating the heat capacity. From this equation we see that $R(t)$ of the linear perturbation is identical to the normalized potential energy auto-correlation function $C_{\rm U}(t)$. 

In terms of the nonlinearity parameter $\Lambda$, the relevant equation for predicting the nonlinear aging curve (2) from the equilibrium time-autocorrelation function is

\begin{equation}
t_{2} = \int_{0}^{t_{1}}\exp\Big[ - {\Lambda}\,\delta U(0)_{2}\,C_{U}(t_{1}^{*}) \Big]dt_{1}^{*}\,.
\end{equation}
Here we have used $\delta U(0)_{1} = 0$ in the general equation involving two jumps \cite{Hecksher2015a}, corresponding to jump 1 being infinitesimal. $\Lambda$ was determined from the two smallest jumps to the reference temperature $T$ = 0.60, i.e., those of magnitude 0.05, using as above for the experimental data the integral criterion of Ref. \onlinecite{Hecksher2015a}. This results in $\Lambda$ = 24.46 .




\end{document}